\documentclass[twocolumn,showkeys,groupedaddress,superscriptaddress,amsmath,amssymb]{revtex4-1}

\usepackage{natbib}

\usepackage{graphics}
\usepackage{graphicx} 
\usepackage{epsfig}
\usepackage{bm} 
\usepackage{amssymb}
\usepackage{amsmath}
\usepackage{amsthm}
\usepackage{wrapfig}
\usepackage{float}
\usepackage{textcomp}
\usepackage{pifont}  
\usepackage[svgnames]{xcolor}
\usepackage{array}
\usepackage[rightcaption]{sidecap}
\usepackage{multirow}
\usepackage{array}
\usepackage{multirow}
\usepackage{rotating}
\usepackage{enumerate}
\usepackage{hyperref}
\hypersetup{
    unicode=true,          
    pdftoolbar=true,        
    pdfmenubar=true,        
    pdffitwindow=false,     
    pdfauthor={Carlos Triguero},     
    pdfsubject={Subject},   
    pdfcreator={Creator},   
    pdfproducer={Producer}, 
    pdfkeywords={keywords}, 
    pdfnewwindow=true,      
    colorlinks=true,       
    linkcolor=blue,          
    citecolor=blue,        
    filecolor=blue,      
    urlcolor=blue           
}
\usepackage{rotating}
\usepackage{verbatim}
\usepackage{lineno}

\usepackage[caption=false]{subfig} 
\def\alphabf{\mbox{\boldmath$\alpha$}}

\def\betabf{\mbox{\boldmath$\beta$}}

\def\gammabf{\mbox{\boldmath$\gamma$}}

\def\w{{\bf w}}

\def\d{{\bf d}}

\def\s{{\bf s}}

\def\J{{\bf J}}

\def\A{{\bf A}}

\def\B{{\bf B}}

\def\Id{{\bf I}}

\def\T{{\bf T}}

\def\Q{{\bf Q}}

\def\h{{\bf h}}

\def\r{{\bf r}}

\definecolor{lightyellow}{cmyk}{0,0,0.3,0}
\definecolor{lightblue}{cmyk}{0.1,0,0,0}
\definecolor{green4}{cmyk}{0.0,0.0,0.7,0.4}
\definecolor{dgreen}{rgb}{0,.4,0}
\definecolor{yellow2}{rgb}{.7,.7,0}
\definecolor{indianred}{rgb}{0.8,0.36,0.36}

\newcommand{\pacodel}[1]{} 

\begin{document}

\title{Origin of scale-free intermittency in  structural first-order phase transitions}



\author{Francisco J. Perez-Reche}
\email{fperez-reche@abdn.ac.uk}
\affiliation{Institute for Complex Systems and Mathematical Biology, SUPA, University of Aberdeen, Aberdeen, AB24 3UE, UK}

\author{Carles Triguero}
\affiliation{Atomistic Simulation Centre, School of Mathematics and Physics, Queen’s University Belfast, Belfast BT7 1NN, Northern Ireland, United Kingdom}

\author{Giovanni Zanzotto}
\affiliation{DPG, Universit\`{a} di Padova, Via Venezia 8, 35131 Padova, Italy}

\author{Lev Truskinovsky}
\affiliation{LMS, CNRS UMR-7649, \'{E}cole Polytechnique, Route de Saclay, 91128 Palaiseau, France}

\date{\today}

\begin{abstract}

A  salient  feature of cyclically driven first-order phase transformations in crystals is   their scale-free avalanche dynamics. This behavior  has been linked to the presence of a  classical critical point but the mechanism leading to criticality without extrinsic tuning remains unexplained. 
Here we show that the source  of  scaling in such systems is an annealed  disorder associated with transformation-induced slip which  co-evolves with the phase transformation, thus ensuring  the crossing of  a critical manifold. Our conclusions are based on a model where annealed disorder emerges in the form of a random field induced by the phase transition. Such disorder leads to super-transient chaotic behavior under thermal loading, obeys a heavy-tailed distribution and exhibits long-range spatial correlations. We show that the universality class is affected by the long-range character of elastic interactions. In contrast, it is not influenced by the heavy-tailed distribution and spatial correlations of disorder.
\end{abstract}


\maketitle


The  ubiquitous presence of scale-free avalanche behavior  during structural transformations in crystals~\cite{Vives1994a,PerezRechePRL2005,Gallardo_PRB2010,Baro_JPhysC2014} is in an  apparent contradiction with the first-order nature of the underlying phase transitions. This question  has attracted considerable attention~\cite{Vives1995Universality,Goicoechea1995,Ahluwalia2001,PerezReche2007PRL,Chandni_PRL2009,PerezReche_CMT2009,Cerruti2008I,Ding-Salje_PRB2013,Ball-Cesana_2015_a} not only due to  its theoretical interest,  but also because of  the widespread use of transforming materials, e.g.~shape memory alloys, in  applications~\cite{Otsuka1998SMA,Bhattacharya2003}. The study of intermittency in such systems  is important  because the   rearrangements of microstructural morphologies associated with avalanches~\cite{Balandraud_PRB2015} can perilously interfere with material and structural response at sub-micron scale preventing reliable control of the pseudo-plastic deformation \cite{Csikor_Science2007,Frick_MatSciEngA2008,Zaiser_JMechBehavMater2013}. 

Avalanche dynamics with power-law statistics ~\cite{Jensen1998,Sornette2000} is an inherent  feature of a broad variety of natural systems   from neural networks~\cite{Plenz_BookCriticalityNeuralNetworks2014} and   animal herds~\cite{Ginelli2015}   to tectonic faults~\cite{Ben-Zion_RevGeophy2008_Review}  and stellar flares~\cite{Aschwanden_BookCriticalityGalaxies}. To explain the mechanism responsible for scale-free regimes in such systems, 
a number of general paradigms have been proposed, including implicit external tuning~\cite{Perkovic1995,Fisher_PhysRep1998}, the involvement of nonlocal restoring fields~\cite{Dickman2000,Durin_review2004}, the inherent complexity of  the  quenched energy landscapes~\cite{Muller-Wyart_annurev-conmatphys2015}, and  the  multiplicative structure of  the endogenous noise~\cite{Weiss2015}.
 
Controversy surrounds, in particular, the scale-free intermittent response of solid materials undergoing first-order phase transitions.  Various proposed mechanisms of scaling in such systems which do not require external tuning to a critical point range from depinning, as in the case of disordered ferromagnets~\cite{Durin_review2004},  to inertia-induced   nucleation~\cite{Ahluwalia2001}  reminiscent of turbulence. A radically different perspective  would be that a classical critical point~\cite{Stanley1971}  is involved, however,  this scenario in athermal systems requires a  particular degree of quenched disorder~\cite{Sethna2001,Sethna_review2004,PerezReche2004RFIMField,Handford_PRE2013a}. The ubiquity of scaling would then mean that either  the critical region is sufficiently wide to be routinely crossed in the course of periodic driving~\cite{Sethna1993,Perkovic1995}, or that there exists  a feedback mechanism  fine tuning the system to  a critical point. The critical point scenario appears to be backed  by  experimental data~\cite{Chandni_PRL2009}  showing that  at least in some types of crystals the power-law avalanching during thermal driving takes place  at  a particular value of the driving parameter without any \emph{a priori} tuning of disorder. 

An important feature of criticality in structural transformations is that  scaling behavior does not emerge until the system is cyclically driven through the phase transition several times,  i.e.~until it is trained~\cite{Carrillo1998,PerezReche2004Cyc}. This is an indication that power-law behavior in such systems may be an outcome of the  development, parallel to the transformation, of a sizable self-induced disorder associated, for instance, with dislocational activity~\cite{PerezReche2004Cyc,PerezReche2007PRL,PerezReche_CMT2009}. Indeed, limited plasticity is known to accompany at least some martensitic phase changes~\cite{Krauss_ActaMetall1963,Pons_ActaMetallMater1990,Simon_ActaMater2010}.  

In this paper, we reinforce the critical point perspective on the avalanche behavior during structural phase transitions and, at the same time, show that it is compatible with the idea of self-organization to criticality through cycling. Our conclusions are based on a simple model  which reconciles the  singular nature of a  critical point  with the occurrence of scaling which does not require extrinsic tuning. We reduce a  classical continuum  description of crystal deformation to a (pseudo)spin model of the random field type with athermal  dynamics~\cite{PerezReche2007PRL,PerezReche_CMT2009} and study  the behavior of  transformation-induced dislocations  as they  co-evolve with the phase transition. We link the scale-free response with the crossing  of  a critical manifold  in the temperature-disorder parameter space. We show that for thermally driven materials such `encounters' with criticality are  robust  under cyclic loading due to the feedback between the primary order parameter (describing the phase transformation) and a secondary order parameter (describing plasticity).  

Our results  offer a novel perspective on the link between  disorder and  scale-free avalanching suggesting that the  presence of a co-evolving field  is a  fundamental mechanism of  robust scaling, in the line of  \cite{Gil-Sornette_PRL1996}. A similar mechanism is expected to be operative in other  systems with local multi-stability and evolving (annealed) disorder, for instance, during quasi-plastic response of  granular media and amorphous solids \cite{Lin-Wyart_PRL2015,Nakayama_2015}. 

 The paper is organized as follows. The model is introduced in Sec.~\ref{sec:Model} and is used in Sec.~\ref{sec:Cycling-Slip} to study intermittent fluctuations of mechanical quantities during thermal cycling. The critical behavior of avalanches is analyzed in Sec.~\ref{sec:critical}. In Sec.~\ref{sec:MarginalStability} we address the problem of  marginal stability in the critical regime. Our conclusions are summarized in Sec.~\ref{sec:conclusions}.

\section{The model}
\label{sec:Model}
\begin{figure}[h]
\includegraphics[width=8cm]{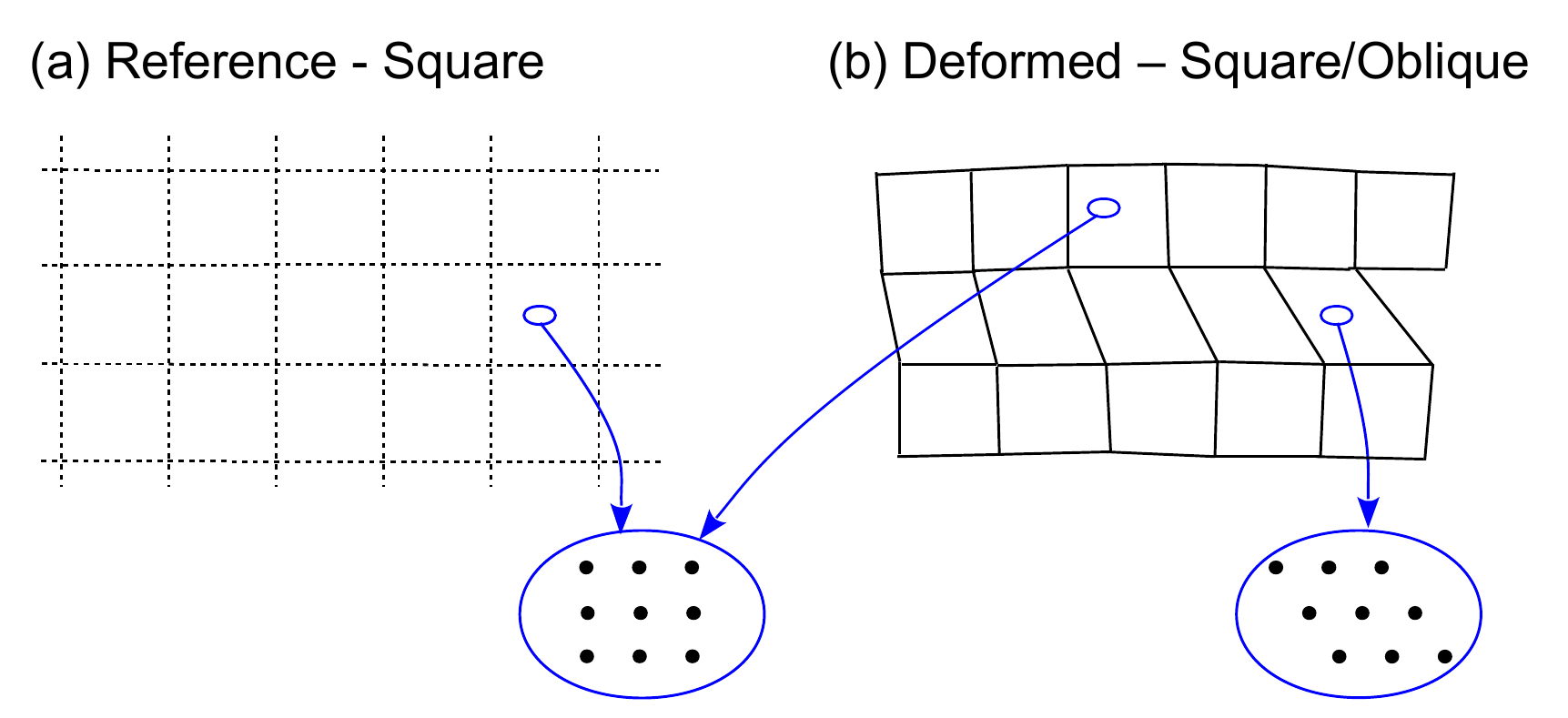}
\caption{\label{Snap-springs_Ensemble} (Color online) Description of a 2D crystalline solid using kinematically compatible elements. Each element represents a region at a mesoscopic scale where the deformation of the lattice is homogeneous, as shown by the insets. When an element is sheared to configurations such as ${\cal S}^+$ in Fig.~\ref{PieceWiseParabolicPotential}, dislocations are created in the square lattice~\cite{PerezReche2007PRL,PerezReche_CMT2009,Conti2004}. }
\end{figure}

We study a first-order structural phase transition in a  2D crystalline solid from a high-temperature square phase  (austenite) to a low-temperature oblique phase (martensite). The deformation is described in terms of a set of $L \times L$ kinematically compatible multistable discrete elements, see  ~\cite{PerezReche2007PRL,PerezReche_CMT2009}. Each element models a homogeneous deformation at a mesoscopic scale \cite{Shenoy_BookBenasque2005} while their ensemble describes heterogeneous deformations at the macroscopic scale (see Fig.~\ref{Snap-springs_Ensemble}).

\subsection{Kinematics}
We take the square lattice in the austenite phase as the reference configuration; the linear size of the mesoscopic elements in this configuration will be used as the unit of length (Fig.~\ref{Snap-springs_Ensemble}(a)). The deformation of the crystal is given by the displacements of the vertices of the mesoscopic elements, $\vec{u}_{i,j}=(u^x_{i,j},u^y_{i,j})$, where $i,j=1,2,\dots,L$. The  field $\vec{u}_{i,j}$ can be expected to be highly inhomogeneous in the low-temperature phase exhibiting a meso-scale mixture of the square phase coexisting with different variants of the oblique phase (Fig.~\ref{Snap-springs_Ensemble}(b)).

\begin{figure}[ht]
\includegraphics[width=7.5cm]{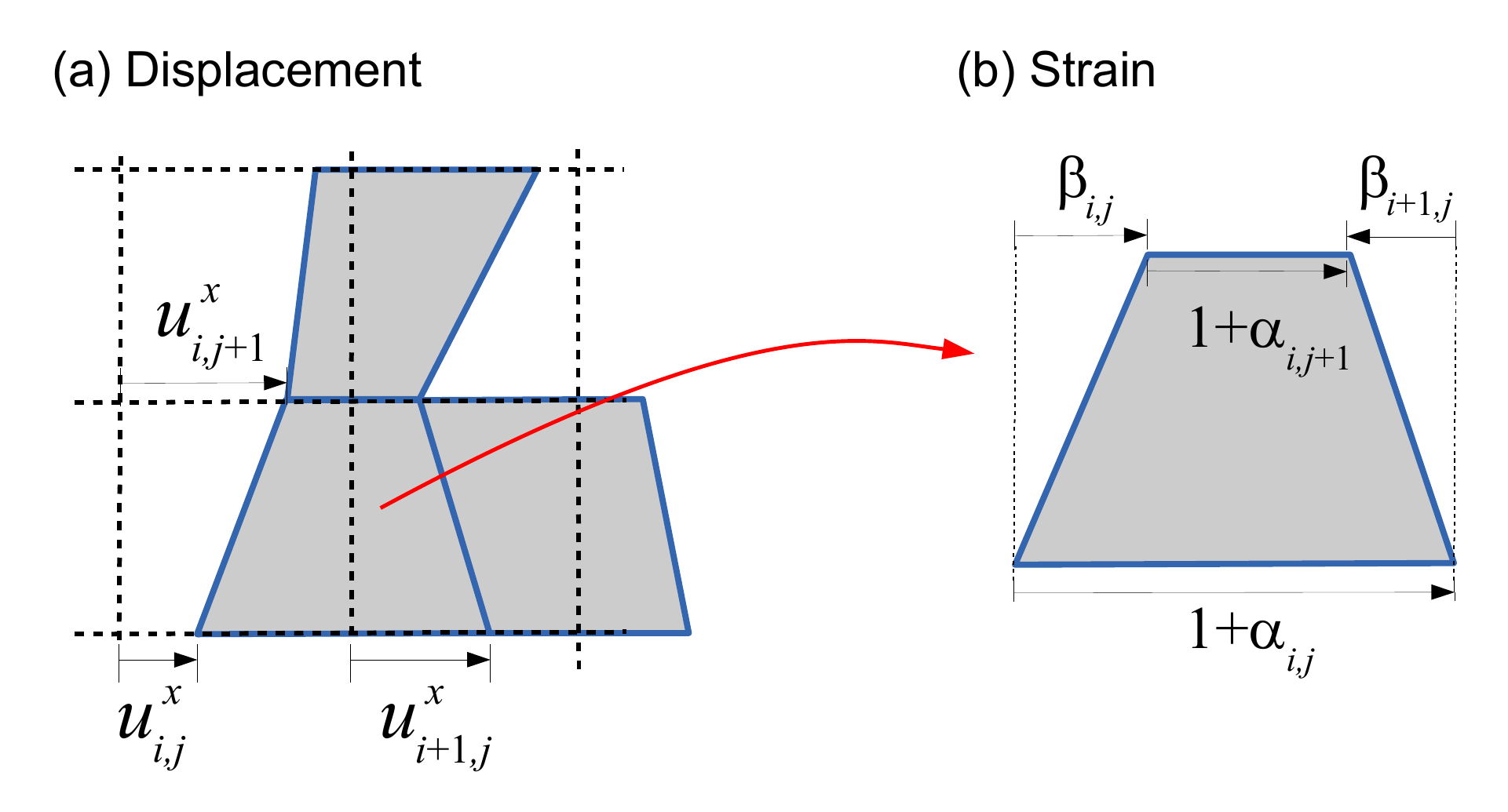}
\caption{\label{Snap-Springs_SetOf3_Displacement} (Color online) Panel (a) indicates the displacement fields $u^x_{i,j},u^x_{i+1,j}$ and $u^x_{i,j+1}$ defining the strain variables $\alpha_{i,j}$ and $\beta_{i,j}$ shown in (b).}
\end{figure}

To keep the model minimal, we assume that the crystal is highly anisotropic and the displacement field is  scalar. This is a common assumption used in the studies of depinning transitions~\cite{Jensen_JPhysA1995} and dislocation dynamics~\cite{Miguel2002,Miguel2001,Ispanovity2014a,Salman-Truskinovsky_PRL2011,Salman-Truskinovsky_IntJEngSci2012}. More precisely, we will only allow displacements in the horizontal direction by setting $u^y_{i,j}\equiv0$, see Fig.~\ref{Snap-Springs_SetOf3_Displacement}(a). Such a kinematic constraint reduces the four variants of oblique martensite in unconstrained conditions to two. It also allows the strain tensor to be reduced to  two fields~\footnote{In general, a complete description of the deformation of a 2D solid requires three strain variables~\cite{PitteriZanzotto2003,Kartha1995}.}: A horizontal strain, 
$$\alpha_{i,j} =
u^x_{i+1,j}-u^x_{i,j}~,\;\; i=1,\dots,L-1,\; j=1,\dots,L~,$$ 
which is a  non-order parameter variable  and a shear strain 
$$\beta_{i,j} =u^x_{i,j+1}-u^x_{i,j}~,\;\; i=1,\dots,L,\; j=1,\dots,L-1~,$$ 
which is the order parameter for the square-oblique transformation (see Fig.~\ref{Snap-Springs_SetOf3_Displacement}(b)). 
They  are not independent due to  the constraint of geometric compatibility~\cite{Kartha1995,Shenoy_BookBenasque2005,Shenoy_PRB2008}. In our model compatibility reduces to the following relations, 
\begin{equation}
\label{eq:El-Compatibility}
\beta_{i,j}-\beta_{i+1,j}=\alpha_{i,j}-\alpha_{i,j+1}~, \;\; i,j=1,\dots,L-1~,
\end{equation}
which can be understood graphically from Fig.~\ref{Snap-Springs_SetOf3_Displacement}(b).
 
In this kinematic description, the stress-free, undeformed austenite square configuration, ${\cal S}$, corresponds to $\alpha_{i,j}= \beta_{i,j} = 0$; the  transformation strains describing the two stress-free variants of oblique martensite, ${\cal O}_1$ and ${\cal O}_2$, are $\alpha_{i,j}= 0$ and $\beta_{i,j} = \pm \epsilon$, where $0<\epsilon<1/2$ is the transformation strain.  Depending on the value of $\epsilon$, it has been common to distinguish weak, almost reversible phase transformations ($\epsilon \sim 0$) from  reconstructive transformations ($\epsilon \sim 1/2$)~\cite{Bhattacharya2004}. In Section~\ref{sec:MaterialTypes} we propose a more precise classification of materials depending on the value of $\epsilon$.

Note that, despite the simplifications introduced by the kinematic constraint, our model respects the multi-variant character of the low-symmetry phase, and includes the  crucial effects of nonlocality (induced by elastic compatibility)  that separate diffusionless structural  transformations from other first-order phase transitions.   

 \subsection{Elastic energy} 
We write  the dimensionless energy of the system in the form 
\begin{equation}
\label{eq:PHI}
\Phi=\sum_{i,j} \phi(\alpha_{i,j},\beta_{i,j})~,
\end{equation}
where 
\begin{equation}
\label{eq:phi_loc}
\phi(\alpha,\beta)=\frac{c}{2}\alpha^2+f(\beta,T)
\end{equation}
is the energy of an element. It will be clear later on that the parameter $c$ describes the  coupling between lattice cells (mesoscopic elements)  with different  values of 
  $\beta$. Within our 2D setting $c$ mimics the  ratios of elastic constants $(C_{11}-C_{12})/(4C_{44})$ or $C_{11}/C_{66}$ ~\cite{Gonzalez-Comas1997,Lloveras_PRL2008}.   

We account for the lattice periodicity through the condition $$f(\beta,T)=f(\beta+d,T),$$ where $d \in \mathbb{Z}$ is an integer-valued variable describing slip~\cite{Ericksen_ArchRationalMechAnal1980,PitteriZanzotto2003,Conti2004,Bhattacharya2004}.

\begin{figure}[h]
\includegraphics[width=8cm]{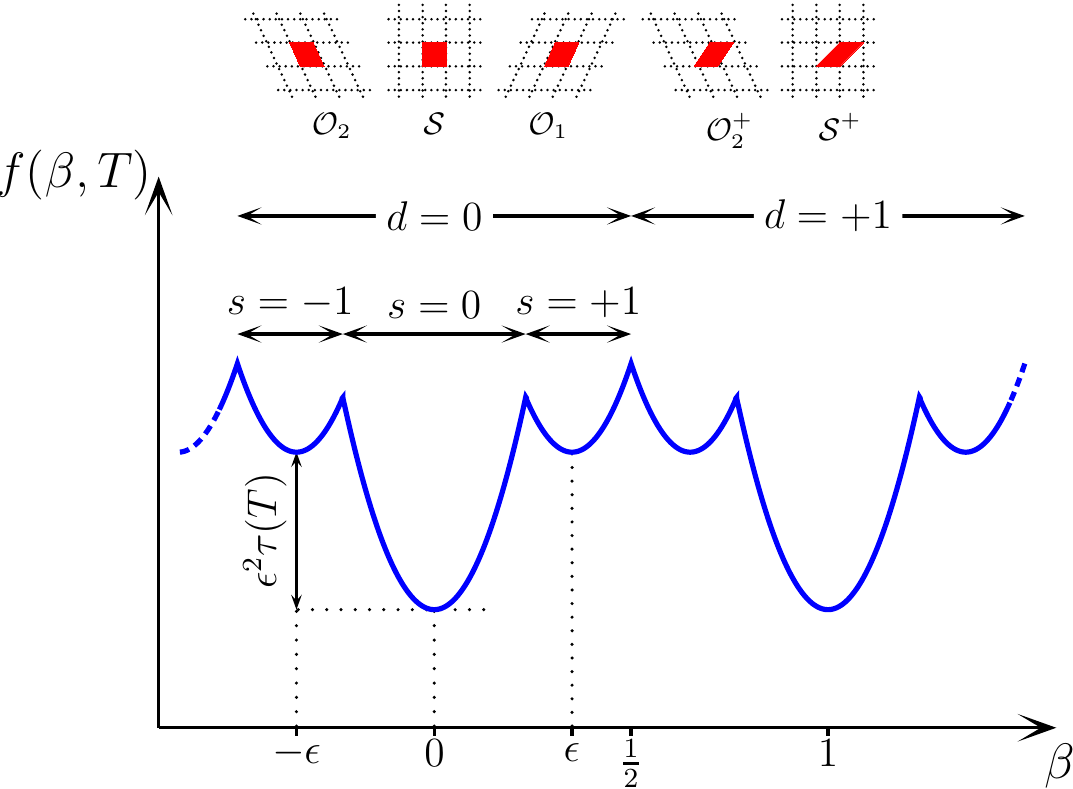}
\caption{\label{PieceWiseParabolicPotential} (Color online) Piece-wise parabolic function $f(\beta,T)$ giving the periodic energy of a mesoscopic element.  The insets on top show in red (grey) the shape of the elements sitting in the bottoms of the wells; the dashed lines indicate the associated lattice structure. The elementary shapes in the domain with $d=0$ correspond to elements in square austenite, ${\cal S}$, and two variants of oblique martensite, ${\cal O}_1$ and ${\cal O}_2$. The rest of shapes correspond to slipped lattices that are identical to those of the basic shapes (compare the dashed lines). }
\end{figure}

For analytical transparency we proceed as in \cite{PerezReche2007PRL,PerezReche_CMT2009} assuming that at a given temperature $T$ the function  $f(\beta,T)$ is  piece-wise parabolic, with minima placed at $\beta = (\epsilon s+d)$, see Fig.~\ref{PieceWiseParabolicPotential}. 
The potential within each well is given by 
\begin{equation}
\label{eq:f}
f(\beta_{i,j};T)=\frac{1}{2}(\beta_{i,j}-(\epsilon s_{i,j}+d_{i,j}))^2 +\epsilon^2\tau(T)s_{i,j}^2~,
\end{equation}
where the function $\tau(T)$ represents the thermal driving.
The choice of the relevant period of $f(\beta,T)$ of an element $i,j$ is specified by the integer-valued variable $d_{i,j}$. Similarly,  the spin variable $s_{i,j}$ selects, for a given $d_{i,j}$, between the austenite  ($s_{i,j}=0$) and two variants of martensite  ($s_{i,j}= \pm1$).
The bottoms of the wells are located at  $\beta_{i,j} = d_{i,j}+\epsilon s_{i,j}$, while the limits of stability of individual energy wells correspond to intersections of adjacent parabolas. The stability conditions for a given well are conveniently expressed in terms of the relative elastic strain,
\begin{equation}
\gamma=\epsilon^{-1}(\beta-(\epsilon s+d))~,
\end{equation}
which measures (in units of $\epsilon$) the distance from bottom of the   corresponding  energy well.
Thus, an element in the austenite phase ($s=0$) is stable if $\gamma_{AM}(\tau)-|\gamma|>0$ and an element in martensite phase ($s=\pm 1$) is stable if $ |\gamma|-\gamma_{\text{MA}}(\tau)>0$ and $\gamma_{\text{S}}(\epsilon)-|\gamma|>0$. Here 
\begin{eqnarray}
\label{eq:gamma_AM}
\gamma_{\text{AM}}(\tau)&=\tau+\frac{1}{2}~,\\
\label{eq:gamma_MA}
\gamma_{\text{MA}}(\tau)&=\tau-\frac{1}{2}~,\\
\label{eq:gamma_S}
\gamma_{\text{S}}(\epsilon)&=\frac{1}{2\epsilon}-1~,
\end{eqnarray}
are the stability limits for austenite-martensite  (AM), martensite-austenite  (MA) and slip  (S) changes, respectively. Note that the dependence on $\tau$ is fully accounted for by the limits for phase transitions, $\gamma_{\text{AM}}(\tau)$ and $\gamma_{\text{MA}}(\tau)$; the dependence on $\epsilon$ is carried by the limit $\gamma_{\text{S}}(\epsilon)$ for slip changes.

\subsection{Thermal driving}

As in typical experiments  \cite{Vives1994a,PerezRechePRL2005,Chandni_PRL2009,Gallardo_PRB2010,Baro_JPhysC2014}, we assume that $\tau(T)$ varies periodically.  Thermal fluctuations are  assumed to be negligible at the time scale of the driving which implies, in particular,  that a  configuration   can  become unstable only as a result of the  variation of $\tau(T)$. This is a reasonable approximation for a wide range of athermal structural transformations~\cite{PerezReche2001}.  Under quasi-static driving~\cite{Puglisi2005a},
the strain variables $\alphabf=\{\alpha_{i,j}\}$ and $\betabf=\{\beta_{i,j}\}$ of an athermal overdamped system vary smoothly while the  configuration fields $\s=\{s_{i,j}\}$  and   $\d=\{d_{i,j}\}$ remain fixed. At the point where the stability condition is violated by at least one of the mesoscopic elements,  an avalanche occurs which typically includes several updates of the configuration fields until the system stabilizes in a new state $(\s, \d)$. We use synchronous dynamics for the updates of $\s$  and  $\d$ during the propagation of the avalanche~\cite{Kuntz1999,Hinrichsen_AdvPhys2000}; quasi-static driving is implemented by keeping $\tau(T)$ constant during the avalanche. Once the avalanche finishes, the variation of $\tau(T)$ resumes and the system evolves elastically along the new equilibrium state until another avalanche begins. The size of an avalanche is defined as the number of transformation events during the avalanche. This includes both phase transition (changing $s$) and slip (changing $d$). We have checked that defining the avalanche size only as the number of phase transition events does not change our conclusions.

We note that,  in the regime of quasi-static driving  and athermal dynamics, the transformation strain $\epsilon$  and the coupling strength  $c$ are the only parameters of the model.

\subsection{Elastic interactions and slip disorder}

During the quiescent periods with fixed $\s$ and $\d$, the system is in equilibrium  with respect to $\alphabf$ and $\betabf$. The equilibria correspond to the stationary points of the energy $\Phi$ (Eq.~\eqref{eq:PHI}) subject to the compatibility conditions (Eq.~\eqref{eq:El-Compatibility}). In Appendix~\ref{App:Min-Energy} we show that the relative elastic strain in such equilibrium states is 
\begin{equation}
\label{eq:elastic-strain}
\gammabf=\J \cdot \s + \h,
\end{equation}  
where $ \h=\J \cdot \d / \epsilon$.  The kernel $\J=\{J_{i,j,k,l}\}$ characterizes the system response (strain redistribution)  induced by a  shear event in a single element~\cite{Picard_EPJE2004,Talamali-Roux_CRMecanique2012}. In particular, the   strain increment in an element $(k,l)$ induced by the phase transition of an element $(i,j)$ is $\delta \gamma_{k,l}=J_{k,l,i,j} \delta s_{i,j}$, where $\delta s_{i,j} = \pm 1$ is the increment of the phase variable $s_{i,j}$. Similarly,  the formation of  a dislocation dipole  (plastic slip event)~\cite{PerezReche_CMT2009,Salman-Truskinovsky_PRL2011,Salman-Truskinovsky_IntJEngSci2012} at $(i,j)$ induces a strain change $\delta \gamma_{k,l}=J_{k,l,i,j} (\delta s_{i,j}  + \epsilon^{-1} \delta d_{i,j})$, where  $\delta d_{i,j}=\pm 1$ and $\delta s_{i,j}=-2 \delta d_{i,j}$, giving $\delta \gamma_{k,l}=\pm (\epsilon^{-1}-2) J_{k,l,i,j}$.

\begin{figure}
\includegraphics[width=8cm]{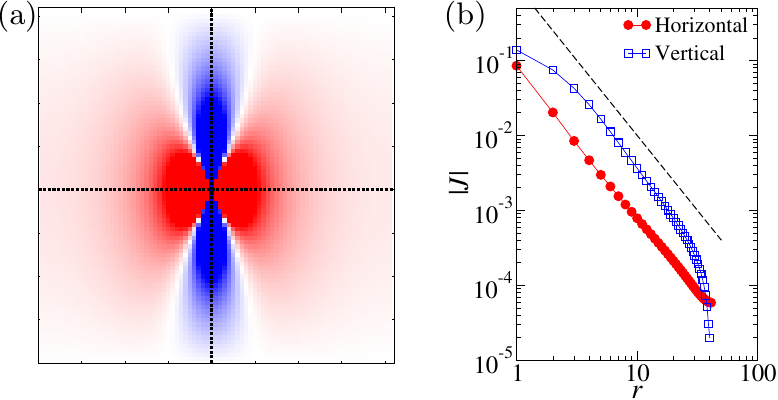}
\caption{\label{Fig:Kernel_c0c2_L51} (Color online) (a) Interaction $\J$ of an element at the center of a system of size $L=51$ with the rest of elements in the system ($c=0.2$). Red, white and blue colors indicate positive, zero, and negative values of $\J$, respectively. (b) Decay of the absolute value of $\J$ with distance, $r$, from the element at the center of the system along the horizontal and vertical directions shown with dotted lines in (a). The dashed line indicates the $r^{-2}$ decay.}
\end{figure}

As shown in Appendix~\ref{App:Min-Energy}, $\J$ in our problem can be calculated explicitly in real space which  contrasts  many other studies where  only the Fourier image of an elastic propagator is known~\cite{Kapetanou_JSTAT2015,Kartha1995,Shenoy_BookBenasque2005,Shenoy_PRB2008,Picard_EPJE2004,Lloveras_PRL2008}). Thus, figure \ref{Fig:Kernel_c0c2_L51}(a) illustrates the elastic interaction of the central  element  with the surrounding elements. It has a dipolar structure and involves both  ferro and anti-ferromagnetic contributions as in, e.g.~\cite{Salman-Truskinovsky_PRL2011,Salman-Truskinovsky_IntJEngSci2012}. The absence of a traditional quadrupolar  structure is a consequence of the extreme anisotropy of our model~\cite{Kapetanou_JSTAT2015,Kartha1995,Shenoy_BookBenasque2005,Shenoy_PRB2008,Picard_EPJE2004,Lloveras_PRL2008}. The assumed anisotropy, however,  does not affect the $r^{-2}$ decay of $\J$ with distance which is typical  for 2D elastic solids ~\cite{Kapetanou_JSTAT2015,Kartha1995,Shenoy_BookBenasque2005,Shenoy_PRB2008,Picard_EPJE2004,Lloveras_PRL2008}, see Fig.~\ref{Fig:Kernel_c0c2_L51}(a).

\begin{figure*}
\label{Fig:Pattern_s}\includegraphics[width=12.0cm]{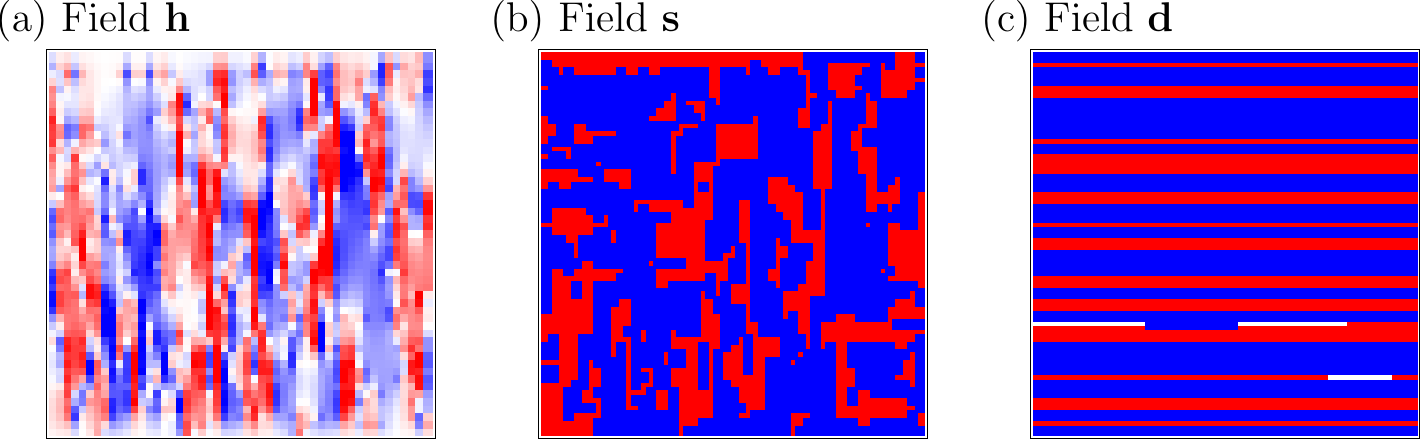}
\caption{\label{Fig:Pattern_h}  (Color online) Spatial configuration of  (a) the slip disorder, $\h$, (b) the field of phases, $\s$, and (c) the slip field, $\d$, in the low-temperature phase after $10^4$ cycles (blue and red correspond to negative and positive values) of a system with $L=51$ undergoing a close-to-reconstructive transition with $\epsilon=0.48$.}
\end{figure*}

The auxiliary field $\h$ introduced in Eq.~\eqref{eq:elastic-strain} allows us to establish a formal link between this model and the more conventional random field models (RFMs) with quenched disorder \cite{Sethna2001,Sethna_review2004,Vives1995Universality}. Indeed,  its  effect on  the phase transition (variables $\s$) is analogous to that of random fields in RFMs. In this analogy, the  mesoscopic elements are identified with the spins in RFMs while the relative elastic strain, $\gammabf$, stands for the effective local field. Even though the field  $\h$, which we formally call  `slip disorder',  is deterministic, it can be highly heterogeneous in materials with high $\epsilon$ (see Fig.~\ref{Fig:Pattern_h}(a)). What makes our model different, however, is the annealed character of such a disorder:  the field $\h$ is caused by the slip $\d$ which is dynamically induced by the phase transition.  The fact that our  thermal driving cannot induce slip directly is a characteristic feature of the model.
   
The competing positive and negative contributions to $\J$ lead to frustration revealed through the formation of heterogeneous phase microstructures at low temperatures (see Fig.~\ref{Fig:Pattern_h}(b)) akin the ones that are typically observed  in martensites~\cite{Otsuka1998SMA,Bhattacharya2003,Shenoy_BookBenasque2005,ren2000,PerezReche2007PRL,PerezReche_CMT2009,Balandraud_PRB2015,Ball-Cesana_2015_a} and ferroelastics~\cite{Salje1993,Ding-Salje_PRB2013}. Microstructures of this type are also typical for other  systems with competing interactions ~\cite{Selke1992}.

In systems with large enough $\epsilon$,  phase heterogeneity  will necessarily lead to the similar heterogeneity of  the field of plastic slip $\d$    (Fig.~\ref{Fig:Pattern_h}(c)).   The heterogeneity of  $\d$,   amplified by the complexity of the  elastic interaction kernel $\J$,     generates  the   slip disorder $\h$ which, in turn,  will contribute to the heterogeneity of  the   transformation.  The study of the outcome of the implied feedback constitutes the main subject of this paper.

\section{Effects of thermal cycling}
\label{sec:Cycling-Slip}
With elastic fields adiabatically eliminated (see  Appendix~\ref{App:Min-Energy}), our dynamical system becomes   an integer-valued automaton in terms of the fields  $(\s,\d)$ which is easy to simulate. We begin thermal loading from an equilibrium state with $\tau > 1/2$ where all elements are  in the austenite phase ($s=0$). A minimal initial slip disorder is introduced by setting $d=1$ in the  element  $(i,j)=(L/2,L/2)$, and it is assumed that $d=0$  everywhere else. We then change the temperature $\tau$ periodically, for different fixed values of $\epsilon$. 

In all  numerical experiments we set  $c=0.2$ which is a representative value for shape-memory alloys for which both ratios   $(C_{11}-C_{12})/(4C_{44})$ and $C_{11}/C_{66}$ are typically   between 0.1 and 2, see   \cite{Gonzalez-Comas1997,Lloveras_PRL2008}. The dependence of the transformation scenario on $c$ is interesting and will be discussed elsewhere. 

\begin{figure}[h]
\includegraphics[width=8.5cm]{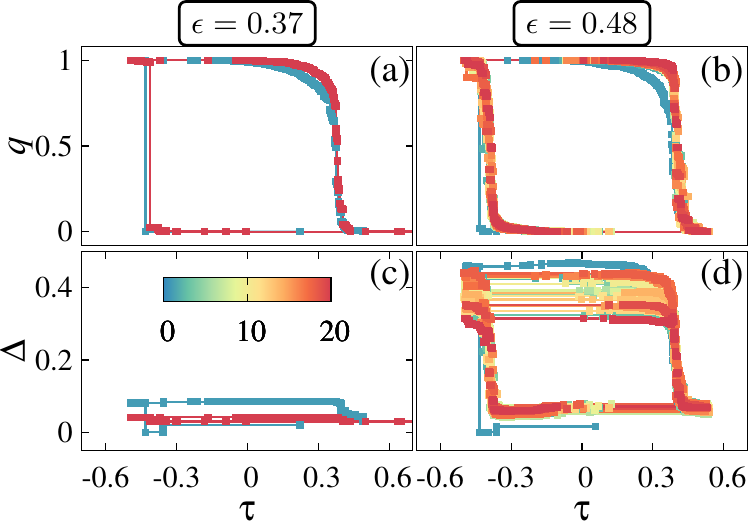}
\caption{\label{Fig:Trainability-Criticality}  
 (Color online) Fraction of elements transformed to martensite, $q(\tau)$, and standard deviation of the slip disorder, $\Delta(\tau)$, for (a,c) $\epsilon=0.37$ and (b,d) $\epsilon=0.48$ in series of 20 cycles. Different colors correspond to different cycles, as indicated by the colorbox in (c).
Simulations correspond to systems of size $L=51$. The initial slip disorder is $\Delta_0(\epsilon)= (\epsilon \, L^2)^{-1}\sum_{i,j} J_{i,j,L/2, L/2}^2 = 7.6 \times 10^{-3}/ \epsilon$.}
\end{figure}

\subsection{Hysteresis loops}

 The behavior of the system depends crucially on the choice of the transformation strain  $\epsilon$ as we  illustrate in Fig.~\ref{Fig:Trainability-Criticality}.   We  follow the evolution of the transformed fraction to  martensite,
\begin{equation}
\label{eq:q}
q(\tau)=L^{-2} \sum_{i,j} s_{i,j}^2~,
\end{equation}
and the co-evolution of the slip disorder $\h$, which is traced through the standard deviation,
\begin{equation}
\label{eq:Deltatau}
\Delta(\tau) =\left(L^{-2} \sum_{i,j} h_{i,j}^2\right)^{1/2}~,
\end{equation}
where we have used the fact that the mean value of $\h$ is zero.

We show two sets of data, for weak transformations with $\epsilon \lesssim 1/3$   and   for close-to-reconstructive transformations with $\epsilon \sim 1/2$.   For weak transformations, the function $q(\tau)$ describes asymmetric cycles with an embedded infinite avalanche (`snap') during cooling runs and small avalanches (`pop') during heating runs (see Fig.~\ref{Fig:Trainability-Criticality}(a)). This is the usual asymmetry of the transformed fraction in transitions to multivariant phases~\cite{Vives1995Universality,Cerruti2008I}. Instead, for close-to-reconstructive transformations the cycles are more symmetric,  rounded and reproducible already after few cycles (see Fig.~\ref{Fig:nc_vs_e_StabilisationCycling}(b)),  in contrast to  what is known for models with quenched disorder~\cite{Vives1995Universality,Cerruti2008I}.

The  disorder  $\Delta(\tau)$  co-evolves with the  phase transition,  increasing  when cooling  and  decreasing when   heating.  The presence of slip  is  much more pronounced  close to $\epsilon = 1/2$  when   the stability limit $\gamma_{\text{S}}(\epsilon)$  is low and dislocations form easily ~\cite{Conti2004,Bhattacharya2004,Balandraud_Zanzotto2007}.   Instead,   for  $\epsilon \lesssim 1/3$   the dislocation activity is negligible because of the large barriers associated with these regimes ~\cite{PerezReche_CMT2009}. 
 
For systems with  small transformation strain, the function $\Delta(\tau)$ exhibits a very short transient with non-periodic dynamics followed by a sudden collapse to a periodic regime with a small hysteresis for $\Delta(\tau)$  (see Fig.~\ref{Fig:Trainability-Criticality}(c)).  The  slip disorder is small and is practically quenched. 

In the regime with large $\epsilon \sim 1/2$, where the slip disorder is annealed, we observe  a lack of reproducibility of the  disorder fluctuations from cycle to cycle and a considerably larger hysteresis for $\Delta(\tau)$. We now show that systems with large $\epsilon$ eventually settle on a periodic regime. The transition to such regime, however, occurs after a significantly larger number of cycles than in the case of weak transformations.

\subsection{Long chaotic transients}

\begin{figure}[h]
\includegraphics[width=0.35\textwidth]{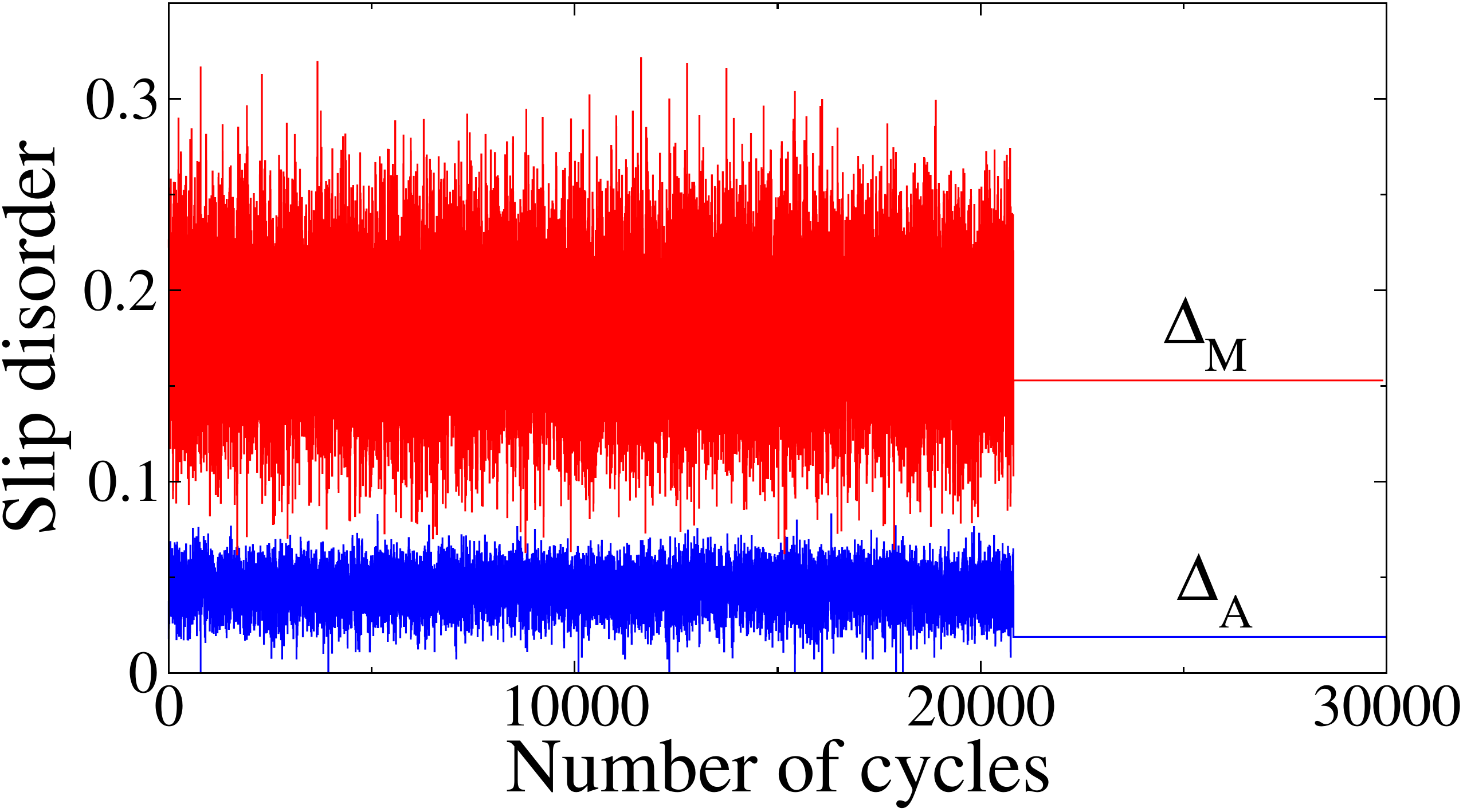}
\caption{\label{Fig:Long_Transient}  (Color online)  Long chaotic transient followed by an abrupt transition to a periodic regime in the case of close-to-reconstructive transformation.The extreme values  $\Delta_\text{M}$ and $\Delta_\text{A}$  in each cycle are presented. Data corresponds to a system of size $L=51$ and $\epsilon=0.431$.}
\end{figure}

\begin{figure}[h]
\includegraphics[width=8.0cm]{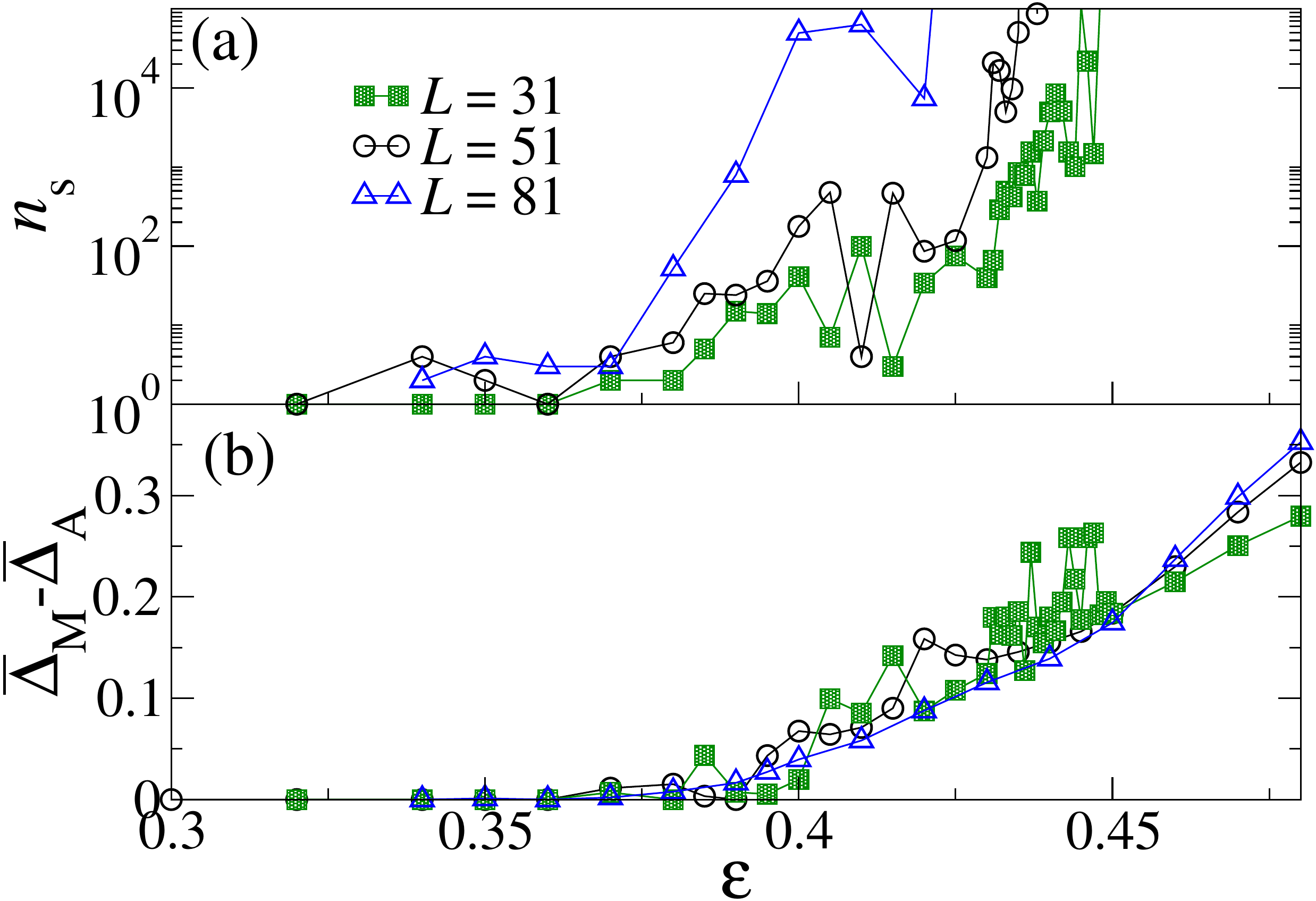}
\caption{\label{Fig:nc_vs_e_StabilisationCycling} (Color online) (a) Log-normal plot of the length $n_\text{s}$   of (super)transients  vs. $\epsilon$ for systems of three different sizes marked by the legend. (b) Mean amplitude $\overline{\Delta}_{\text{M}}-\overline{\Delta}_{\text{A}}$ of the slip disorder vs. $\epsilon$ for the same system sizes as in (a).}
\end{figure}

To understand  the delayed synchronization of slip disorder with thermal cycling in the case of close-to-reconstructive transformations, we studied  the behavior of the extremal values of $\Delta$ corresponding to the austenite and martensite phases, $\Delta_{\text{A}}$ and $\Delta_{\text{M}}$, respectively. In Fig.~\ref{Fig:Long_Transient} we see that the behavior of these two quantities is irregular for a large  number of cycles before it abruptly becomes periodic. This is observed 
  in the regimes where disorder is annealed (for $\epsilon \gtrsim 0.3$)  and the  duration of stochastic transients, $n_\text{s}$,  increases exponentially with the system size and $\epsilon$, see  Fig.~\ref{Fig:nc_vs_e_StabilisationCycling}(a). We also observe a sensitive dependence of $n_\text{s}$ on the  initial conditions (e.g. on the location of the initial slip). 
  
The indicated  features of the dynamics are  typical for super-transient chaotic phenomena ~\cite{Tel-YingChengLai_PhysRep2008,Politi-Torcini_StableChaos2010}  and have been also observed  in cyclically driven  amorphous solids~\cite{Regev-Lookman-Charles_PRE2013,Fiocco-Foffi-Sastry_PRE2013_CyclicGlass}. In view of these analogies, we can argue   that the  basin of attraction of the synchronized behavior in our model is narrow  and  finding this attractor  becomes increasingly difficult as the  system   gains access to a  larger configurations space by generating slip.   In the thermodynamic limit, the synchronized behavior in such systems is usually unreachable and therefore  super-transients  become a central object of interest (as in the case of turbulence in pipe flows \cite{Hof2006}).

 To study the statistics  of the  annealed slip  disorder $\h$ during  super-transients, we can exploit the statistical stationarity of the signal by pooling over long series of cycles (at least $N_{\text{cyc}} \sim 10^4$). For a given quantity $a$ (e.g. $\Delta(\tau)$) taking values $\{a_n\}$ in a series of $N_{\text{cyc}}$ thermal cycles, we consider averages over cycles, $\bar{a}=N_{\text{cyc}}^{-1} \sum_n a_n$.  Figure \ref{Fig:nc_vs_e_StabilisationCycling}(b) shows the mean amplitude of the slip disorder fluctuations averaged over thermal cycles, $\overline{\Delta}_{\text{M}}-\overline{\Delta}_{\text{A}}$. We see once again that, despite the fact that transformation-induced slip takes place  for $\epsilon \gtrsim 0.3$, its hysteretic response (shown in Fig.~\ref{Fig:Trainability-Criticality}(d))  only unlocks for $\epsilon \gtrsim 0.35$  when $\overline{\Delta}_{\text{M}}-\overline{\Delta}_{\text{A}}$ starts to deviate from zero. Note that $\overline{\Delta}_{\text{M}}-\overline{\Delta}_{\text{A}}$ exhibits a more irregular dependence on $\epsilon$ for smaller system sizes while  the overall behavior is only weakly dependent on the system size.
 
\subsection{Dynamical regimes - Material types}
\label{sec:MaterialTypes}
 Our numerical experiments allow us to identify four distinct types of materials, depending on the value of the transformation strain $\epsilon$. They are associated with qualitatively different dynamic responses to the periodic driving.   Regime I corresponds to systems with $\epsilon \lesssim 0.3$ when the phase transition is weak in the sense that it does not induce slip deformation.  Regime I$^\prime$ corresponds to systems with $0.3 \lesssim \epsilon \lesssim 0.35$ which can generate some slip disorder during the first transformation cycle but it remains quenched afterwards. Regime II corresponds to values of the transformation strain in the interval $0.35 \lesssim \epsilon \lesssim 0.45$. In this regime, slip disorder co-evolves with the phase transition indefinitely, describing hysteresis loops in which disorder increases significantly during cooling runs and decreases equally significantly  during heating runs.  The fluctuations of disorder between cycles in this regime are first chaotic, however, the system  eventually synchronizes with the driving.  The  synchronization takes place after a number of cooling-heating cycles  which increases exponentially with increasing $\epsilon$ and the system size. Regime II$^\prime$ corresponds to systems with $\epsilon \gtrsim 0.45$ which, as shown below, exhibit scale-free avalanche dynamics.  In this regime, the chaotic super-transients for slip disorder are extremely long ($n_{\text s}>10^5$) and we were not able to reach synchronization in our experiments. 
 
 Using the language of materials science, we can say that solids operating in Regimes I and I$^\prime$ are not trainable. Training observed in many martensites can be  effective in regimes II and II$^\prime$ and the rest of the paper will be devoted to a more detailed study of the corresponding properties.

\section{Statistics during supertransients in regimes II and II$^\prime$}
\label{sec:critical}
 To study  statistics  of the  `unlocked'  disorder $\h$ in Regimes II and II$^\prime$, we focus on  super-transients and  use sufficiently long series of cycles to perform averaging  (at least $10^4$). Given the relative symmetry of cooling and heating in  these regimes  (see Fig.~\ref{Fig:Trainability-Criticality}(a)),  we mostly report results for  cooling runs.  Some  results on heating runs are also presented  to confirm their statistical similarity with those occurring during cooling runs.  
 
\begin{figure}
\includegraphics[width=6.0cm]{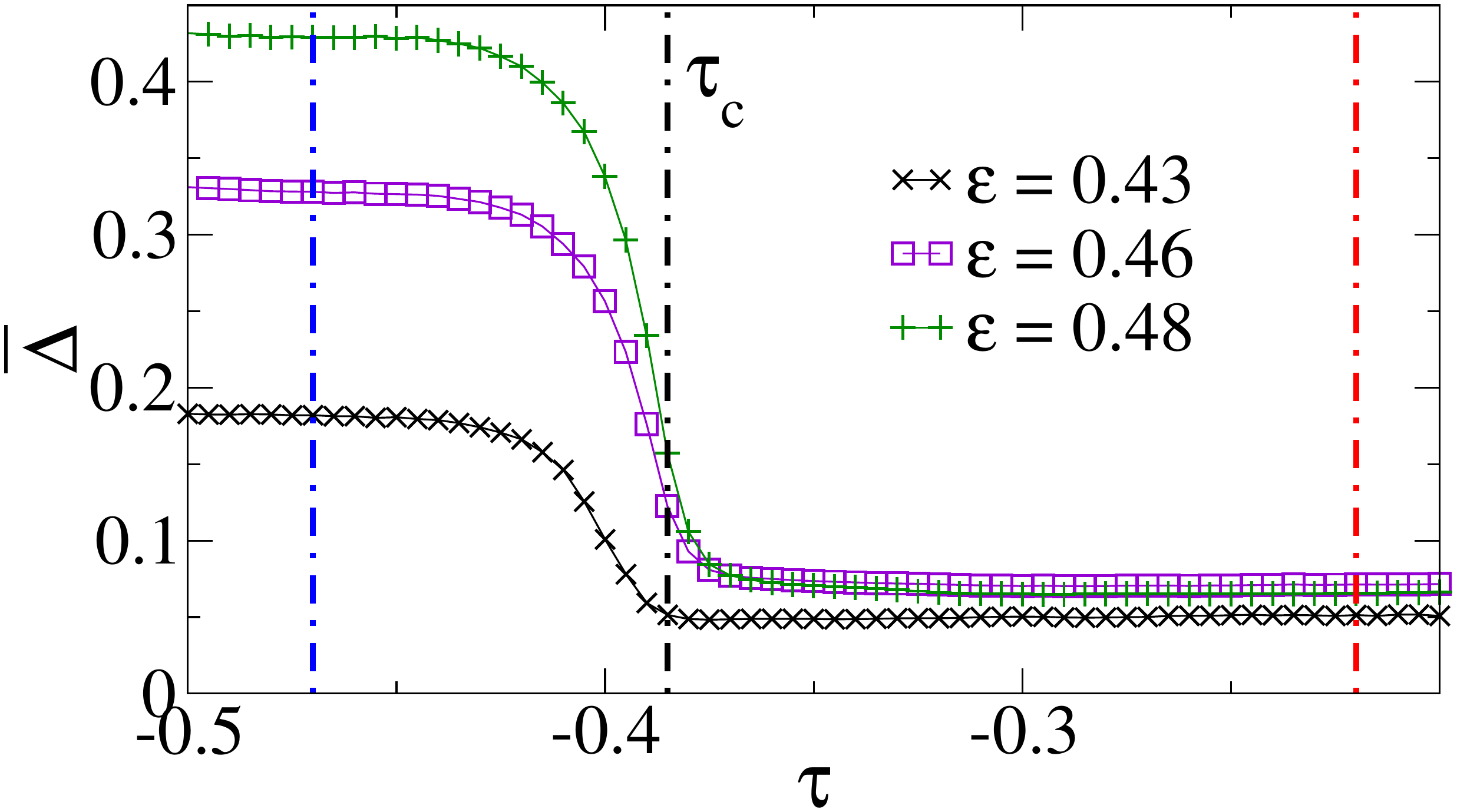}
\caption{\label{Fig:Mean_Delta_vs_tau}  (Color online)  Temperature dependence of the mean standard deviation of the slip disorder, $\overline{\Delta}(\tau)$, during cooling runs  (cycles $400-10^4$) for  $L=51$ and several values of $\epsilon$, as marked by the legend. }
\end{figure}

 We first show that  for materials with $\epsilon \gtrsim 0.35$ which are in regimes II and II$^\prime$,  the standard deviation averaged over cycles, $\overline{\Delta}(\tau) $,  smoothly  increases with  decreasing $\tau$ (see Fig.~\ref{Fig:Mean_Delta_vs_tau}). The build up of disorder, however,  accelerates considerably  around  $\tau_{\text{c}} \simeq -0.385$ where most of the  phase transition from austenite to martensite phase takes place. The precise value $\tau_{\text{c}} \simeq -0.385$ is estimated in Sec.~\ref{sec:Avalanches}.

\subsection{Correlated non-Gaussian disorder}
\label{sec:NonGaussian}
  
\begin{figure}
\includegraphics[width=6.0cm]{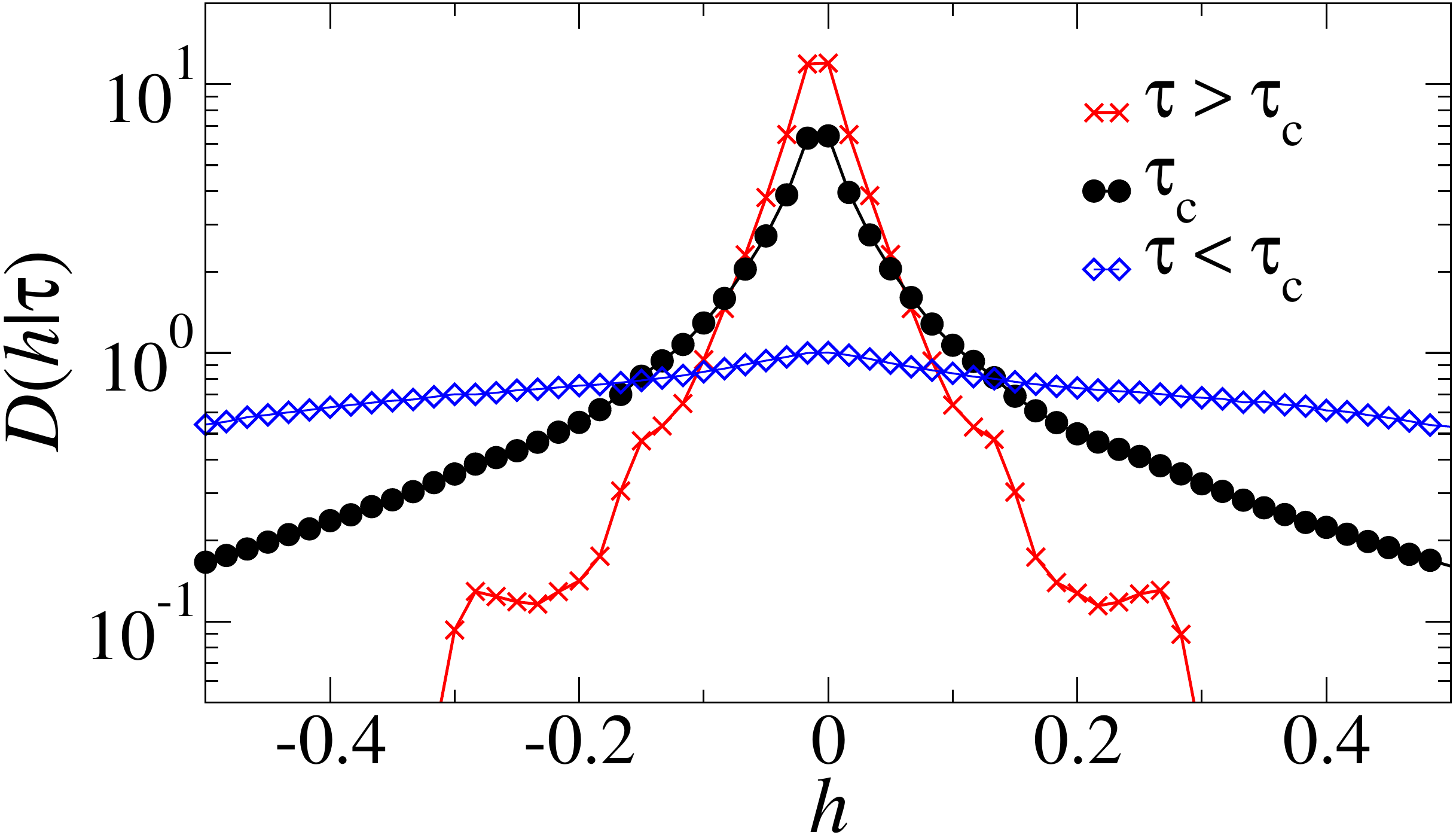}
\caption{\label{Fig:Dh_Given_tau}  (Color online) Log-normal plot of the distribution $D(h|\tau)$ for evolving disorder at three values of $\tau$ (marked by dot-dashed vertical lines in Fig.~\ref{Fig:Mean_Delta_vs_tau}). Data correspond to cooling runs (cycles $400-10^4$) in a system with $L=51$ and $\epsilon=0.48$. }
\end{figure}

\begin{figure}[h]
\includegraphics[width=6cm]{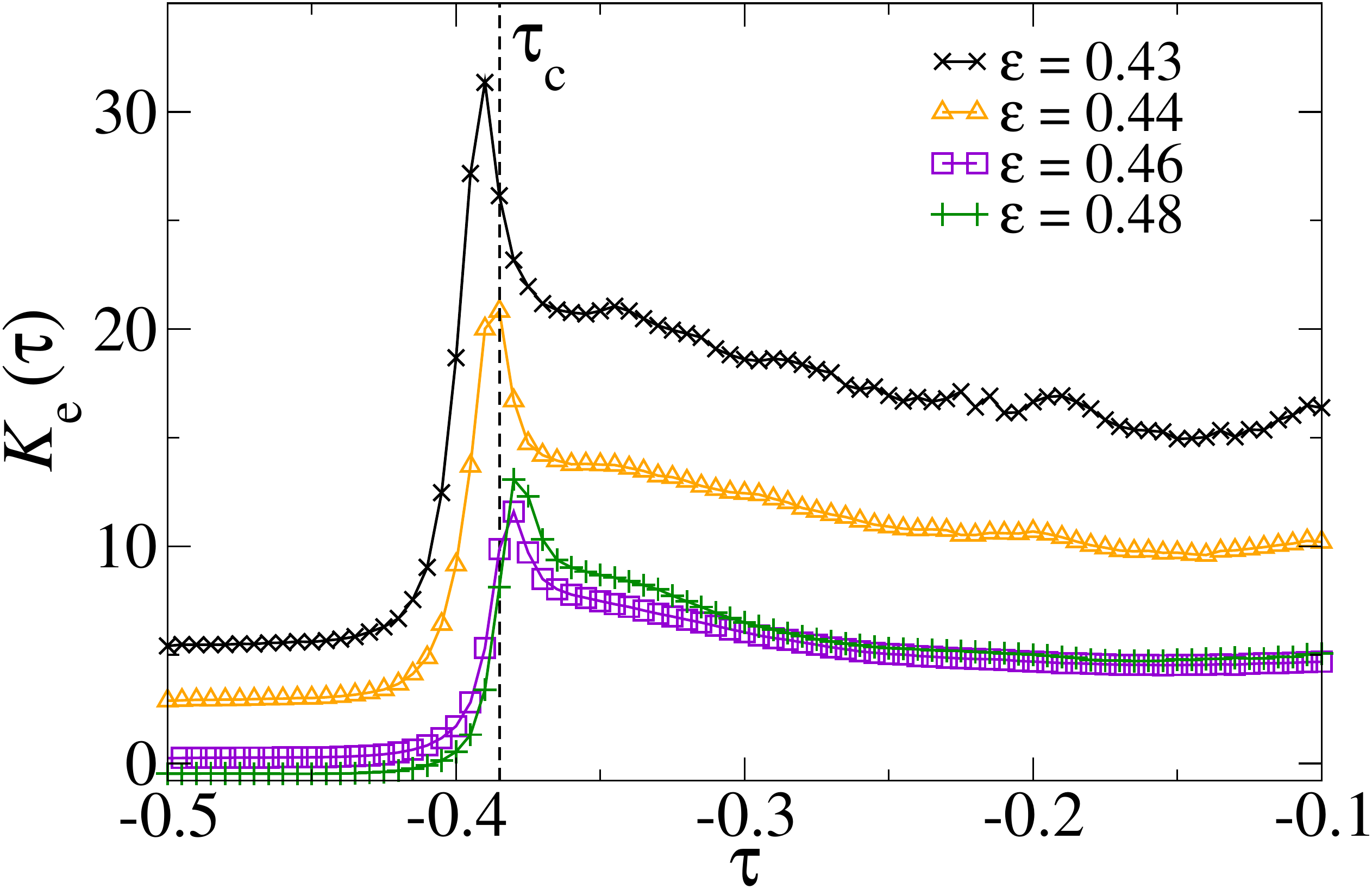}
\caption{\label{Fig:Kutosis_h} (Color online) Temperature dependence of the excess kurtosis, $K_{\text{e}}$, corresponding to slip disorder during cycles $400-10^4$ in a system of size $L=51$ and several values of $\epsilon$.}
\end{figure}

The probability distribution of slip disorder $D(h|\tau)$ in these regimes  is a heavy-tailed distribution for all $\tau$ with relatively narrower peak and wider tails than a Gaussian distribution with the same mean and standard deviation, see Fig.~\ref{Fig:Dh_Given_tau}.  The non-Gaussian character of $D(h|\tau)$ can be systematically studied as a function of $\tau$ in terms of the excess kurtosis~\cite{Riley_Hobson_Bence_MathsBook2006} averaged over cycles:
\begin{equation}
K_{\text{e}}(\tau) =\overline{\frac{\nu_4(\tau)}{\Delta^4(\tau)}-3}~.
\end{equation}
Here $\nu_4(\tau)=L^{-2} \sum_{i,j} h_{i,j}^4$ is the 4\textsuperscript{th} central moment of the slip disorder field. Figure \ref{Fig:Kutosis_h} shows that $K_{\text{e}}$ exhibits a peak corresponding to a maximal deviation   from a Gaussian distribution during the phase transition  around $\tau_c$ where  the slip activity is maximal.  

The  non-Gaussian nature of the slip disorder  reveals itself through the emergence of large correlated domains, see Fig.~\ref{Fig:Pattern_h}(a).  Similar spatial correlations have been  reported for dislocation patterns in martensites \cite{Groger_PRB2008} and plastic activity in amorphous materials~\cite{Talamali-Roux_CRMecanique2012,Picard_PRE2005}.

\begin{figure}
\includegraphics[width=6.0cm]{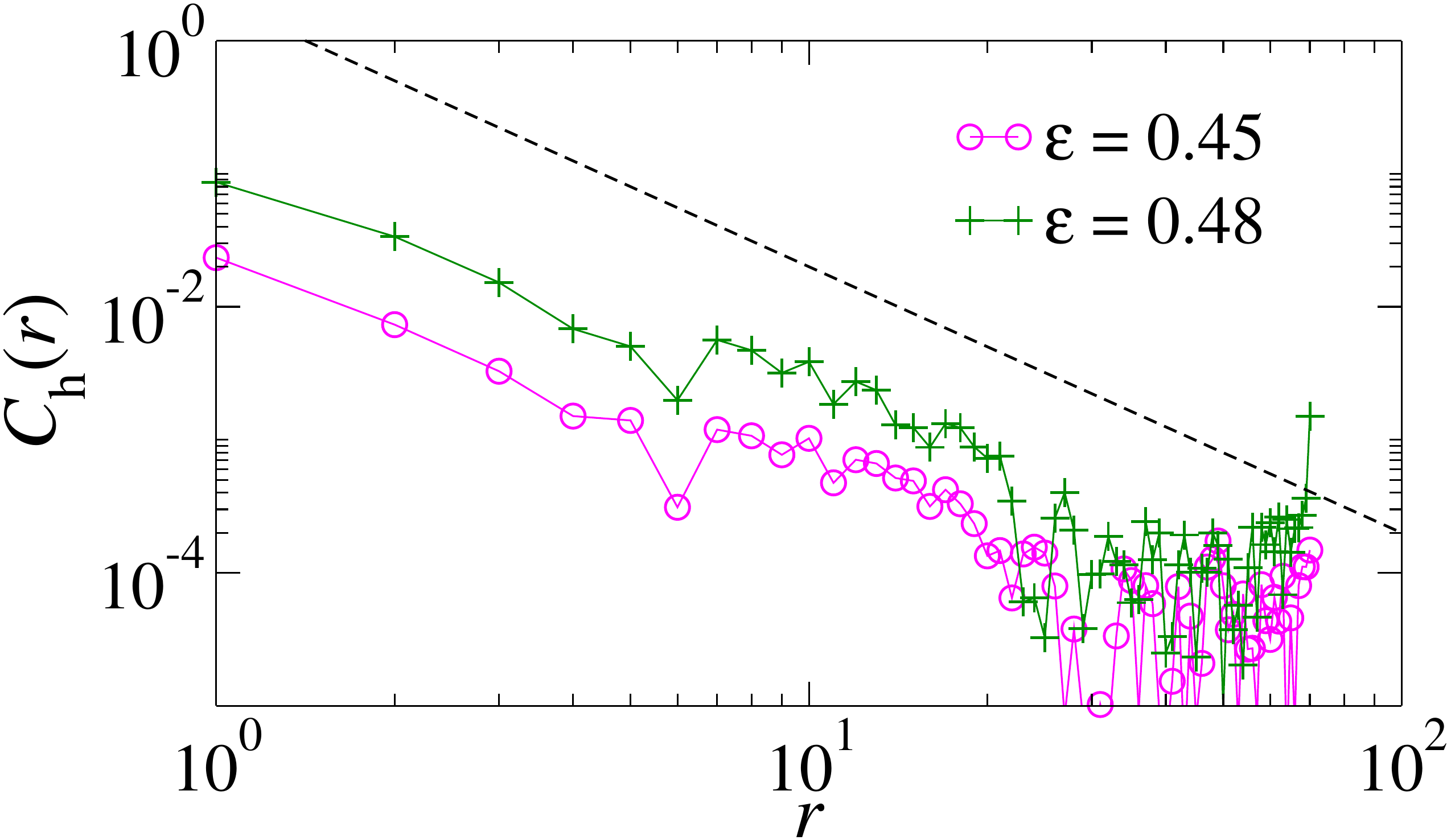}
\caption{\label{Fig:Correlation_h_vs_r}  (Color online) Spatial correlation function of disorder, $C_{\text{h}}(r)$ in systems of size $L=51$ in regime II$^\prime$. $C_{\text{h}}(r)$ decreases in a similar way for all values of $\tau$; the dashed line illustrates the $r^{-2}$ decay.}
\end{figure}

For a given value of $\tau$, we define  the spatial correlations of $\h$  averaged over  cycles as follows:
\begin{equation}
\label{eq:Correlation_h}
C_{\text{h}}(r)=\left| \overline{\frac{1}{\text{card}({\cal P}_r)} \sum_{{\cal P}_r} h_{i,j}h_{i^{\prime},j^{\prime}}} \right|~.
\end{equation}
Note that we take here the absolute value of correlations, denoted by $|\cdot|$. We denoted by ${\cal P}_r$  the set of pairs of elements at $(i,j)$ and $(i^{\prime},j^{\prime})$ that are a distance $r$ from each other; $\text{card}({\cal P}_r)$ is the number of such pairs. Figure \ref{Fig:Correlation_h_vs_r} shows that  $C_{\text{h}}(r)$  decays as $r^{-2}$   in the whole range of loading. Such correlations, that reproduce the decay of the elastic kernel,  reflect the scale free nature of the elastic interactions (described by $\J$)  which mediate the evolution of the field $\h$. The influence  of $\J$ on the correlations of the slip disorder is obvious  from the definition of $\h=\J \cdot \d / \epsilon$: since the slip field $\d$ is dominated by large homogeneous domains (see Fig.~\ref{Fig:Pattern_h}), the $r^{-2}$ decay in $\h$ is predominantly associated with $\J$.

\subsection{Avalanches}
\label{sec:Avalanches}

\begin{figure}
\includegraphics[width=8cm]{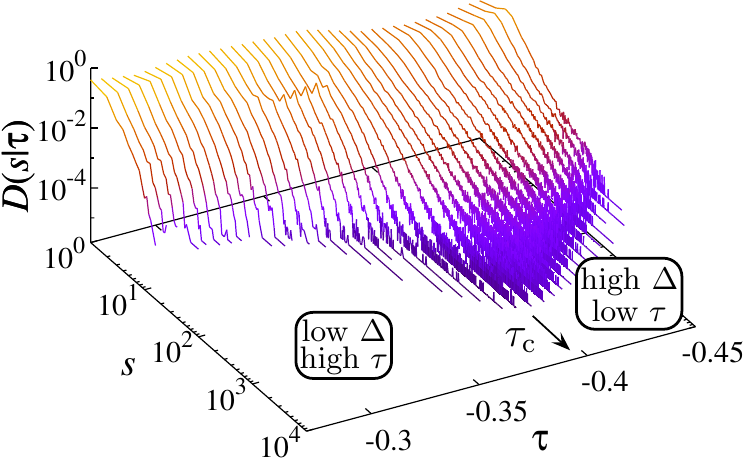}
\caption{\label{Fig:Ds_vs_s_tau} (Color online) Avalanche size distribution at given temperature, $D(s|\tau)$, for cooling runs in a system with $\epsilon=0.48$ and $L=51$. Note that $\tau$ decreases from left to right while  $\Delta$ is increasing  as shown in Fig.~\ref{Fig:Mean_Delta_vs_tau}.}
\end{figure}

\begin{figure}
\includegraphics[width=6cm]{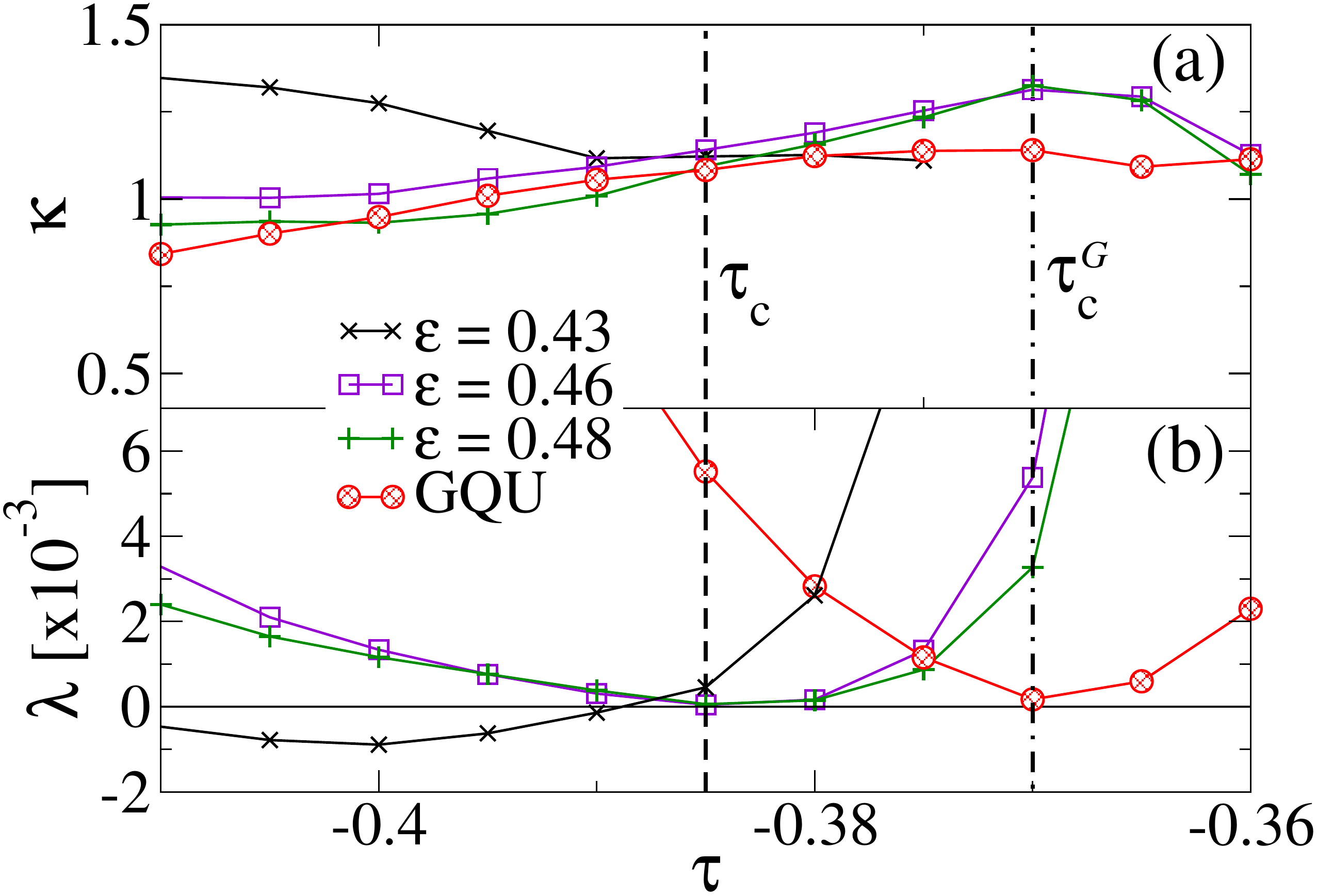}
\caption{\label{Fig:kappa_lambda_Cooling}  (Color online)   Temperature dependence of (a) the power-law exponent  $\kappa$ and (b) the cut-off parameter $\lambda$ of the avalanche size distribution for heating runs. Different symbol types correspond to different values of $\epsilon$, as marked by the legend. Dashed circles show the parameters of fits to $D(S|\tau)$ in systems with Gaussian, quenched and uncorrelated (GQU) random variables, $\h$. Systems of size $L=51$.}
\end{figure}

\begin{figure}
\includegraphics[width=6cm]{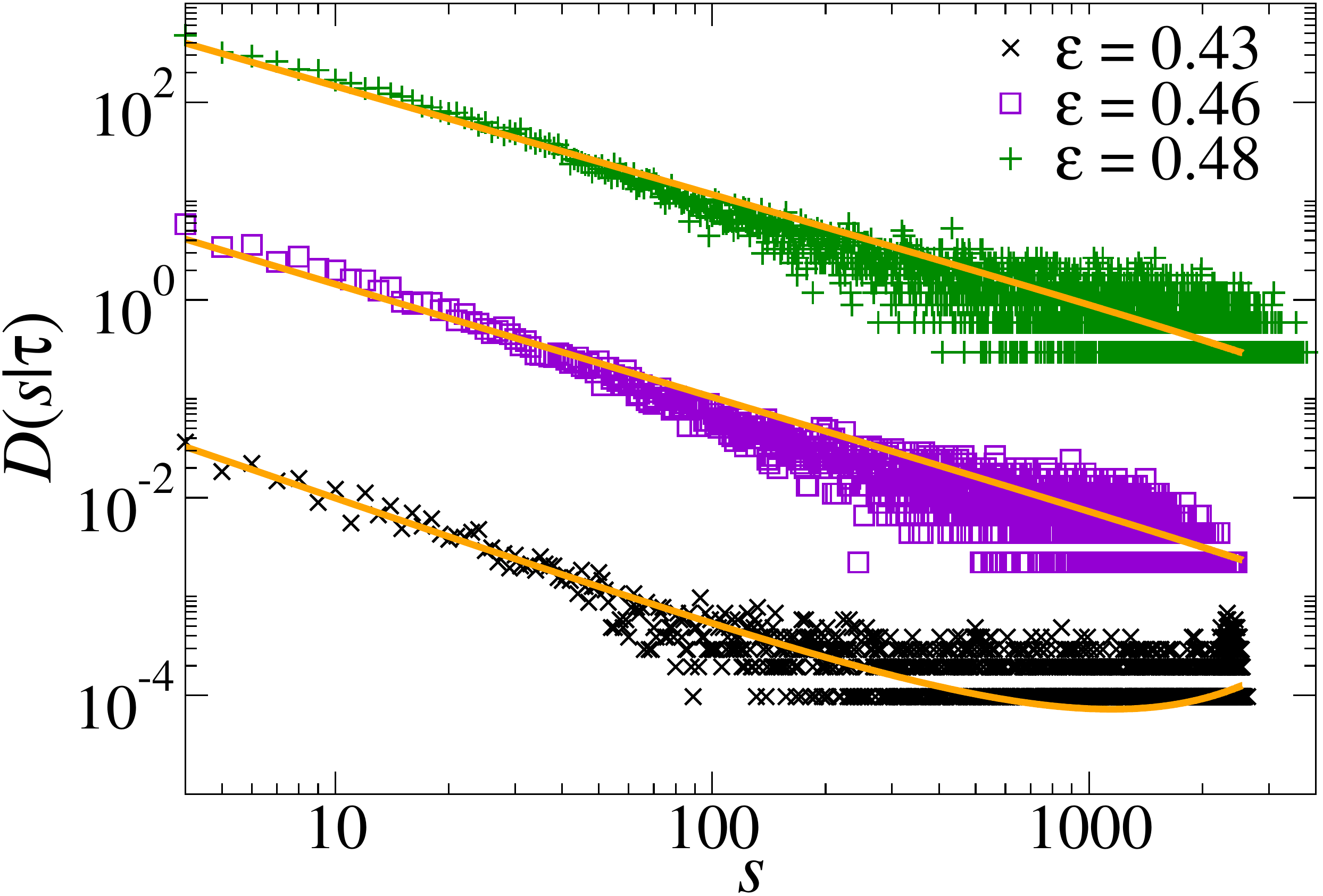}
\caption{\label{Fig:Ds_tau_minimumLambda}  (Color online)   Avalanche size distribution $D(s|\tau)$ for three different values of $\epsilon$ corresponding to the value of $\tau$ at which the cut-off parameter $\lambda$ assumes its minimum value (see Fig.~\ref{Fig:kappa_lambda_Cooling}(b)). The continuous lines show the results of the maximum likelihood fit for each distribution. The distributions for $\epsilon=0.46$ and $\epsilon=0.48$ have been shifted vertically for clarity. Systems of size $L=51$.}
\end{figure}

If instead of slip disorder we now focus on the transformation itself, we observe that for materials in regime II$^\prime$ ($\epsilon \gtrsim 0.45$) the avalanche size distribution $D(s|\tau)$   exhibits extended power-law scaling  around  $\tau_{\text{c}}$, see Fig.~\ref{Fig:Ds_vs_s_tau}. A maximum likelihood fit~\cite{Baro_PRE2012_PLML,Riley_Hobson_Bence_MathsBook2006} with
$$D(S|\tau) \sim S^{-\kappa} e^{-\lambda S}$$ 
gives  at this point  $\lambda=0$ and $\kappa \simeq 1.1$  (see Fig.~\ref{Fig:kappa_lambda_Cooling}). 
The power-law behavior at $\tau_c$ is illustrated in Figure~\ref{Fig:Ds_tau_minimumLambda} which shows examples of the fits to $D(s|\tau)$ in regime II$^\prime$ for $\epsilon=0.46$ and $0.48$.
The behavior at other values of $\tau$  is  subcritical  (`pop') with $\lambda>0$. In contrast, for materials with $ \epsilon \lesssim 0.45$, the  dislocation activity  is too  weak to ensure  `criticality' and   during each cycle  the parameter $\lambda$  assumes negative values. 

The fit to the data for $\epsilon =0.43$ (regime II) in Fig.~\ref{Fig:Ds_tau_minimumLambda} shows that $\lambda<0$ is associated with a characteristic peak in $D(s|\tau)$ at large avalanche sizes. This indicates the occurrence of snap events analogous to an infinite avalanche in weakly disordered RFMs~\cite{Handford_PRE2013a,Sethna2001,Sethna_review2004,Vives1995Universality}. 

\begin{figure}[h]
\includegraphics[width=6cm]{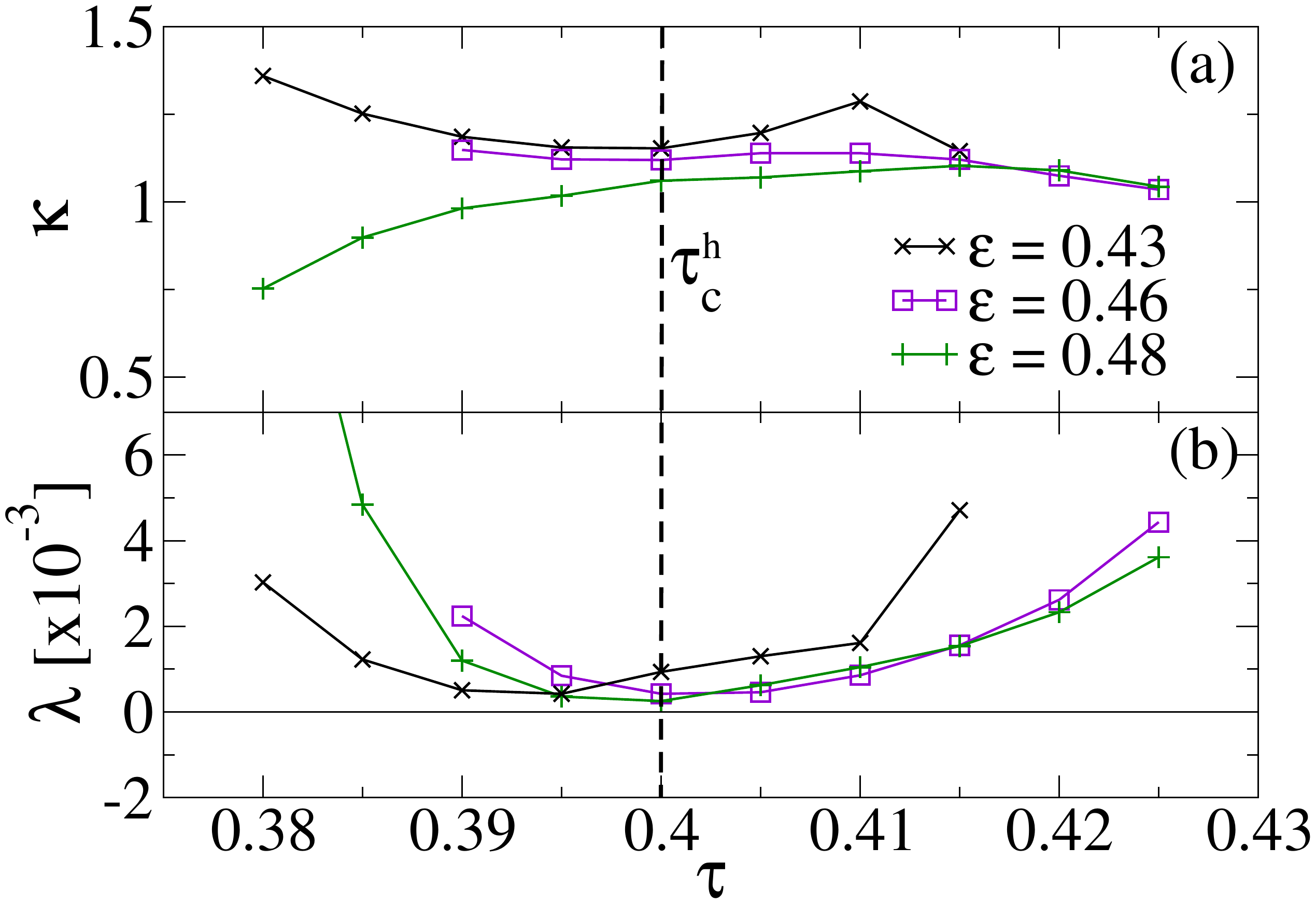}
\caption{\label{Fig:kappa_lambda_Heating} (Color online) Temperature dependence of (a) the power-law exponent, $\kappa$, and (b) the cut-off, $\lambda$, of the avalanche size distribution for heating runs. Different symbol types correspond to different values of $\epsilon$, as marked by the legend. The vertical dashed line indicates our estimate for the critical driving parameter for heating runs.}
\end{figure}

Figure ~\ref{Fig:kappa_lambda_Heating} illustrates  the fitting of a function $D(S|\tau) \sim S^{-\kappa} e^{-\lambda S}$ to the avalanche size distribution for heating runs. The parameter $\lambda$ approaches the zero value corresponding to a power-law distribution at $\tau_c^{\text{h}} \simeq -0.4$. The value of $\kappa$ at $\tau_c^{\text{h}}$ is compatible with the value $\kappa = 1.1$ obtained for cooling runs, suggesting that the system belongs to the same universality class for cooling and heating runs. Note that $\lambda$ does not get so close to zero during heating runs as for cooling runs (cf. Fig.~\ref{Fig:kappa_lambda_Cooling}). This suggests a slight asymmetry between cooling and heating runs  with the behavior being closer to critical during cooling runs. Besides that, the value of $\tau_c^{\text{h}}$ where $\lambda$ is closer to zero shows a stronger dependence on $\epsilon$ for heating runs than for cooling runs. 

The observed  scaling behavior  for systems in regime II$^\prime$ at $\tau_c \simeq -0.385$ is an indication of the closeness to a critical manifold in the space spanned by structural parameters (transformation strain), driving parameters (temperature in our case) and annealed disorder (variance of slip). We conjecture that for structural transformations in the regime  II$^\prime$ which generate enough slip,  the periodically driven system  ratchets in the space of parameters until it eventually approaches or even crosses such  manifold. The fact that the ensuing scaling without extrinsic tuning of disorder and that criticality only occurs at a critical value of the driving parameter, explain the  experiments presented in~\cite{Chandni_PRL2009}.  
Also note that the scale-free behavior is robust with respect to $\epsilon$ provided $\epsilon$ is large enough for the system to be in regime II$^\prime$ (i.e. for the system to generate enough disorder).
Each  encounter with the critical manifold, implied by this scenario,  would mean that the  system  passes close to the classical critical manifold in the corresponding  system with \emph{quenched} disorder~\cite{Handford_PRE2013a,Sethna2001,Sethna_review2004,Vives1995Universality}.  In the language of dynamical systems, the observed scaling may also mean  that  around  $\tau_{\text{c}}$ the extremely long chaotic transients pass in the vicinity of a chaotic saddle~\cite{Tel-YingChengLai_PhysRep2008}. 

\vskip15pt
\subsection{Universality}

We now investigate the effect of correlated non-Gaussian disorder, $\h$, and long-range interaction, $\J$, on the universality class of the observed criticality. First, we analyze the effect of disorder by simplifying the model assuming that  the disorder is quenched; to prevent  the development of additional  disorder  we  set $\gamma_{\text{S}}(\epsilon) = \infty$,  which makes  the parameter  $\epsilon$   irrelevant. Assuming that $\h$  is Gaussian  quenched uncorrelated  (GQU) disorder with zero mean and standard deviation $\Delta^{\text{G}}$, we recover a  long-range version of an athermal random field Blume-Emery-Griffiths model (RFBEG) \cite{Vives1995Universality}  with interactions given by  the kernel $\J$ and a parametric dependence on $\tau$. Numerical simulations of this model driven through $\tau$ reveal a  critical point  at  $\Delta_{\text{c}}^{\text{G}} = 0.05$ and $\tau_{\text{c}}^{\text{G}}=-0.37$ (for  the cooling paths).  Within numerical errors, the value of the  exponent $\kappa^{\text{G}}$ evaluated at $\tau_{\text{c}}^\text{G}$ coincides with the value of $\kappa$ at $\tau_\text{c}$ (see red circles in Fig.~\ref{Fig:kappa_lambda_Cooling}).   Instead,  the maximum likelihood fits to the integrated avalanche size distribution in our model with long-range interaction, $D(S)=\int_{-\infty}^{\infty} D(S|\tau) \text{d}\tau$, give an exponent $\kappa^\prime=1.67$ which differs from the analogous value $\kappa^\prime = 1.3$ obtained in  a RFBEG with  short-range interactions ~\cite{Vives1995Universality}. The latter value $\kappa^\prime = 1.3$ is in fact compatible with the exponent for the integrated avalanche size distribution produced by the model in Ref.~\cite{PerezReche2007PRL} where the kernel was assumed to be anisotropic but short-ranged.

From the comparison with the quenched-disorder version of our model and models with short-range interactions, we  conclude that the universality class of our system is controlled by  long-range  interactions but it is largely insensitive to the correlations in disorder.  This   is  in qualitative agreement with predictions of the equilibrium random-field Ising model with   interactions $J(r) \sim r^{-(d+b)}$ and disorder correlations $C_{\text{h}}(r) \sim r^{-(d-a)}$ which  exhibits mean-field behavior for dimensionality $d>d_{\text{u}}=\min(6,3b)+\max(0,a)$~\cite{Bray_JPhysC1986_LongRangeRF}. Our model with $a=b=0$ and  $d=2>d_{\text{u}}=0$ shows similar insensitivity to the correlations of disorder. The universality class, however, is not the same since the value  $\kappa=1.1$ for our model significantly deviates from the value $\kappa^{MF}=3/2$ in mean-field theories~\cite{Dahmen2009,Sethna1993}. The difference may be associated with the topology of interactions:  mean-field models imply a tree-like topology ~\cite{Handford_PRE2013a,Sabhapandit2000,Broker-Grassberger_PRE1997,Kinouchi_PRE1999} while our model and other models predicting similar values for $\kappa$ ~\cite{Pazmandy_PRL1999,LeDoussal_PRB2012_SK-Analytics,Talamali-Roux_PRE2011,Ispanovity2014a}  preserve crucial  short  interaction  loops (while remaining long-range).

Our results correspond to a 2D system but, based on the above considerations, we can expect that the model can capture the universality class of at least some realistic martensites with $d=3 >d_{\text{u}}=0$. In order to compare our results with experimental data, we assume that the size of avalanches is proportional to their energy. Experimental results on the energy distribution of avalanches are available for the integrated distribution in the whole range of loading~\cite{Planes_JAlloysCompounds2013}. In particular, the exponent $\kappa^\prime=1.67$ for the integrated avalanche size distribution predicted by our model is close to the value $\kappa^\prime_{\text{exp}} = 1.6$ (denoted as $\epsilon$ in~\cite{Bonnot_PRB2008_AE-FePd,Planes_JAlloysCompounds2013}) for the avalanche energy distribution in materials undergoing a thermally-driven transformation from face-centered-cubic (fcc) austenite to face-centered-tetragonal (fct) martensite~\cite{Bonnot_PRB2008_AE-FePd,Planes_JAlloysCompounds2013}. Experiments suggest that the value of the exponent depends on the number of martensite variants. It is encouraging that the exponent predicted by our model with two variants is close to the exponent for an fcc-fct transition with three variants of martensite. It is also interesting that the experimentally obtained exponent for single-crystal materials is the same for cooling and heating runs~\cite{Bonnot_PRB2008_AE-FePd}, in agreement with our predictions for systems in regime II$^\prime$. Extending our theoretical framework to phase transitions with more variants could provide a link between the number of variants of martensite and the universality class.

\section{Marginal stability}
\label{sec:MarginalStability}

To understand the  robustness of the critical behavior in our model,  we  now focus on the  \emph{reproductive number} $R$, defined as the mean number of elements becoming unstable after the transformation of every element during avalanches~\cite{Pazmandy_PRL1999,Kinouchi_PRE1999}.  Marginal stability corresponds to $R=1$  when, in average, each transforming element triggers a transformation in exactly one element. This regime separates unstable ($R>1$) and stable states ($R<1$) in the configuration space. Marginal stability appears to be a generic signature of criticality \cite{Muller-Wyart_annurev-conmatphys2015}.
The question is whether the observed scaling in the neighborhood of $\tau_{\text{c}}$ is linked to the fact that the system reaches the state of marginal stability in the sense that it approaches the threshold $R=1$.

 An analytical  study of the dependence of $R$ on $\tau$ is a challenging task because it requires discriminating the numbers of elements becoming unstable  at different stages of an avalanche. Therefore, we first study  the number $R^{(1)}({\bf r}|\tau)$ of elements becoming unstable after an element triggers an avalanche  at ${\bf r}=(k,l)$ which is more amenable to analysis. The reproductive number changes during avalanches, being larger at the beginning of avalanches than at the end when activity fades out~\cite{Handford_PRE2013a}. Accordingly, the knowledge of $R^{(1)}({\bf r}|\tau)$ should give an upper bound to the mean reproductive number in complete avalanches, $\bar{R}(\tau)$. At the end of this section we compare the semi-analytical  estimates for $R^{(1)}({\bf r}|\tau)$ with the actual function $\bar{R}(\tau)$ computed numerically.
 
\subsection{Instability mechanisms}
\label{sec:StabilityOnsetAvalanche}

In our model, an instability takes place when the local strain reaches one of the three thresholds, $\gamma_{AM}(\tau)$, $\gamma_{\text{MA}}(\tau)$ and  $\gamma_{\text{S}}(\epsilon)$ given in Eqs.~\eqref{eq:gamma_AM}-\eqref{eq:gamma_S}. 

The conditions for instabilities can be described in terms of the `stability' fields (analogous to those defined in \cite{Pazmandy_PRL1999, Muller-Wyart_annurev-conmatphys2015}):
\begin{align}
\label{eq:mu_def}
\mu_{\text{AM}} &= \gamma_{AM}(\tau)-|\gamma|~,\\
\mu_{\text{MA}} &= |\gamma|-\gamma_{\text{MA}}(\tau)~,\\
\mu_{\text{S}} &= \gamma_{\text{S}}(\epsilon)-|\gamma|~,
\end{align}
which quantify the distance to the stability limits $\{\gamma_t; t=\text{AM, MA, S}\}$. Unstable elements with respect to a transition of type $t$ have $\mu_t<0$. 
While there are  several types of instabilities in our system,  we argue below that  the dominating contribution  to the reproductive number is associated with the phase transition events.

\begin{figure}
\includegraphics[angle=0,width=0.35\textwidth]{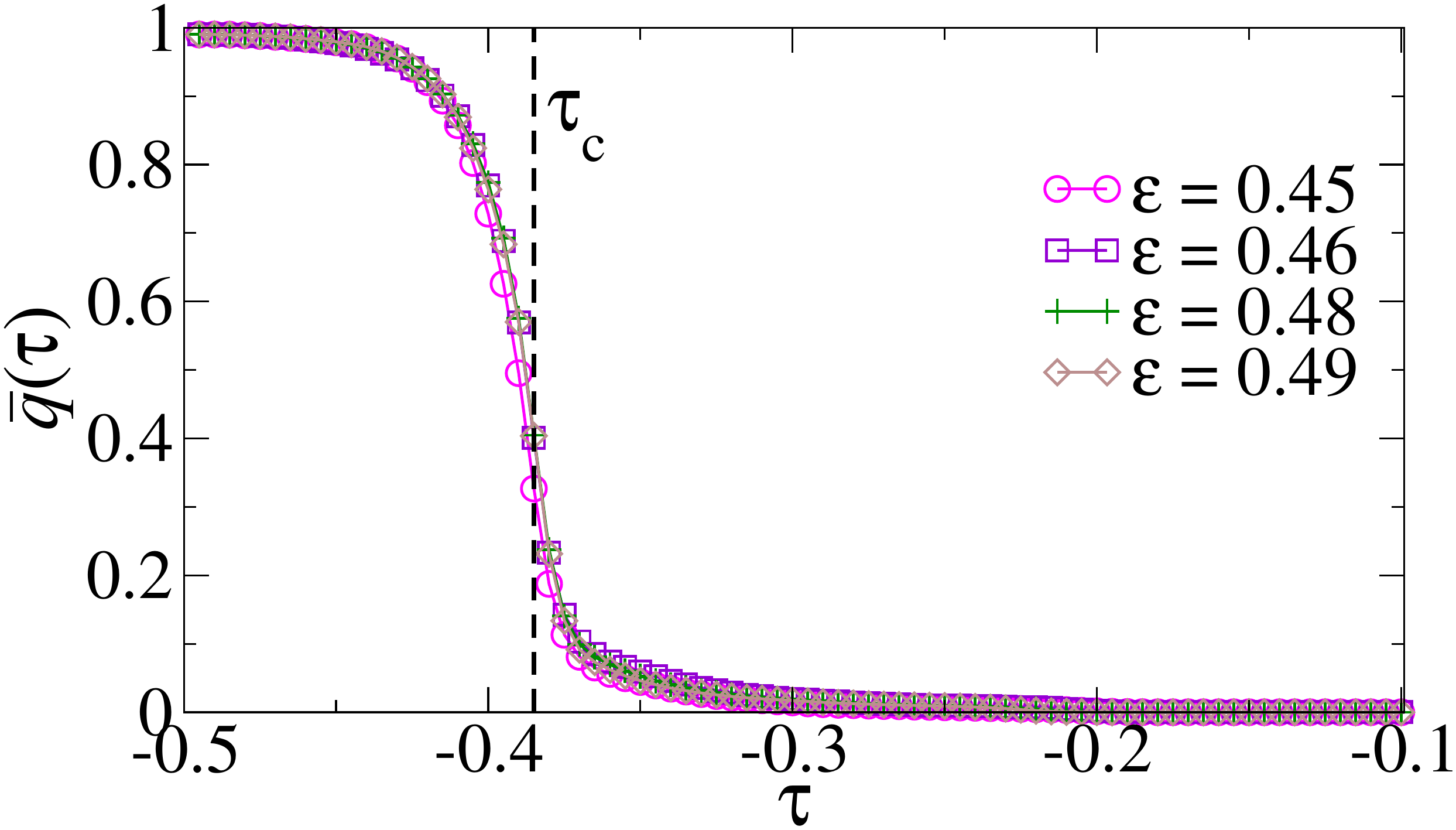}
\caption{\label{Fig_q_vs_tau}  (Color online) Mean fraction of elements in martensite for cooling runs in systems of size $L=51$. Data are obtained by averaging the value of $q$ at given $\tau$ over $10^5$ cycles during super-transients. Different symbols correspond to different values of $\epsilon$ in the regime with super-transient behavior, as marked by the legend.}
\end{figure}

The number of elements becoming unstable after the phase transformation of an element at a site ${\bf r}=(k,l)$ can be computed as a sum of three terms:
\begin{equation}
\label{eq:R_3Contributions}
R^{(1)}({\bf r}|\tau)=R^{(1)}_{\text{AM}}({\bf r}|\tau)+R^{(1)}_{\text{MA}}({\bf r}|\tau)+R^{(1)}_{\text{S}}({\bf r}|\tau)~.
\end{equation}
The right hand side of this  equation  contains  the mean number of elements reaching each of the three stability limits, $\{\gamma_t; t=\text{AM, MA, S}\}$.  Each of  these contributions  can be split in turn  into the product of two terms:
\begin{align}
\label{eq:Rt_AM}
R_{\text{AM}}^{(1)}({\bf r}|\tau)&=(1-\bar{q}(\tau))\, U_{\text{AM}}^{(1)}({\bf r}|\tau)~,\\
\label{eq:Rt_MA}
R_{\text{MA}}^{(1)}({\bf r}|\tau)&=\bar{q}(\tau)\, U_{\text{MA}}^{(1)}({\bf r}|\tau)~,\\
\label{eq:Rt_S}
R_{\text{S}}^{(1)}({\bf r}|\tau)&=\bar{q}(\tau)\, U_{\text{S}}^{(1)}({\bf r}|\tau)~.
\end{align}
Here, $\bar{q}(\tau)$ is the fraction of elements in martensite at given $\tau$ averaged over transformation cycles,  see Fig.~\ref{Fig_q_vs_tau}. The quantities $U_{t}^{(1)}({\bf r}|\tau)$ give the maximum number of elements becoming unstable after an element at ${\bf r}$ undergoes a configuration change of type $t=\text{AM, MA, S}$.  

Now, we can write %
\begin{equation}
\label{eq:Ut_definition}
U_{t}^{(1)}({\bf r}|\tau) =  \sum_{{\bf r^\prime} \neq {\bf r}} \int_0^{|\delta \gamma_{{\bf r,r^\prime}}|} D_t(\mu_t|\tau) \text{d} \mu_t~,
\end{equation}
for $t=\text{AM, MA, S}$. Here $\{D_t(\mu_t|\tau);\,t=\text{AM, MA, S}\}$  are the probability density functions (PDFs) for the stability fields $\mu_t$  at given $\tau$. 

The  upper limit  $\delta \gamma_{{\bf r,r^\prime}}$ of the integral in Eq.~\eqref{eq:Ut_definition} corresponds to the change of relative elastic strain experienced by an element at ${\bf r^\prime}=(i,j)$ after the configuration change of an element at site ${\bf r}=(k,l)$. Based on Eq.~\eqref{eq:elastic-strain}, we can write:
\begin{equation}
\label{eq:delta_gamma_rrp}
|\delta \gamma_{{\bf r,r^\prime}}|=
\begin{cases}
|J_{{\bf r,r^\prime}}|~& \text{if phase change at ${\bf r}$,}\\
(\epsilon^{-1}-2)|J_{{\bf r,r^\prime}}|~& \text{if slip at ${\bf r}$}~, 
\end{cases}
\end{equation}
where $\{J_{{\bf r,r^\prime}}\}=\{J_{i,j,k,l}\}$ is a more economical notation for the kernel elements. Elastic strain changes associated with elements undergoing slip are smaller by a factor $(\epsilon^{-1}-2)$ compared to those associated with phase changes. For systems with relatively large  $\epsilon \gtrsim 0.35$  where slip disorder is unlocked (i.e. for systems in regimes II and II$^\prime$), the number of elements becoming unstable after a phase change (Eq.~\eqref{eq:Ut_definition}) is then significantly larger than after a slip event. The phase transition events are therefore the main contributors to $U_{t}^{(1)}({\bf r}|\tau)$ and $R_{t}^{(1)}({\bf r}|\tau)$.

Note also that avalanches are necessarily triggered by one or more elements changing phase; avalanches cannot start with a slip event in our thermally-driven model. Indeed, the absence of thermally activated events implies that an avalanche can only be induced by changing the driving parameter $\tau$ which in turn  can only induce phase transitions (recall that in Eqs.\eqref{eq:gamma_AM}-\eqref{eq:gamma_S} only the stability limits for phase transitions $\gamma_{\text{AM}}(\tau)$ and $\gamma_{\text{MA}}(\tau)$ depend on $\tau$). Therefore in what follows we focus exclusively on the analysis of instabilities triggered by elements undergoing a phase change. 

\subsection{Simplified model}
\label{sec:Analytical-R1}

We start with the analysis of  a simplified analytical model to study the main factors responsible for the dependence of $R^{(1)}(\r|\tau)$ on the driving parameter $\tau$.

Note that the definition of stability limits implies compact,  $\tau$-dependent support for $D_t(\mu_t|\tau)$.  More precisely, the stability fields can  take values only in a finite interval:
$$\mu_t \in [0,\mu_t^{\max}(\tau)],$$ 
where 
\begin{equation}
\label{eq:sigma_t_max}
\mu_t^{\max}(\tau)=
\begin{cases}
\gamma_{\text{AM}}(\tau)=\tau+\frac{1}{2}~& \text{if $t=$AM}\\
\gamma_{\text{S}}(\epsilon)-\gamma_{\text{MA}}(\tau)=\frac{\epsilon^{-1}-1}{2}-\tau~& \text{if $t=$MA, S}~.
\end{cases}
\end{equation}
Now, since the interaction kernel $J_{\r,\r^\prime}$ decays with distance, $r=|\r-\r^\prime|$, as $J_0/r^2$, we know that  $|\delta \gamma_{\r,\r^\prime}| \sim J_0/r^2$ (see Eq.~\eqref{eq:delta_gamma_rrp}).  We can then approximate the cumulative probability density function $F_t(|\delta \gamma_{{\bf r,r^\prime}}|\,|\tau)=\int_0^{|\delta \gamma_{{\bf r,r^\prime}}|} D_t(\mu_t|\tau) \text{d} \mu_t$  by the expression
\begin{align}
\widetilde{F}_t(|\delta \gamma_{{\bf r,r^\prime}}|\;|\tau)=
\begin{cases}
F_t(J_0r^{-2}|\tau)~& \text{if $r>\sqrt{J_0\mu_t^{\max}(\tau)}$}\\
1~& \text{if $r \leq \sqrt{J_0\mu_t^{\max}(\tau)}$}~.\\
\end{cases}
\end{align}
Hence, the maximum number of unstable elements triggered by a  localized phase change from austenite to martensite can be approximated by the expression
\begin{align}
\label{eq:Ut_Approx1}
U_t^{(1)}({\bf r}|\tau) &\sim \widehat{U}_t^{(1)}(\tau) \equiv \int_1^L \text{d}r r \widetilde{F}_t(|\delta \gamma_{{\bf r,r^\prime}}|\,|\tau) \nonumber\\
&= \frac{1}{2} \Big[J_0 \mu_t^{\max}(\tau)-1 \nonumber \\
&\quad + J_0 \int_0^{1/\mu_t^{\max}(\tau)} \text{d}x \frac{F_t(x|\tau)}{x^2}\Big]~,
\end{align}
where we  assumed large system size and  replaced $L^{-1}$ by zero  in the lower limit of the last integral.   If  we now, following \cite{Pazmandy_PRL1999,Muller-Wyart_annurev-conmatphys2015},   assume  that for relatively small $\mu$ we can approximate the distribution of `stabilities' by  a power-law, 
\begin{equation}
\label{eq:Dt-powerlaw}
D_t(\mu_t|\tau) \sim C_{t}(\tau) \mu_t^{\alpha_t(\tau)},
\end{equation}
we obtain an explicit approximation  for the maximum number of elements triggered by an elements undergoing a phase transformation:
\begin{equation}
\small
\label{eq:Ut_pseudogap_general}
\widehat{U}_t^{(1)}(\tau) \sim 
\begin{cases}
\begin{aligned}
       & \frac{1}{2} \Big[ J_0\mu_t^{\max}(\tau)-1 \\
       & +  \frac{J_0 C_t(\tau)}{\alpha_t(\tau)} (\mu_t^{\max}(\tau))^{-\alpha_t(\tau)}\Big]
    \end{aligned}
\vspace{0.7cm} &  \text{if } \alpha_t>0\\
\infty & \text{if }   \alpha_t \leq 0~.
\end{cases} 
\end{equation}
From this expression we see that $\widehat{U}_t^{(1)}(\tau)$ is finite if $\alpha_t(\tau)$ is positive. The condition $\alpha_t(\tau)>0$ is typical for systems with long-range interaction~\cite{Lin-Wyart_EPL2014} and it is less restrictive than the stability condition, $\alpha_t(\tau) \geq 1$, for models with mean-field interaction such as the Sherrington-Kirkpatrick model~\cite{Pazmandy_PRL1999}. 
\begin{figure}
\subfloat{\label{Fig_Dsigma_AM_tau}\includegraphics[angle=0,width=0.3\textwidth]{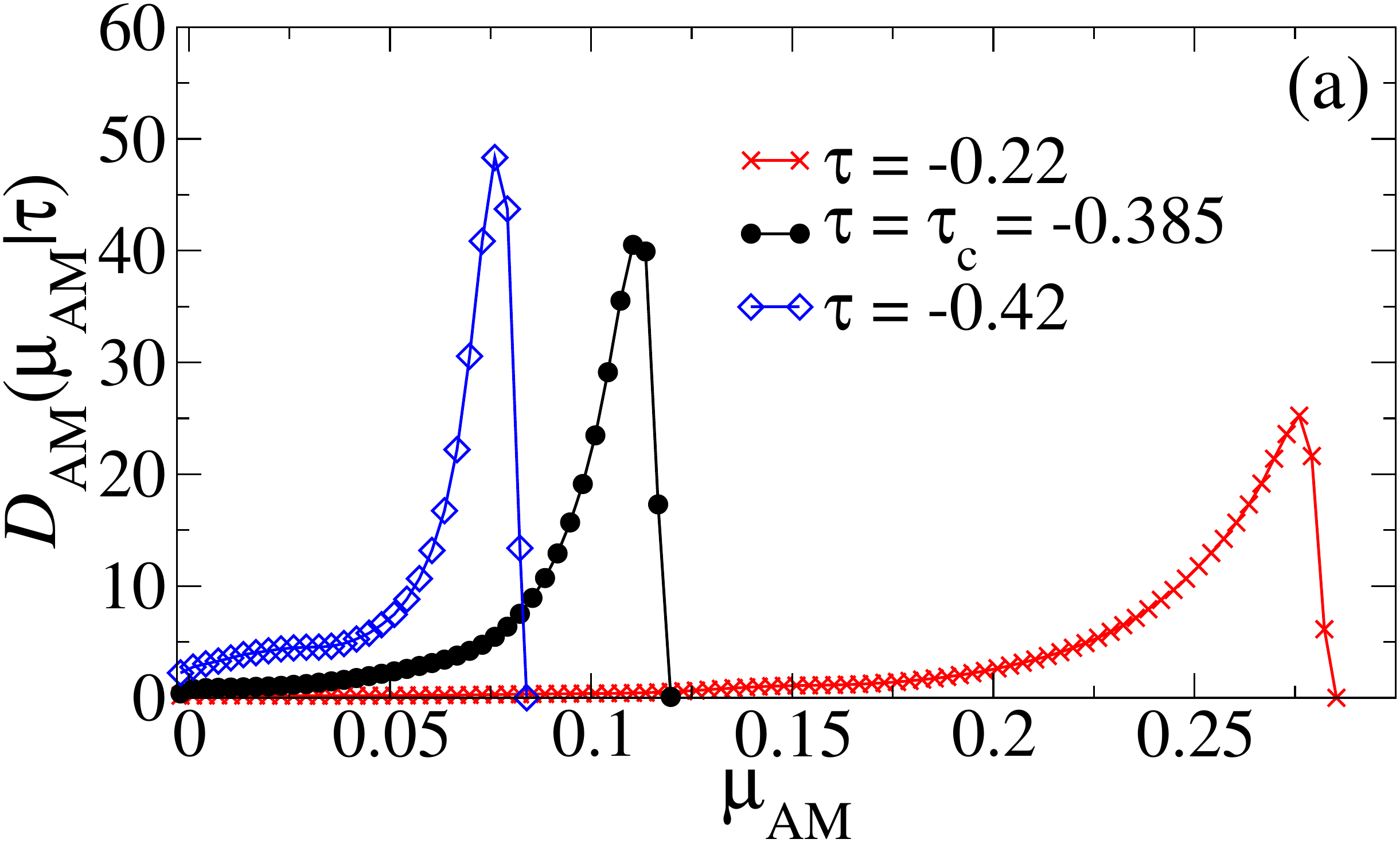}}\\
\vspace{-0.3cm}
\subfloat{\label{Fig_Dsigma_MA_tau}\includegraphics[angle=0,width=0.3\textwidth]{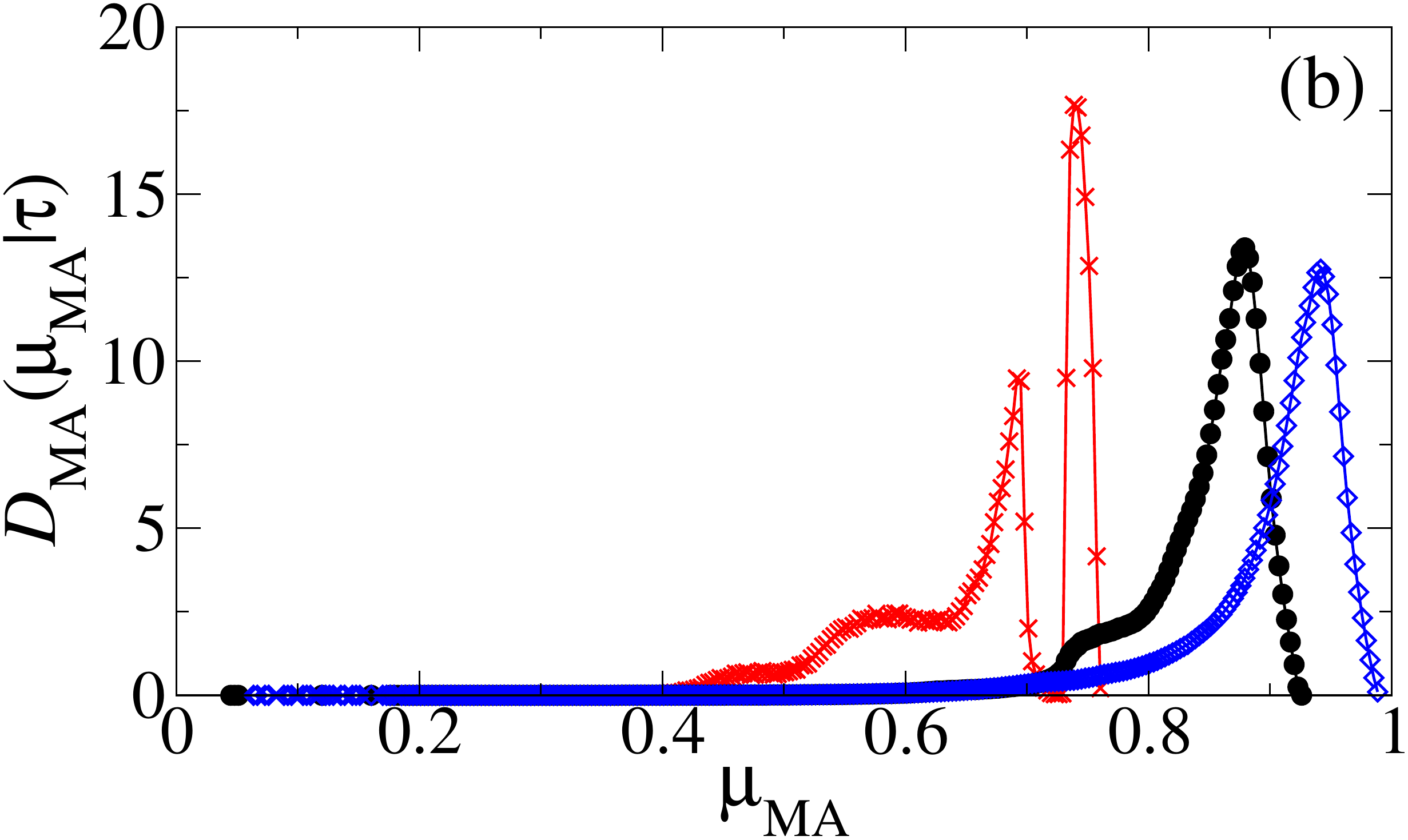}}\\
\vspace{-0.3cm}
\subfloat{\label{Fig_Dsigma_S_tau}\includegraphics[angle=0,width=0.3\textwidth]{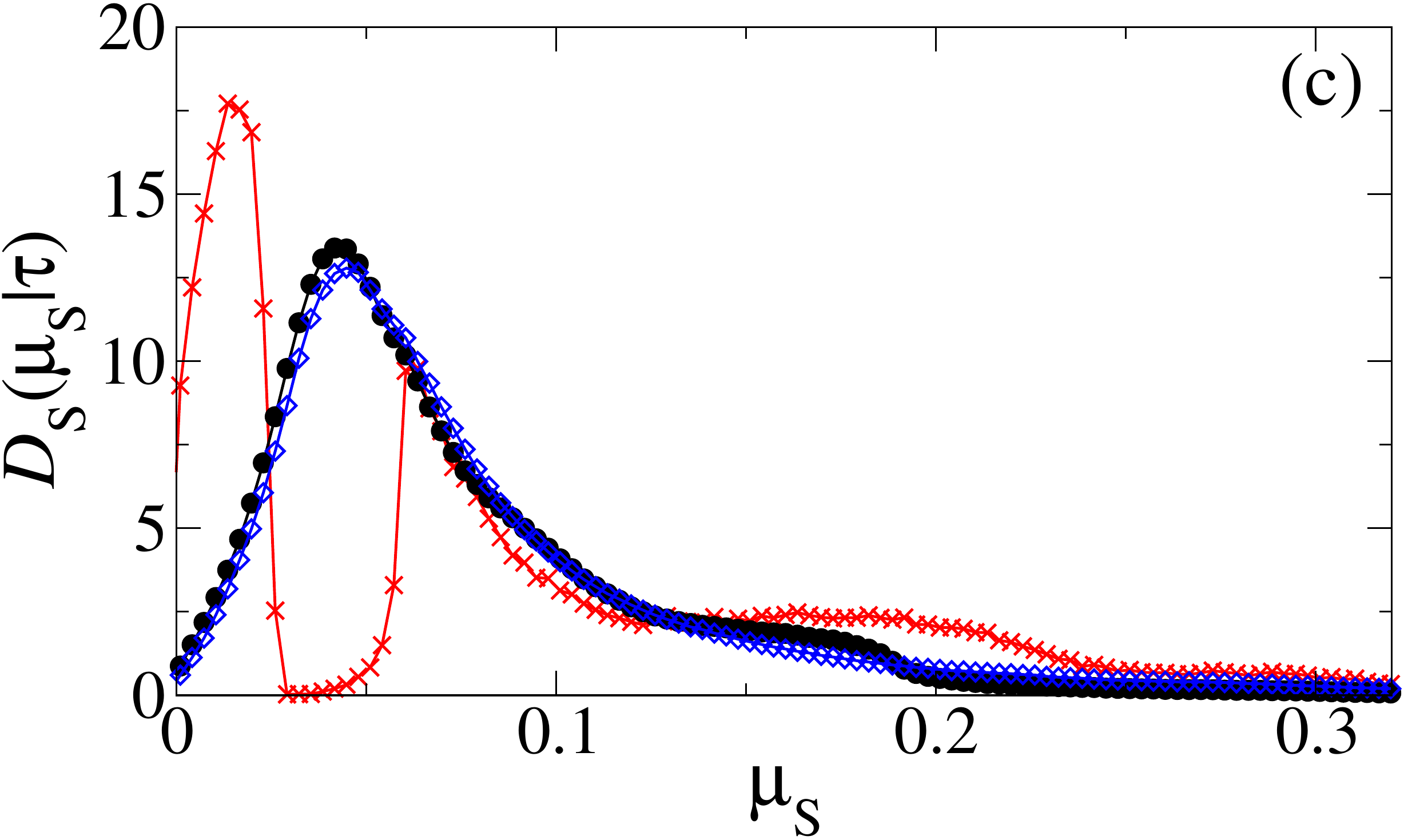}}
\caption{\label{Fig_Plambda_Plasticity_tauc} (Color online) Probability density functions for the stability fields
(a) $\mu_{\text{AM}}$, (b) $\mu_{\text{MA}}$ and (c) $\mu_{\text{S}}$ in a system with $L=51$ and $\epsilon=0.48$. Different symbol types correspond to different values of the driving parameter, $\tau$,  as marked by the legend.}
\end{figure}

This simple model illustrates  the competing influences on  the reproductive number  $R^{(1)}(\r|\tau)$  of the phase fractions described by  $\bar{q}(\tau)$, and  the distribution  of `stabilities',  $D_t(\mu_t|\tau)$, through the functions $\widehat{U}_t^{(1)}(\tau)$ (see Eqs.~\eqref{eq:R_3Contributions}-\eqref{eq:Ut_definition}).  Below we give numerical support for the power-law approximation of  the distribution of stabilities $D_t(\mu_t;\tau)$ (see Eq.~\eqref{eq:Dt-powerlaw}). We then find approximate analytical expressions for $\widehat{U}_t^{(1)}(\tau)$ and compare them with numerical results.

\subsection{Distribution of stabilities}
\label{sec:Numerical-R1}

\begin{figure}
\begin{center}
\subfloat{\label{Fig_Dsigma_given_tau_AM}\includegraphics[angle=0,width=0.3\textwidth]{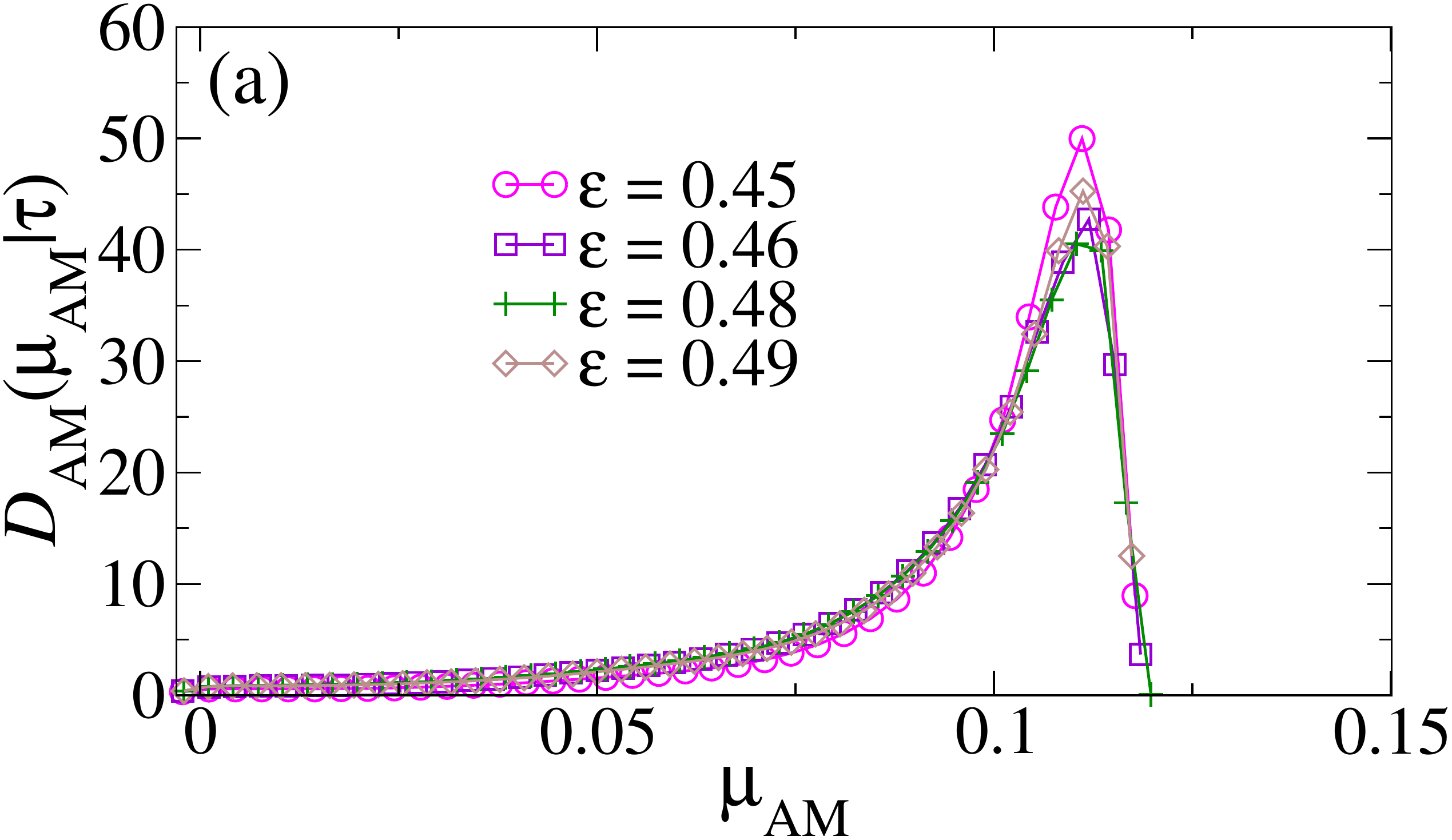}}\\
\subfloat{\label{Fig_Dsigma_given_tau_S}\includegraphics[angle=0,width=0.3\textwidth]{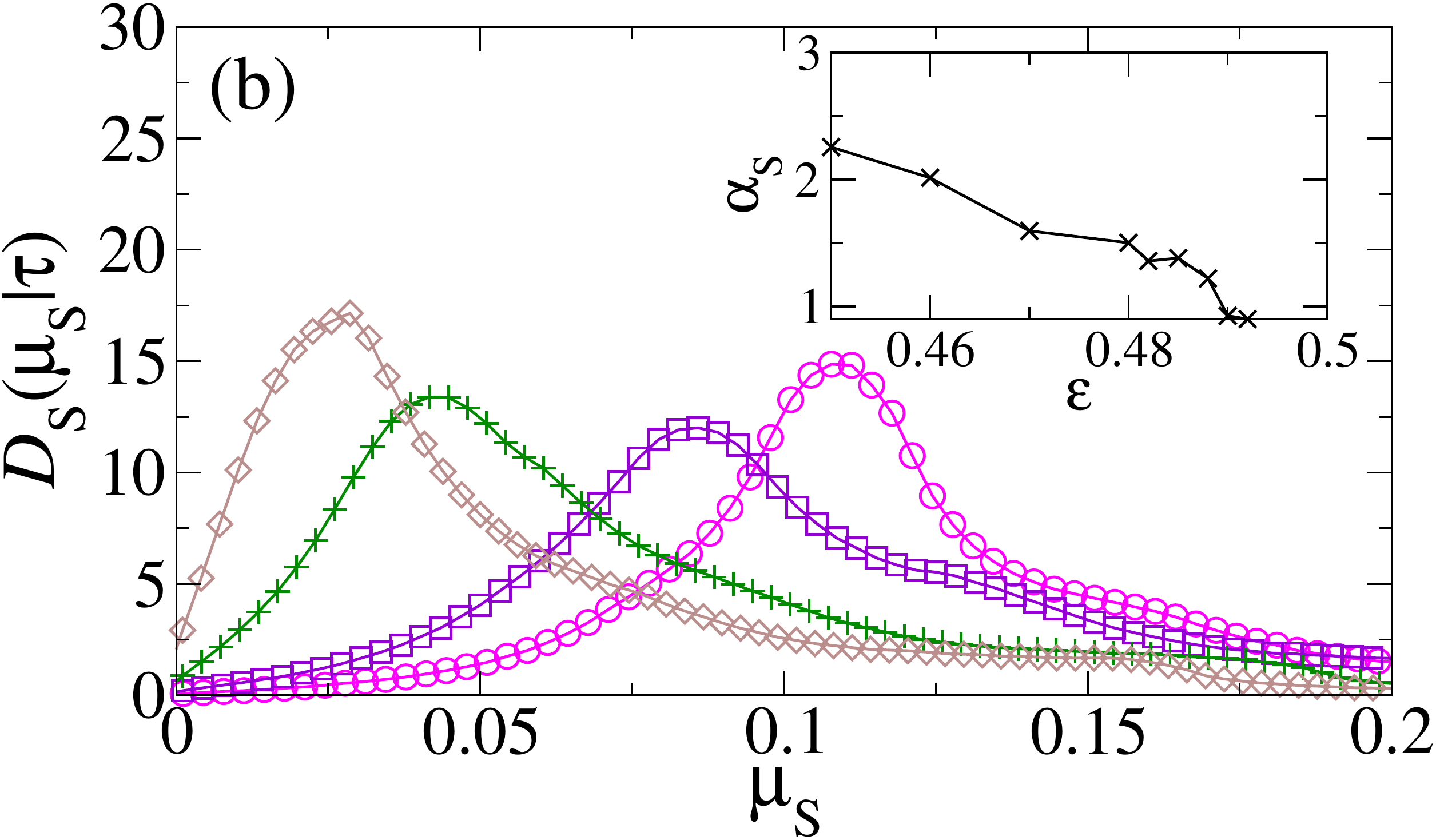}}
\end{center}
\caption{\label{Fig_Dsigma_given_tau} (Color online) Probability density functions (a) $D_{\text{AM}}(\mu_{\text{AM}}|\tau)$ and (b) $D_{\text{S}}(\mu_{\text{S}}|\tau)$ for stability variables at $\tau_{\text{c}}=-0.385$  during cooling runs in systems of size $L=51$. The inset shows the exponent $\alpha_{\text{S}}$ obtained by fitting a power-law $D_{\text{S}}(\mu_{\text{S}}|\tau) \sim C_{\text{S}}(\tau) \mu_{\text{S}}^{\alpha_{\text{S}}(\tau)}$ for small values of $\mu_{\text{S}}$.}
\end{figure}

The numerically obtained stability distributions $D_t(\mu_t|\tau)$ for cooling runs (during super-transients) are illustrated in  Fig.~\ref{Fig_Plambda_Plasticity_tauc}   for several values of $\tau$  and  $\epsilon=0.48$. Figure \ref{Fig_Dsigma_given_tau} shows the distributions $D_{\text{AM}}(\mu_{\text{AM}}|\tau)$ and $D_{\text{S}}(\mu_{\text{S}}|\tau)$ at $\tau_c$ for several values of $\epsilon$. 

One can see that for the values of the driving parameter close to $\tau_c$, where most of the activity occurs (see Fig.~\ref{Fig_q_vs_tau}),  the two stability distributions for phase transitions, $D_{\text{AM}}(\mu|\tau)$ and $D_{\text{MA}}(\mu|\tau)$, can be approximated by a power-law for sufficiently small values of $\mu$ (see Figs.~\subref{Fig_Dsigma_AM_tau}, \subref{Fig_Dsigma_MA_tau} and \subref{Fig_Dsigma_given_tau_AM}). Moreover,  $D_{\text{AM}}(\mu_{\text{AM}}|\tau)$ increases with $\mu_{\text{AM}}$ up to the limit $\mu_{\text{AM}}^{\max}(\tau)=\tau+1/2$, which means that   a power-law dependence  
$$D_{\text{AM}}(\mu_{\text{AM}}|\tau) \sim C_{\text{AM}}(\tau) \mu_{\text{AM}}^{\alpha_{\text{AM}}(\tau)}$$ 
gives a reasonable approximation  for most of the interval $[0,\mu_{\text{AM}}^{\max}(\tau)]$. The leading term in  Eq.~\eqref{eq:Ut_pseudogap_general} is then
\begin{equation} 
\label{eq:UAM-Approx}
\widehat{U}_{\text{AM}}^{(1)}(\tau) \sim (\tau+1/2)^{-{\alpha_{\text{AM}}(\tau)}}~.
\end{equation}

\begin{figure}
\subfloat{\label{Fig_Mean_U_AM}\includegraphics[angle=0,width=0.3\textwidth]{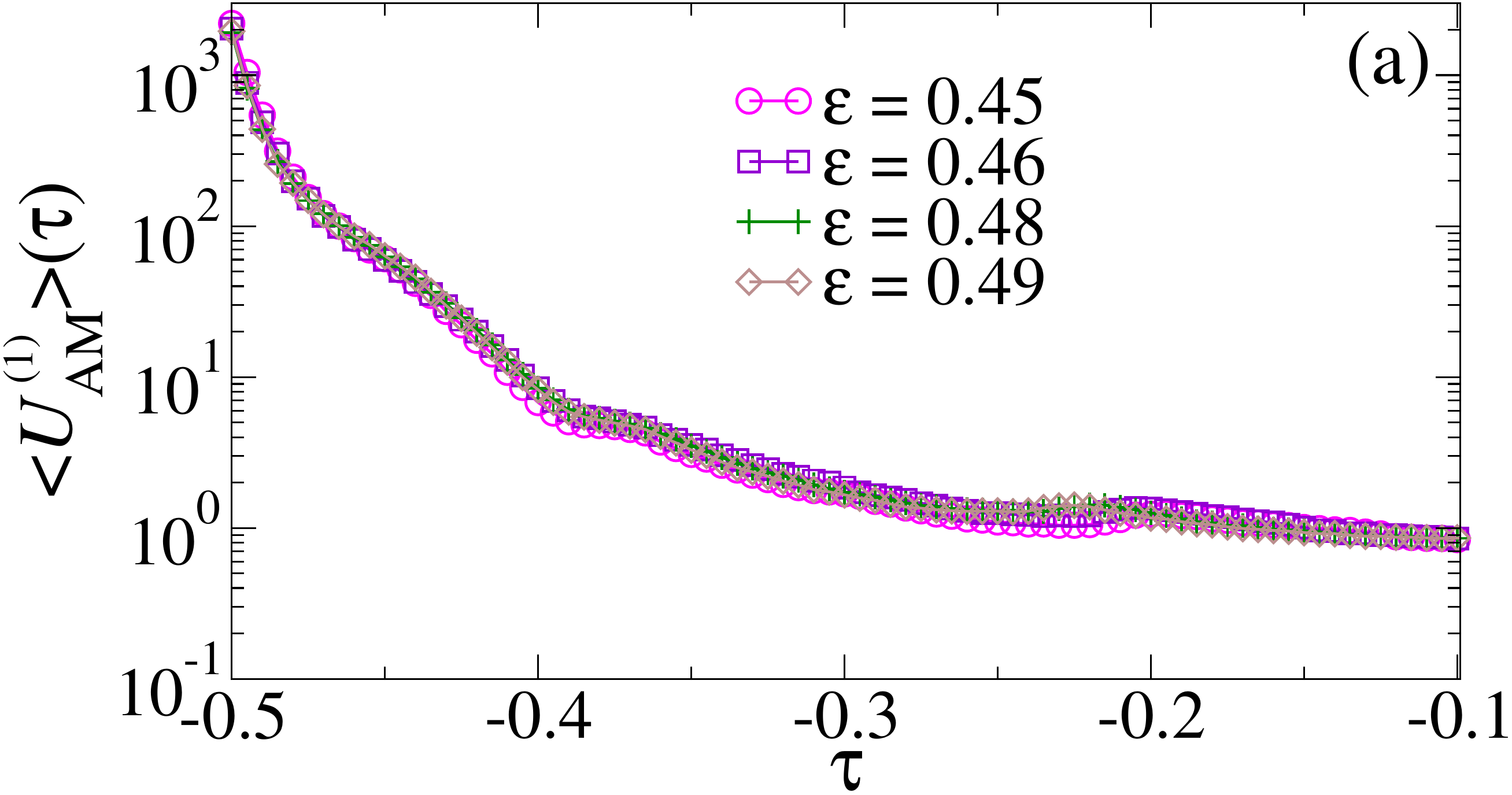}}\\
\subfloat{\label{Fig_Mean_U_S}\includegraphics[angle=0,width=0.3\textwidth]{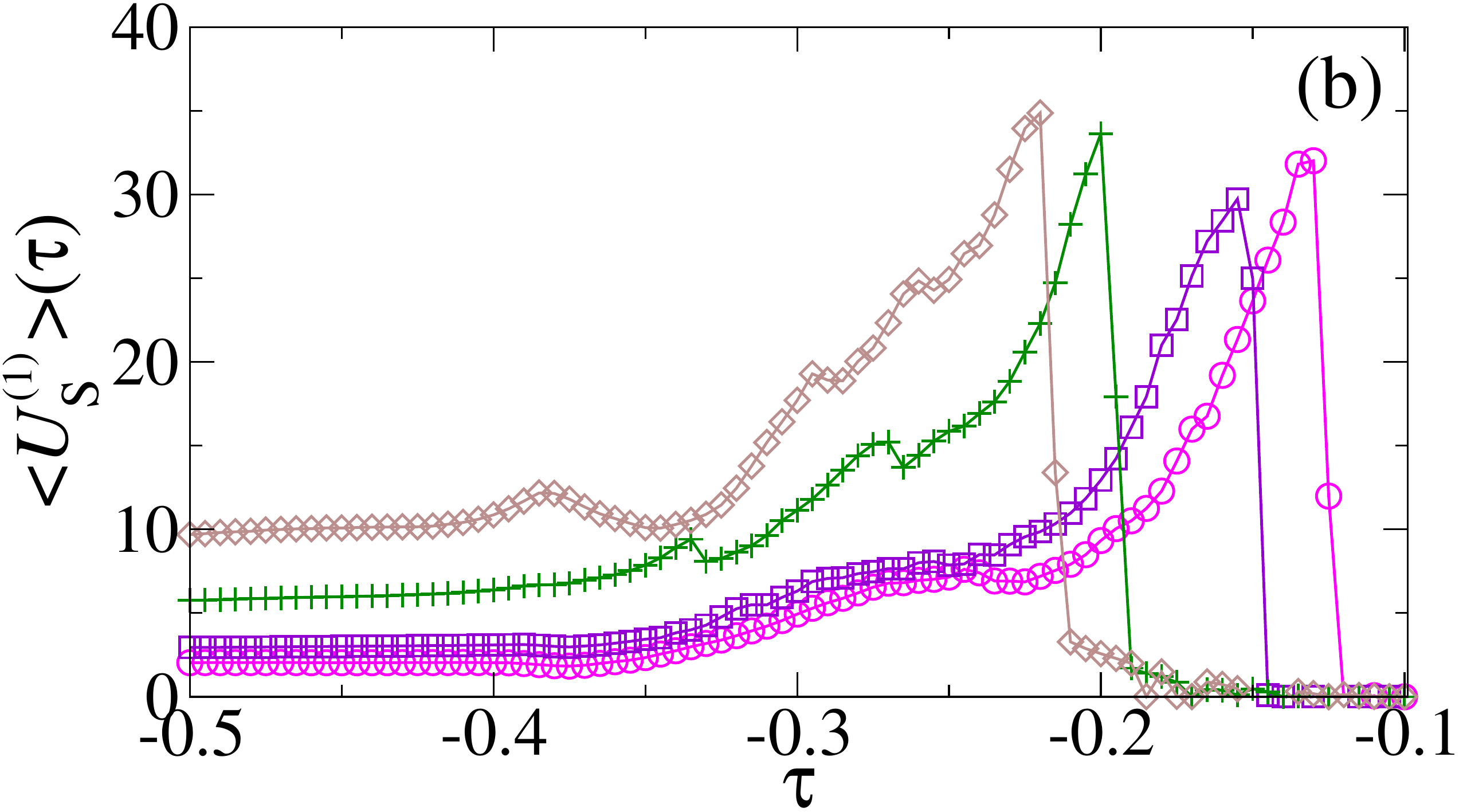}}
\caption{\label{Fig_Mean_U_t} (Color online) Mean maximum number of elements becoming unstable with respect to the (a) AM and (b) S stability limits as a consequence of the phase transition from austenite to martensite of any element in the system. Note the log-normal scales in (a). The system size is $L=51$.}
\end{figure}

This approximation predicts a strong increase of $\widehat{U}_{\text{AM}}^{(1)}(\tau)$ when decreasing $\tau$ which is supported by exact numerical results for $U_{\text{AM}}^{(1)}({\bf r}|\tau)$. The numbers $U_{t}^{(1)}({\bf r}|\tau)$ can be determined exactly from Eq.~\eqref{eq:Ut_definition} using the numerical results for the PDFs of the stability variables,  $D_t(\mu_t|\tau)$, and the interaction kernel $\{J_{{\bf r,r^\prime}}\}$. In order to compare $U_{\text{AM}}^{(1)}({\bf r}|\tau)$ with Eq.~\eqref{eq:UAM-Approx}, we use the mean value of $U_{\text{AM}}^{(1)}({\bf r}|\tau)$ obtained by averaging over all the elements in the system:
 $$\langle U_{\text{AM}}^{(1)} \rangle(\tau)=L^{-2} \sum_{\r} U_{\text{AM}}^{(1)}(\r|\tau).$$
 Figure \subref{Fig_Mean_U_AM} shows that $\langle U_{\text{AM}}^{(1)} \rangle(\tau)$ indeed exhibits a strong increase for decreasing $\tau$, as predicted by our approximation in Eq.~\eqref{eq:UAM-Approx}.

Figures ~\subref{Fig_Dsigma_S_tau} and \subref{Fig_Dsigma_given_tau_S} also support  a power-law approximation for  
$$D_{\text{S}}(\mu_{\text{S}}|\tau) \sim C_{\text{S}}(\tau) \mu_{\text{S}}^{\alpha_{\text{S}}(\tau)}, $$
 at least  for small values of $\mu$ and $\tau$ around $\tau_c$. The exponent $\alpha_{\text{S}}(\tau)$  decreases with increasing $\epsilon$ (see the inset in Fig.~\subref{Fig_Dsigma_given_tau_S}) which  means that the relative fraction of elements in martensite that are close to the stability limit for slip increases with increasing $\epsilon$. This is in agreement with the fact that phase transformations in systems with   larger value of  $\epsilon$    generate more  plasticity. A power-law approximation is not as accurate for  $\widehat{U}_{\text{S}}^{(1)}(\tau)$ as it is for $\widehat{U}_{\text{AM}}^{(1)}(\tau)$ since the decay of $D_{\text{S}}(\mu_{\text{S}}|\tau)$ for large $\mu_{\text{S}}$ extends over a wider interval than the decay of $D_{\text{AM}}(\mu_{\text{AM}}|\tau)$. In spite of that, at low temperatures (expansion around $\tau = -1/2$),  the  power-law approximation, 
$$\widehat{U}_{\text{S}}^{(1)}(\tau) \sim ({\alpha_{\text{S}}(\tau)})^{-1}+1/2+\tau,$$
  gives the leading contribution. This conclusion is based on the  fact that  $C_{\text{S}}(\tau_c) \sim 10^3 \gg 1$ for every $\epsilon$  in regime II$^\prime$  (estimated from the curves in Fig.~\subref{Fig_Dsigma_given_tau_S}). The increasing trend of the approximated $\widehat{U}_{\text{S}}^{(1)}(\tau)$ for increasing $\tau$ is in qualitative agreement with the numerical results for 
$$\langle U_{\text{S}}^{(1)}\rangle(\tau)=L^{-2} \sum_{\r} U_{\text{S}}^{(1)}(\r|\tau)$$ 
presented in  Fig.~\subref{Fig_Mean_U_S}.

For cooling runs, the number $U_{\text{MA}}^{(1)}(\r|\tau)$ of elements undergoing the inverse transition from martensite to austenite is negligible  since $D_{\text{MA}}(\mu_{\text{MA}}|\tau)$ takes very small values in the interval of integration in \eqref{eq:Ut_definition}, see. Fig.~\subref{Fig_Dsigma_MA_tau}. From Eq.~\eqref{eq:Rt_MA} we then conclude that $R_{\text{MA}}^{(1)}({\bf r}|\tau)$ is negligible. This means that during cooling runs elements do not become unstable with respect to the inverse transition from martensite to austenite.

To summarize, our approximate  calculations suggest that the behavior of the function $\widehat{U}_t^{(1)}(\tau)$  can be qualitatively captured based on the knowledge  of the dependence of the limits of stability of the energy wells on the loading parameter $\tau$. From Eqs.~\eqref{eq:R_3Contributions}-\eqref{eq:Rt_S} it is then clear that in our model the behavior of the reproductive number is largely controlled by the interplay between  the   phase fractions and the limits of  stability.

\subsection{Reproductive number}

We can now describe the degree of  stability of our configurations through the reproductive number averaged over all the elements,
$$ \langle R_t^{(1)}\rangle(\tau) = L^{-2} \sum_{\r} R_{t}^{(1)}({\bf r}|\tau)~.$$
The analysis is based on Eqs.~\eqref{eq:Rt_AM}-\eqref{eq:Rt_S} which express $\langle R_t^{(1)}\rangle(\tau)$ in terms of  the mean fraction of elements in the martensitic phase, $\bar{q}(\tau)$, and the maximum number of triggered elements, $\langle U_{t}^{(1)}(\tau) \rangle$.

\begin{figure}
\subfloat{\label{Fig_Mean_R_AM}\includegraphics[angle=0,width=0.3\textwidth]{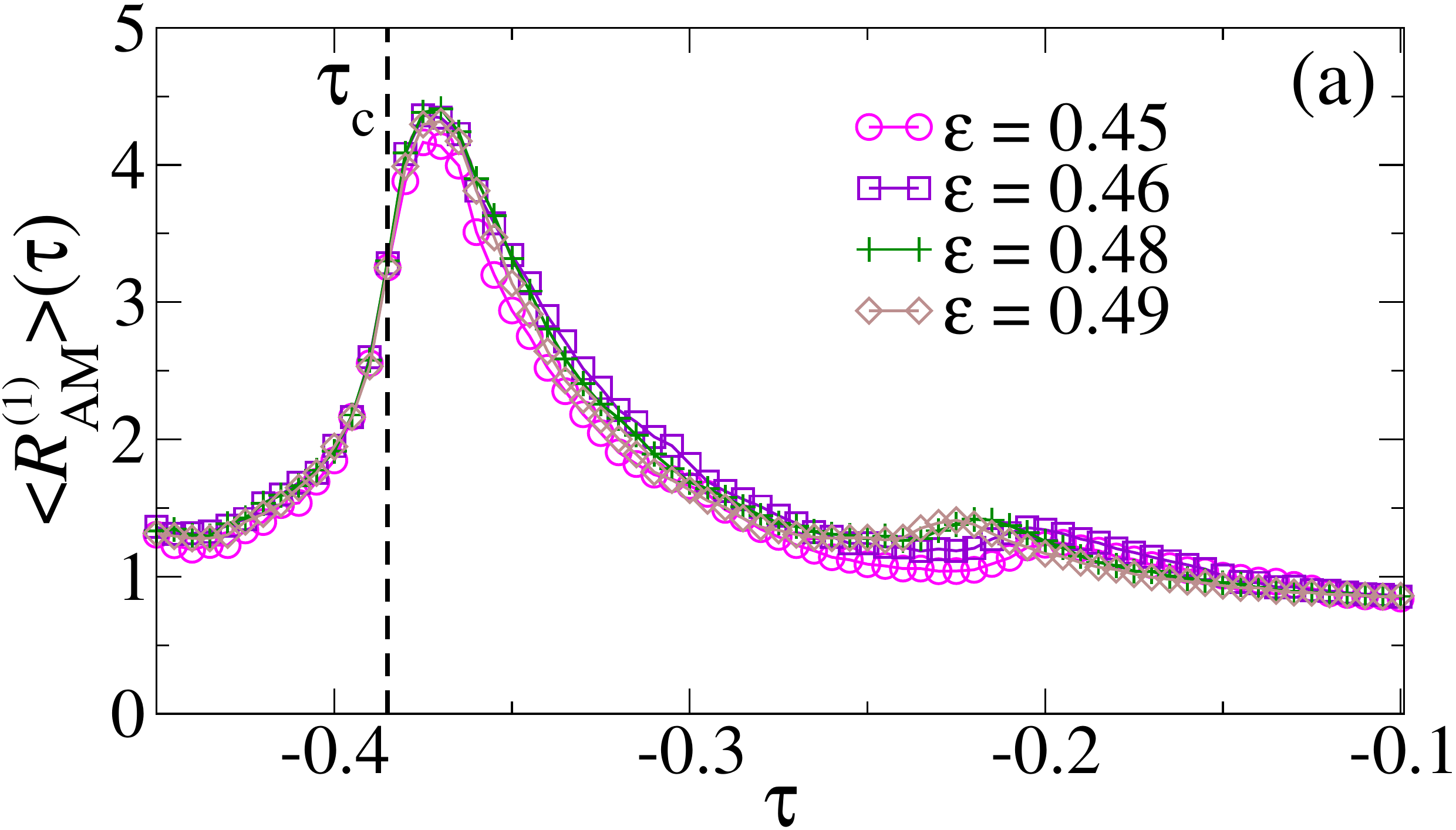}}\\
\subfloat{\label{Fig_Mean_R_S}\includegraphics[angle=0,width=0.3\textwidth]{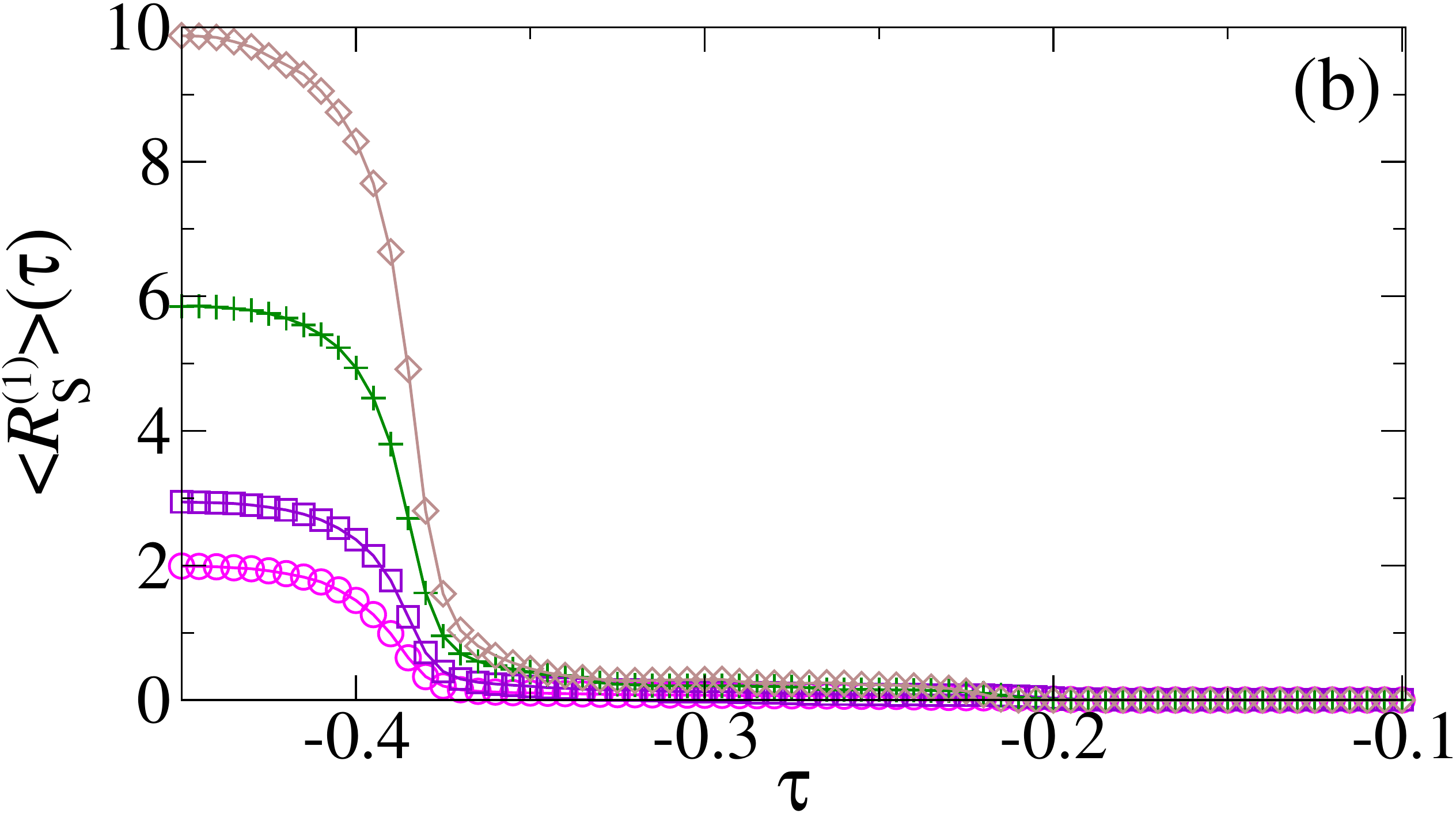}}
\caption{\label{Fig_Mean_R_single} (Color online) Contributions to the mean reproductive number of elements reaching the (a) AM and (b) S stability limits as a consequence of the phase transition from austenite to martensite of any triggering elements in a system of size $L=51$. }
\end{figure}

For elements undergoing an austenite-martensite phase change, the fast growth of $\langle U_{\text{AM}}^{(1)}(\tau) \rangle$ for decreasing $\tau$ competes with the decay of $1-\bar{q}(\tau)$ (cf. Figs.~\subref{Fig_Mean_U_AM} and \ref{Fig_q_vs_tau}). As a result the reproductive number averaged over elements, $\langle R_{\text{AM}}^{(1)}\rangle(\tau)=(1-\bar{q}(\tau))\langle U_{\text{AM}}^{(1)}\rangle(\tau)$, exhibits a peak  for values of $\tau$ close to $\tau_c=-0.385$ (see Fig.~\subref{Fig_Mean_R_AM}). This peak is an indication that a state of marginal stability may be reached around $\tau_{\text{c}}$, however,  no quantitative conclusion can be stated at this point since $\langle R_{\text{AM}}^{(1)}\rangle(\tau)$ overestimates the reproductive number, $R$.

For elements undergoing slip, $\langle U_{\text{S}}^{(1)}\rangle (\tau)$ displays a weak dependence on $\tau$ in the interval with $\tau \lesssim -0.3$ where $\bar{q}>0$. The behavior of $\langle R_{\text{S}}^{(1)}\rangle(\tau)=\bar{q}(\tau)\langle U_{\text{S}}^{(1)}\rangle(\tau)$ is therefore
dominated by the increase of $\bar{q}(\tau)$ with decreasing $\tau$, see Fig.~\ref{Fig_Mean_R_S}.

Note that both  $\langle U_{\text{S}}^{(1)} \rangle (\tau)$ and $\langle R_{\text{S}}^{(1)} \rangle (\tau)$ increase with $\epsilon$ (see Figs.~\subref{Fig_Mean_U_S} and \subref{Fig_Mean_R_S}). This agrees with the fact that the ability of a system to generate transformation-induced slip increases when it approaches the reconstructive limit, $\epsilon=1/2$, where the stability limit for slip vanishes ($\gamma_{\text{S}} = 0$, see Eq.~\eqref{eq:gamma_S}). It is interesting, however, that the specific amount of slip generated by systems with $\epsilon \gtrsim 0.45$ (i.e. those in regime II$^\prime$ which exhibit critical behavior) does not have a significant effect on elements undergoing phase transition. This is clear from Figs.~\subref{Fig_Mean_U_AM} and \subref{Fig_Mean_R_AM} which show a  negligible dependence of $\langle U_{\text{AM}}^{(1)} \rangle(\tau)$ and $\langle R_{\text{AM}}^{(1)}\rangle (\tau)$ on $\epsilon$. Behind this is the fact that slip deformations produce very small elastic strain changes, $|\delta \gamma_{{\bf r,r^\prime}}|$, and despite being potentially numerous, are unlikely to trigger phase changes from austenite to martensite. Note that $\bar{q}(\tau)$ is also largely independent of $\epsilon$ in the regime II$^\prime$, see Fig.~\ref{Fig_q_vs_tau}.

\begin{figure}
\includegraphics[width=6cm]{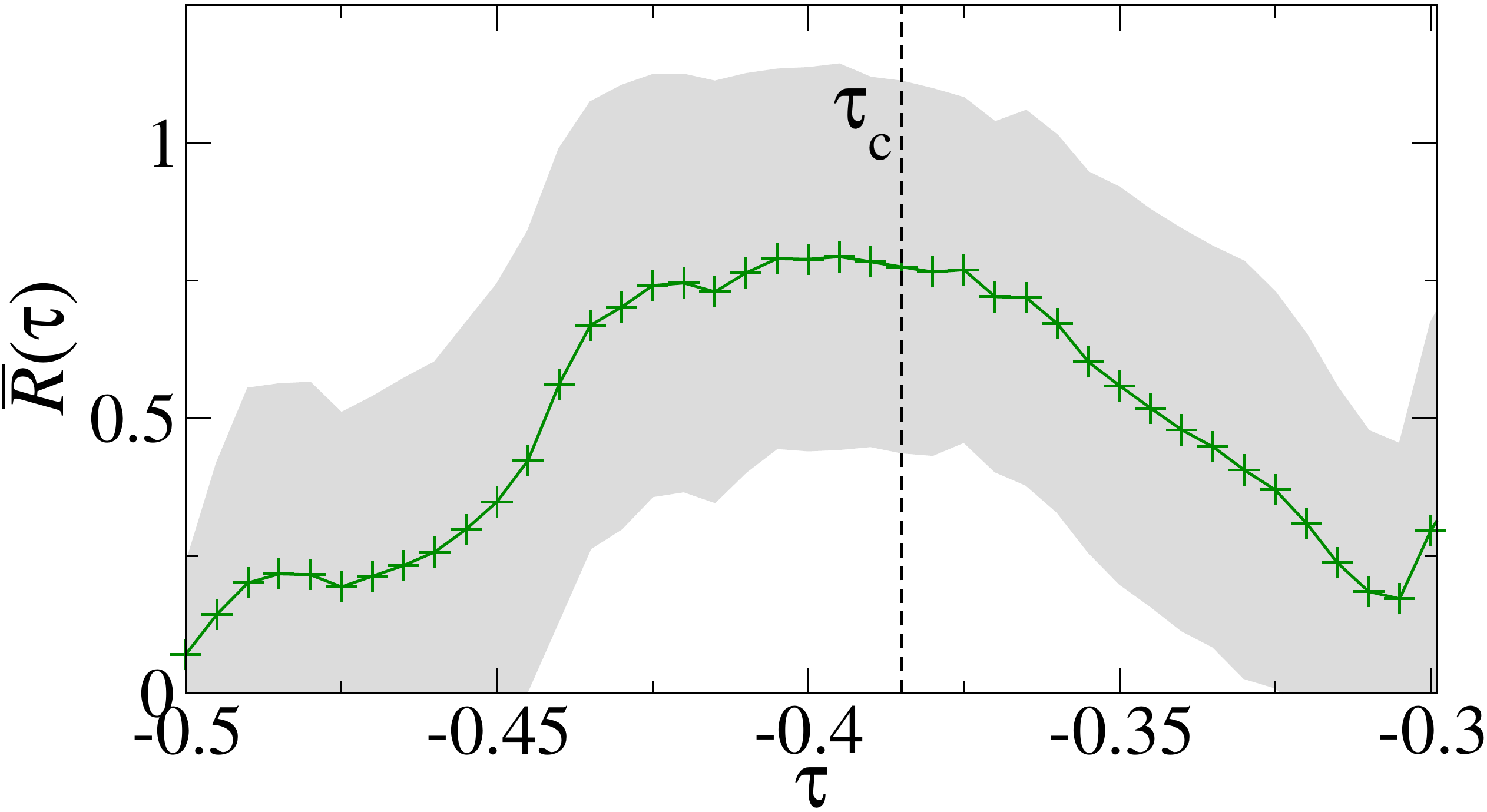}
\caption{\label{Fig:Reproductive_Rate}  (Color online)   Mean  reproductive number $\bar{R}(\tau)$  for individual avalanches during cooling runs in long series of thermal cycles for a system of size $L=51$ with $\epsilon=0.48$. The shaded region indicates the standard deviation of  $\bar{R}(\tau)$.}
\end{figure}

To  check our semi-analytic results numerically, we  define the  mean value of $R$ in individual avalanches (during cooling runs)  as 
$$\bar {R}(\tau)= \frac{1}{{\cal T} -1} \sum_{t=1}^{ {\cal T} -1} \frac{N_{\text{AM}}(t+1|\tau)}{N_{\text{AM}}(t|\tau)}~,$$ 
where $N_{\text{AM}}(t|\tau)$ is the number of elements undergoing the austenite-martensite transformation at time step $t$ during an avalanche at driving field $\tau$. ${\cal T}$ is the duration of the avalanche which in our synchronous dynamics is given by the number of simultaneous updates of transforming elements during an avalanche.
The results, shown in Fig.~\ref{Fig:Reproductive_Rate}, confirm the existence of a peak around  $\tau_{\text{c}}$ and the closeness of this peak to the value $\bar {R}(\tau)=1$ implies that the stability around this point is only marginal.  Comparison of Fig.~\subref{Fig_Mean_R_S} and Fig.~\ref{Fig:Reproductive_Rate} confirms the claim  that the reproductive number at the onset of an avalanche, $\langle R_{\text{AM}}^{(1)}\rangle (\tau)$, predicts the general trend of the mean reproductive number in individual avalanches, $\bar{R}(\tau)$,  while  overestimating the  numerical value. 
 
The fact that $\bar{R}(\tau)$ exhibits a peak and approaches marginal stability only locally in temperature, suggests that the  mechanism of  self-organization to criticality in our model  is  different  from the one in systems where $R=1$  over   an extended  range of values  of the driving parameter~\cite{Muller-Wyart_annurev-conmatphys2015}. Such extended marginality has been reported, for instance, for the Sherrington-Kirkpatrick model~\cite{Pazmandy_PRL1999,LeDoussal-Muller-Wiese_EPL2010} and for models of amorphous solids~\cite{Talamali-Roux_PRE2011,Lin-Wyart_EPL2014} and dislocation dynamics~\cite{Miguel2001,Salman-Truskinovsky_PRL2011,Salman-Truskinovsky_IntJEngSci2012}.
Instead of getting locked at a critical manifold as in these systems, ours  approaches it at times just to be carried away again  by the periodic driving.  The observed  behavior is also different from the critical self-organized avalanche oscillations under monotone driving studied in ~\cite{Papanikolaou_Nature2012}.  The specificity of our system  may be due to the particular nature  of  thermal driving  which is not conjugate to the order parameter (deformation in our case). In contrast, the driving parameter in systems exhibiting extended marginality is conjugate to the order parameter. For instance, the Sherrington-Kirkpatrick model is driven by a magnetic field which is conjugate to the magnetization;  the driving in models for deformation of amorphous solids or dislocation dynamics is the stress which is conjugate to the deformation.
 
\section{Conclusions}
\label{sec:conclusions}
We have presented a simple model showing  that a  system with periodically driven structural phase transformations may  operate in a regime where  a critical manifold is approached robustly.  The model  shows that the   intrinsic tuning to criticality of the first-order phase transition can be provided by a  co-evolving non-order parameter field describing  correlated  transformation-induced defects. The proposed mechanism is in line with  the abstract sweeping paradigm~\cite{Gil-Sornette_PRL1996, Andersen1997} which postulates  that a suitable coupling between the primary and secondary order parameters  can lead to robust scaling.  Our model specifies  how such  coupling can operate  in  solids  undergoing structural transformations. In this way, the model explains the scale-free avalanche behavior through recurrent encounters with a critical manifold. 

The primary order parameter in our framework is associated with the martensitic transformation (field $\s$) while  the secondary corresponds to the transformation-induced slip (field $\d$). The slip field plays the role of a non-Gaussian, spatially-correlated annealed random field,  which affects the phase transition. The degree of coupling between the phase transformation and the slip increases with the transformation strain $\epsilon$. Interpreting the slip as a disorder acting on the phase transition variables is possible since slip is unlikely to trigger phase changes. 
 
We distinguish four regimes depending on the value of $\epsilon$. Regime I corresponds to systems with $\epsilon \lesssim 0.3$ in which the phase transition is weak in the sense that it does not induce any plasticity.  Regime I$^\prime$ corresponds to systems with $0.3 \lesssim \epsilon \lesssim 0.35$ which can generate some slip  at the beginning  but dislocations remain trapped afterwards. Regime II corresponds to values of the transformation strain in the interval $0.35 \lesssim \epsilon \lesssim 0.45$ where  slip disorder co-evolves with the phase transition.  In this regime the avalanche size distribution becomes progressively closer to a power-law  as $\epsilon$ increases. However,  there is still  a characteristic peak at large avalanche sizes.
 
In the  most interesting regime II$^\prime$, corresponding  to   $\epsilon \gtrsim 0.45$, the system  generates enough disorder to approach a critical manifold at specific values of the driving parameter ($\tau_c$ and $\tau_c^{\text{h}}$, for cooling and heating runs, respectively). The intermittent  closeness to   the critical manifold is robust in the sense that it is not sensitive to the specific value of $\epsilon$. In this regime, the chaotic super-transients for slip disorder are extremely long ($n_{\text s}>10^5$) and it is of interest to study  their possible transition to a stable cycle.  Another  challenge is to perform   the  finite-size scaling analysis sharpening our  numerical estimates for the critical exponents and specifying the scaling functions. An important challenge in this respect from a computational viewpoint is to efficiently deal with long-range interactions. 
 
Our results suggest that many  shape-memory alloys with close-to-reconstructive transitions~\cite{Balandraud_Zanzotto2007} (i.e. materials operating in regimes II or II$^\prime$) should not  be treated as purely elastic.  Instead,  they  can be expected to  exhibit  `breathing' microscopic  disorder under cyclic thermal loading, with very long shakedown transients and critical avalanching only at critical values of the temperature. The proposed coupling between phase transition and disorder therefore appear to be unifying  conflicting interpretations of criticality in such  first-order phase transitions ~\cite{PerezReche2007PRL,Chandni_PRL2009}. However,  more detailed  experimental studies are necessary to affirm this interpretation. In particular, it would be interesting to check if the electrical resistivity results in Ref.~\cite{Chandni_PRL2009} are corroborated by acoustic emission or calorimetry experiments in thermally-driven martensites. An even more fundamental experimental challenge would be to directly identify the slip disorder in these materials and to study its evolution during thermal cycling.

\begin{acknowledgments}
FJPR acknowledges the financial support from the Carnegie Trust. LT acknowledges the financial support from the French ANR-2008 grant EVOCRIT.
\end{acknowledgments}

\vskip10pt

\appendix

\section{Minimization of the energy and elastic kernel}
\label{App:Min-Energy}

In this Appendix, we present details about the minimization of the energy of the ensemble of mesoscopic elements subject to the compatibility constraints and derive the interaction kernel $\J$. 

Our starting point is the   functional:
\begin{equation}
\label{eq:Phi_tilde}
\tilde{\Phi} = \Phi-\sum_{i,j=1}^{L-1} \zeta_{i,j} (\alpha_{i,j
+1}-\alpha_{i,j}-\beta_{i+1,j}+\beta_{i,j})~,
\end{equation}
where $\Phi$ is the energy of the system given by Eq.~\eqref{eq:PHI} and $\{\zeta_{i,j}\}$ are Lagrange multipliers to account for the  compatibility constraints given by Eq.~\eqref{eq:El-Compatibility}. 

The stationary points of \eqref{eq:Phi_tilde} are given by the following
equations:

\begin{equation}
\label{eq:alpha_min}
\begin{aligned}
c
\alpha_{i,j}+\zeta_{i,j}(1-\delta_{j,L})-\zeta_{i,j-1}(1-\delta_{j,1})=0,& \\i=1,\dots,L-1,\; j=1,\dots,L &
\end{aligned}
\end{equation}

\begin{equation}
\label{eq:beta_min}
\begin{aligned}
\beta_{i,j}-w_{i,j}-\zeta_{i,j}(1-\delta_{i,L})+\zeta_{i-1,j}(1-\delta_{i,1})=0,& \\ \quad i=1,\dots,L,\; j=1,\dots,L-1&
\end{aligned}
\end{equation}

\begin{equation}
\label{eq:lambda_min}
\begin{aligned}
\alpha_{i,j+1}-\alpha_{i,j}-\beta_{i+1,j}+\beta_{i,j}=0,& \\ i,j=1,
\dots,L-1,&
\end{aligned}
\end{equation}
where $w_{i,j}=\epsilon s_{i,j}+d_{i,j}$ and $\delta_{i,j}$ is the Kronecker delta which is 1 if $i=j$ and 0 otherwise. The set of equations \eqref{eq:alpha_min}-\eqref{eq:lambda_min} defines a closed system for the strain variables: $\{\alpha_{i,j},\, i=1,\dots,L-1,\; j=1,\dots,L\}$ and
$\{\beta_{i,j},\, i=1,\dots,L,\; j=1,\dots,L-1\}$. 

For $i \leq L-1$, Eq.~\eqref{eq:beta_min} can be solved recursively for
$\zeta_{i,j}$ as a function of $\{\beta_{i,j}\}$:
\begin{equation}
\label{eq:lambda_beta}
\zeta_{i,j}=\sum_{k=1}^i (\beta_{k,j}-w_{k,j}), \quad i,j=1,
\dots,L-1~.
\end{equation}

For $i=L$, one obtains $\beta_{L,j}-w_{L,j}=-\zeta_{L-1,j}$ which,
using Eq.~\eqref{eq:lambda_beta}, gives
\begin{equation}
\label{eq:closure_beta}
\sum_{k=1}^L (\beta_{k,j}-w_{k,j}) =0~.
\end{equation}

Introducing Eq.~\eqref{eq:lambda_beta} into Eq.~\eqref{eq:alpha_min} allows
$\alpha_{i,j}$ to be given in terms of $\beta_{i,j}$ for $i,j=1,\dots,L-1$ as follows:
\begin{equation}
\label{eq:alpha_Global_1}
\begin{aligned}
c
\alpha_{i,j}=&(1-\delta_{j,1})\sum_{k=1}^i(\beta_{k,j-1}-w_{k,j-1})\\
&-(1-\delta_{j,L})\sum_{k=1}^i(\beta_{k,j} - w_{k,j}).
\end{aligned}
\end{equation}

Using Eq.~\eqref{eq:alpha_Global_1} to express $\alpha_{i,j+1}$ and
$\alpha_{i,j}$ as a function of the strains $\{\beta_{i,j}\}$ in the
compatibility condition, Eq.~\eqref{eq:lambda_min}, leads to a set of
equations for the equilibrium branches that relate $\beta_{i,j}$ to
$w_{i,j}$ for $i,j=1,\dots,L-1$ as follows:
\begin{equation}
\label{eq:XXX}
\begin{aligned}
&\sum_{k=1}^i (2-\delta_{j+1,1}-\delta_{j,L})
(\beta_{k,j}-w_{k,j})- \\
-&\sum_{k=1}^i(1-\delta_{j+1,L})(\beta_{k,j+1}-w_{k,j+1})-\\
-&\sum_{k=1}^i(1-\delta_{j,1})(\beta_{k,j-1}-w_{k,j-1})=c(\beta_{i+1,j}-\beta_{i,j})~.
\end{aligned}
\end{equation}

At this stage, it is technically convenient to sort the elements in lexicographic order in such a way that the Cartesian coordinates $(i,j)$ are given in terms of a single label $p=i+(j-1)L$ that takes values $p=1,2, \dots L \times (L-1)$. With this notation, a second-order tensor with components $\{a_{i,j}\}$ can be represented as a vector with components $a_{p=i+(j-1)L}=a_{i,j}$. Similarly, the components $\{a_{i,j,k,l}\}$ of a fourth-order tensor are represented by a matrix with elements $a_{p=i+(j-1)L,q=k+(l-1)L}=a_{i,j,k,l}$.

Combining expression \eqref{eq:XXX} and the closure relation
\eqref{eq:closure_beta} gives a determined system of equations for the
strain fields $\{\beta_{i,j},\, i=1,\dots,L,\; j=1,\dots,L-1\}$ which can be recast in matrix form as follows:
\begin{equation}
\label{eq:Matrix_form_1}
\A \, (\betabf-\w)=c\B \, \betabf~.
\end{equation}
Here, $\w$ and $\betabf$ are lexicographic vectors with $L(L-1)$ components. $\A$ is an $[L(L-1)] \times [L(L-1)]$ block tridiagonal
matrix (of $(L-1)\times (L-1)$ blocks),
\begin{equation}
\label{eq:Mat_A}
\A =
\begin{pmatrix}
\T_1 & -\T_2 & {\bf 0} & {\bf 0} & \cdots & {\bf 0} \\
-\T_2 & \T_1 & -\T_2 & {\bf 0} & \cdots & {\bf 0} \\
{\bf 0} & -\T_2 & \T_1 & -\T_2 & \cdots & {\bf 0} \\
\vdots & \vdots & \vdots & \vdots & \ddots & \vdots \\
{\bf 0} & {\bf 0} & {\bf 0} & {\bf 0} & -\T_2 & \T_1 
\end{pmatrix}~,
\end{equation} 
where $\T_1$ and $\T_2$ are $L \times L$ matrices defined as follows:
\begin{align}
\label{eq:Mat_T}
\T_1 &=
\begin{pmatrix}
2 & 0 & 0 & \cdots & 0 & 0 \\
2 & 2 & 0 & \cdots & 0 & 0 \\
\vdots & \vdots & \vdots & \ddots & \vdots & \vdots \\
2 & 2 & 2 & \cdots & 2 & 0 \\
1 & 1 & 1 & \cdots & 1 & 1 
\end{pmatrix},\\
\T_2 &=
\begin{pmatrix}
1 & 0 & 0 & \cdots & 0 & 0 \\
1 & 1 & 0 & \cdots & 0 & 0 \\
\vdots & \vdots & \vdots & \ddots & \vdots & \vdots \\
1 & 1 & 1 & \cdots & 1 & 0 \\
0 & 0 & 0 & \cdots & 0 & 0 
\end{pmatrix}~.
\end{align} 

The matrix $\B$ in Eq.~\eqref{eq:Matrix_form_1} is a block-diagonal
matrix with $[L(L-1)] \times [L(L-1)]$ elements made of $(L-1) \times
(L-1)$ blocks:
\begin{equation}
\label{eq:Mat_B}
\B =
\begin{pmatrix}
\Q & {\bf 0} & {\bf 0} & \cdots & {\bf 0} \\
{\bf 0} & \Q & {\bf 0} & \cdots & {\bf 0} \\
\vdots & \vdots & \vdots & \ddots & \vdots \\
{\bf 0} & {\bf 0} & {\bf 0} & \cdots & \Q 
\end{pmatrix}~,
\end{equation} 
where, $\Q$ is an $L \times L$ matrix defined as follows:
\begin{equation}
\label{eq:Mat_Q}
\Q =
\begin{pmatrix}
-1 & 1 & 0 & \cdots & 0 & 0 \\
0 & -1 & 1 & \cdots & 0 & 0 \\
\vdots & \vdots & \vdots & \ddots & \vdots & \vdots \\
0 & 0 & 0 & \cdots & -1 & 1 \\
0 & 0 & 0 & \cdots & 0 & 0 
\end{pmatrix}~.
\end{equation} 
The last row of the matrices $\T_1$, $\T_2$, and $\Q$ ensures that the
closure relation \eqref{eq:closure_beta} is satisfied.

Finally, straightforward algebraic manipulations with Eq.~\eqref{eq:Matrix_form_1}, that use the relation between the elastic strain and the relative elastic strain, $\gammabf = (\betabf-\w)\epsilon^{-1}$, lead to Eq.~\eqref{eq:elastic-strain} with the kernel $\J=\left(\Id-c\A^{-1}\B\right)^{-1}-\Id$. Here $\Id$ is the unit matrix with elements $\{I_{p,q}= \delta_{p,q} ;\; p,q=1, \dots L \times (L-1)\}$.


\begin{thebibliography}{94}%
\makeatletter
\providecommand \@ifxundefined [1]{%
 \@ifx{#1\undefined}
}%
\providecommand \@ifnum [1]{%
 \ifnum #1\expandafter \@firstoftwo
 \else \expandafter \@secondoftwo
 \fi
}%
\providecommand \@ifx [1]{%
 \ifx #1\expandafter \@firstoftwo
 \else \expandafter \@secondoftwo
 \fi
}%
\providecommand \natexlab [1]{#1}%
\providecommand \enquote  [1]{``#1''}%
\providecommand \bibnamefont  [1]{#1}%
\providecommand \bibfnamefont [1]{#1}%
\providecommand \citenamefont [1]{#1}%
\providecommand \href@noop [0]{\@secondoftwo}%
\providecommand \href [0]{\begingroup \@sanitize@url \@href}%
\providecommand \@href[1]{\@@startlink{#1}\@@href}%
\providecommand \@@href[1]{\endgroup#1\@@endlink}%
\providecommand \@sanitize@url [0]{\catcode `\\12\catcode `\$12\catcode
  `\&12\catcode `\#12\catcode `\^12\catcode `\_12\catcode `\%12\relax}%
\providecommand \@@startlink[1]{}%
\providecommand \@@endlink[0]{}%
\providecommand \url  [0]{\begingroup\@sanitize@url \@url }%
\providecommand \@url [1]{\endgroup\@href {#1}{\urlprefix }}%
\providecommand \urlprefix  [0]{URL }%
\providecommand \Eprint [0]{\href }%
\providecommand \doibase [0]{http://dx.doi.org/}%
\providecommand \selectlanguage [0]{\@gobble}%
\providecommand \bibinfo  [0]{\@secondoftwo}%
\providecommand \bibfield  [0]{\@secondoftwo}%
\providecommand \translation [1]{[#1]}%
\providecommand \BibitemOpen [0]{}%
\providecommand \bibitemStop [0]{}%
\providecommand \bibitemNoStop [0]{.\EOS\space}%
\providecommand \EOS [0]{\spacefactor3000\relax}%
\providecommand \BibitemShut  [1]{\csname bibitem#1\endcsname}%
\let\auto@bib@innerbib\@empty
\bibitem [{\citenamefont {Vives}\ \emph {et~al.}(1994)\citenamefont {Vives},
  \citenamefont {Ort{\'{i}}n}, \citenamefont {Ma{\~{n}}osa}, \citenamefont
  {R{\`{a}}fols}, \citenamefont {P{\'{e}}rez-Magran{\'{e}}},\ and\
  \citenamefont {Planes}}]{Vives1994a}%
  \BibitemOpen
  \bibfield  {author} {\bibinfo {author} {\bibfnamefont {E.}~\bibnamefont
  {Vives}}, \bibinfo {author} {\bibfnamefont {J.}~\bibnamefont {Ort{\'{i}}n}},
  \bibinfo {author} {\bibfnamefont {L.}~\bibnamefont {Ma{\~{n}}osa}}, \bibinfo
  {author} {\bibfnamefont {I.}~\bibnamefont {R{\`{a}}fols}}, \bibinfo {author}
  {\bibfnamefont {R.}~\bibnamefont {P{\'{e}}rez-Magran{\'{e}}}}, \ and\
  \bibinfo {author} {\bibfnamefont {A.}~\bibnamefont {Planes}},\ }\href
  {\doibase 10.1103/PhysRevLett.72.1694} {\bibfield  {journal} {\bibinfo
  {journal} {Phys. Rev. Lett.}\ }\textbf {\bibinfo {volume} {72}},\ \bibinfo
  {pages} {1694} (\bibinfo {year} {1994})}\BibitemShut {NoStop}%
\bibitem [{\citenamefont {P{\'{e}}rez-Reche}\ \emph
  {et~al.}(2004{\natexlab{a}})\citenamefont {P{\'{e}}rez-Reche}, \citenamefont
  {Tadi{\'{c}}}, \citenamefont {Ma{\~{n}}osa}, \citenamefont {Planes},\ and\
  \citenamefont {Vives}}]{PerezRechePRL2005}%
  \BibitemOpen
  \bibfield  {author} {\bibinfo {author} {\bibfnamefont {F.-J.}\ \bibnamefont
  {P{\'{e}}rez-Reche}}, \bibinfo {author} {\bibfnamefont {B.}~\bibnamefont
  {Tadi{\'{c}}}}, \bibinfo {author} {\bibfnamefont {L.}~\bibnamefont
  {Ma{\~{n}}osa}}, \bibinfo {author} {\bibfnamefont {A.}~\bibnamefont
  {Planes}}, \ and\ \bibinfo {author} {\bibfnamefont {E.}~\bibnamefont
  {Vives}},\ }\href {\doibase 10.1103/PhysRevLett.93.195701} {\bibfield
  {journal} {\bibinfo  {journal} {Phys. Rev. Lett.}\ }\textbf {\bibinfo
  {volume} {93}},\ \bibinfo {pages} {195701} (\bibinfo {year}
  {2004}{\natexlab{a}})},\ \Eprint {http://arxiv.org/abs/0406098}
  {arXiv:0406098 [cond-mat]} \BibitemShut {NoStop}%
\bibitem [{\citenamefont {Gallardo}\ \emph {et~al.}(2010)\citenamefont
  {Gallardo}, \citenamefont {Manchado}, \citenamefont {Romero}, \citenamefont
  {{Del Cerro}}, \citenamefont {Salje}, \citenamefont {Planes}, \citenamefont
  {Vives}, \citenamefont {Romero},\ and\ \citenamefont
  {Stipcich}}]{Gallardo_PRB2010}%
  \BibitemOpen
  \bibfield  {author} {\bibinfo {author} {\bibfnamefont {M.~C.}\ \bibnamefont
  {Gallardo}}, \bibinfo {author} {\bibfnamefont {J.}~\bibnamefont {Manchado}},
  \bibinfo {author} {\bibfnamefont {F.~J.}\ \bibnamefont {Romero}}, \bibinfo
  {author} {\bibfnamefont {J.}~\bibnamefont {{Del Cerro}}}, \bibinfo {author}
  {\bibfnamefont {E.~K.~H.}\ \bibnamefont {Salje}}, \bibinfo {author}
  {\bibfnamefont {A.}~\bibnamefont {Planes}}, \bibinfo {author} {\bibfnamefont
  {E.}~\bibnamefont {Vives}}, \bibinfo {author} {\bibfnamefont
  {R.}~\bibnamefont {Romero}}, \ and\ \bibinfo {author} {\bibfnamefont
  {M.}~\bibnamefont {Stipcich}},\ }\href {\doibase 10.1103/PhysRevB.81.174102}
  {\bibfield  {journal} {\bibinfo  {journal} {Phys. Rev. B - Condens. Matter
  Mater. Phys.}\ }\textbf {\bibinfo {volume} {81}},\ \bibinfo {pages} {174102}
  (\bibinfo {year} {2010})}\BibitemShut {NoStop}%
\bibitem [{\citenamefont {Bar{\'{o}}}\ \emph {et~al.}(2014)\citenamefont
  {Bar{\'{o}}}, \citenamefont {Mart{\'{i}}n-Olalla}, \citenamefont {Romero},
  \citenamefont {Gallardo}, \citenamefont {Salje}, \citenamefont {Vives},\ and\
  \citenamefont {Planes}}]{Baro_JPhysC2014}%
  \BibitemOpen
  \bibfield  {author} {\bibinfo {author} {\bibfnamefont {J.}~\bibnamefont
  {Bar{\'{o}}}}, \bibinfo {author} {\bibfnamefont {J.-M.}\ \bibnamefont
  {Mart{\'{i}}n-Olalla}}, \bibinfo {author} {\bibfnamefont {F.~J.}\
  \bibnamefont {Romero}}, \bibinfo {author} {\bibfnamefont {M.~C.}\
  \bibnamefont {Gallardo}}, \bibinfo {author} {\bibfnamefont {E.~K.~H.}\
  \bibnamefont {Salje}}, \bibinfo {author} {\bibfnamefont {E.}~\bibnamefont
  {Vives}}, \ and\ \bibinfo {author} {\bibfnamefont {A.}~\bibnamefont
  {Planes}},\ }\href {\doibase 10.1088/0953-8984/26/12/125401} {\bibfield
  {journal} {\bibinfo  {journal} {J. Phys. Condens. Matter}\ }\textbf {\bibinfo
  {volume} {26}},\ \bibinfo {pages} {125401} (\bibinfo {year}
  {2014})}\BibitemShut {NoStop}%
\bibitem [{\citenamefont {Vives}\ \emph {et~al.}(1995)\citenamefont {Vives},
  \citenamefont {Goicoechea}, \citenamefont {Ort{\'{i}}n},\ and\ \citenamefont
  {Planes}}]{Vives1995Universality}%
  \BibitemOpen
  \bibfield  {author} {\bibinfo {author} {\bibfnamefont {E.}~\bibnamefont
  {Vives}}, \bibinfo {author} {\bibfnamefont {J.}~\bibnamefont {Goicoechea}},
  \bibinfo {author} {\bibfnamefont {J.}~\bibnamefont {Ort{\'{i}}n}}, \ and\
  \bibinfo {author} {\bibfnamefont {A.}~\bibnamefont {Planes}},\ }\href
  {\doibase 10.1103/PhysRevE.52.R5} {\bibfield  {journal} {\bibinfo  {journal}
  {Phys. Rev. E}\ }\textbf {\bibinfo {volume} {52}},\ \bibinfo {pages} {R5}
  (\bibinfo {year} {1995})}\BibitemShut {NoStop}%
\bibitem [{\citenamefont {Goicoechea}\ and\ \citenamefont
  {Ort{\'{i}}n}(1995)}]{Goicoechea1995}%
  \BibitemOpen
  \bibfield  {author} {\bibinfo {author} {\bibfnamefont {J.}~\bibnamefont
  {Goicoechea}}\ and\ \bibinfo {author} {\bibfnamefont {J.}~\bibnamefont
  {Ort{\'{i}}n}},\ }\href {\doibase 10.1051/jp4:1995210} {\bibfield  {journal}
  {\bibinfo  {journal} {Le J. Phys. IV}\ }\textbf {\bibinfo {volume} {05}},\
  \bibinfo {pages} {C2} (\bibinfo {year} {1995})}\BibitemShut {NoStop}%
\bibitem [{\citenamefont {Ahluwalia}\ and\ \citenamefont
  {Ananthakrishna}(2001)}]{Ahluwalia2001}%
  \BibitemOpen
  \bibfield  {author} {\bibinfo {author} {\bibfnamefont {R.}~\bibnamefont
  {Ahluwalia}}\ and\ \bibinfo {author} {\bibfnamefont {G.}~\bibnamefont
  {Ananthakrishna}},\ }\href {\doibase 10.1103/PhysRevLett.86.4076} {\bibfield
  {journal} {\bibinfo  {journal} {Phys. Rev. Lett.}\ }\textbf {\bibinfo
  {volume} {86}},\ \bibinfo {pages} {4076} (\bibinfo {year}
  {2001})}\BibitemShut {NoStop}%
\bibitem [{\citenamefont {P{\'{e}}rez-Reche}\ \emph {et~al.}(2007)\citenamefont
  {P{\'{e}}rez-Reche}, \citenamefont {Truskinovsky},\ and\ \citenamefont
  {Zanzotto}}]{PerezReche2007PRL}%
  \BibitemOpen
  \bibfield  {author} {\bibinfo {author} {\bibfnamefont {F.-J.}\ \bibnamefont
  {P{\'{e}}rez-Reche}}, \bibinfo {author} {\bibfnamefont {L.}~\bibnamefont
  {Truskinovsky}}, \ and\ \bibinfo {author} {\bibfnamefont {G.}~\bibnamefont
  {Zanzotto}},\ }\href {\doibase 10.1103/PhysRevLett.99.075501} {\bibfield
  {journal} {\bibinfo  {journal} {Phys. Rev. Lett.}\ }\textbf {\bibinfo
  {volume} {99}},\ \bibinfo {pages} {075501} (\bibinfo {year}
  {2007})}\BibitemShut {NoStop}%
\bibitem [{\citenamefont {Chandni}\ \emph {et~al.}(2009)\citenamefont
  {Chandni}, \citenamefont {Ghosh}, \citenamefont {Vijaya},\ and\ \citenamefont
  {Mohan}}]{Chandni_PRL2009}%
  \BibitemOpen
  \bibfield  {author} {\bibinfo {author} {\bibfnamefont {U.}~\bibnamefont
  {Chandni}}, \bibinfo {author} {\bibfnamefont {A.}~\bibnamefont {Ghosh}},
  \bibinfo {author} {\bibfnamefont {H.~S.}\ \bibnamefont {Vijaya}}, \ and\
  \bibinfo {author} {\bibfnamefont {S.}~\bibnamefont {Mohan}},\ }\href
  {\doibase 10.1103/PhysRevLett.102.025701} {\bibfield  {journal} {\bibinfo
  {journal} {Phys. Rev. Lett.}\ }\textbf {\bibinfo {volume} {102}},\ \bibinfo
  {pages} {025701} (\bibinfo {year} {2009})}\BibitemShut {NoStop}%
\bibitem [{\citenamefont {P{\'{e}}rez-Reche}\ \emph {et~al.}(2009)\citenamefont
  {P{\'{e}}rez-Reche}, \citenamefont {Truskinovsky},\ and\ \citenamefont
  {Zanzotto}}]{PerezReche_CMT2009}%
  \BibitemOpen
  \bibfield  {author} {\bibinfo {author} {\bibfnamefont {F.~J.}\ \bibnamefont
  {P{\'{e}}rez-Reche}}, \bibinfo {author} {\bibfnamefont {L.}~\bibnamefont
  {Truskinovsky}}, \ and\ \bibinfo {author} {\bibfnamefont {G.}~\bibnamefont
  {Zanzotto}},\ }\href {\doibase 10.1007/s00161-009-0096-2} {\bibfield
  {journal} {\bibinfo  {journal} {Contin. Mech. Thermodyn.}\ }\textbf {\bibinfo
  {volume} {21}},\ \bibinfo {pages} {17} (\bibinfo {year} {2009})}\BibitemShut
  {NoStop}%
\bibitem [{\citenamefont {Cerruti}\ and\ \citenamefont
  {Vives}(2008)}]{Cerruti2008I}%
  \BibitemOpen
  \bibfield  {author} {\bibinfo {author} {\bibfnamefont {B.}~\bibnamefont
  {Cerruti}}\ and\ \bibinfo {author} {\bibfnamefont {E.}~\bibnamefont
  {Vives}},\ }\href {\doibase 10.1103/PhysRevB.77.064114} {\bibfield  {journal}
  {\bibinfo  {journal} {Phys. Rev. B}\ }\textbf {\bibinfo {volume} {77}},\
  \bibinfo {pages} {064114} (\bibinfo {year} {2008})}\BibitemShut {NoStop}%
\bibitem [{\citenamefont {Ding}\ \emph {et~al.}(2013)\citenamefont {Ding},
  \citenamefont {Lookman}, \citenamefont {Zhao}, \citenamefont {Saxena},
  \citenamefont {Sun},\ and\ \citenamefont {Salje}}]{Ding-Salje_PRB2013}%
  \BibitemOpen
  \bibfield  {author} {\bibinfo {author} {\bibfnamefont {X.}~\bibnamefont
  {Ding}}, \bibinfo {author} {\bibfnamefont {T.}~\bibnamefont {Lookman}},
  \bibinfo {author} {\bibfnamefont {Z.}~\bibnamefont {Zhao}}, \bibinfo {author}
  {\bibfnamefont {A.}~\bibnamefont {Saxena}}, \bibinfo {author} {\bibfnamefont
  {J.}~\bibnamefont {Sun}}, \ and\ \bibinfo {author} {\bibfnamefont {E.~K.~H.}\
  \bibnamefont {Salje}},\ }\href {\doibase 10.1103/PhysRevB.87.094109}
  {\bibfield  {journal} {\bibinfo  {journal} {Phys. Rev. B}\ }\textbf {\bibinfo
  {volume} {87}},\ \bibinfo {pages} {094109} (\bibinfo {year}
  {2013})}\BibitemShut {NoStop}%
\bibitem [{\citenamefont {Ball}\ \emph {et~al.}(2015)\citenamefont {Ball},
  \citenamefont {Cesana},\ and\ \citenamefont {Hambly}}]{Ball-Cesana_2015_a}%
  \BibitemOpen
  \bibfield  {author} {\bibinfo {author} {\bibfnamefont {J.~M.}\ \bibnamefont
  {Ball}}, \bibinfo {author} {\bibfnamefont {P.}~\bibnamefont {Cesana}}, \ and\
  \bibinfo {author} {\bibfnamefont {B.}~\bibnamefont {Hambly}},\ }\href
  {https://doaj.org/article/3288008af5c543e6a7685a23d0f1b846} {\bibfield
  {journal} {\bibinfo  {journal} {MATEC Web Conf.}\ }\textbf {\bibinfo {volume}
  {33}},\ \bibinfo {pages} {02008} (\bibinfo {year} {2015})}\BibitemShut
  {NoStop}%
\bibitem [{\citenamefont {Otsuka}\ and\ \citenamefont
  {Wayman}(1998)}]{Otsuka1998SMA}%
  \BibitemOpen
  \bibinfo {editor} {\bibfnamefont {K.}~\bibnamefont {Otsuka}}\ and\ \bibinfo
  {editor} {\bibfnamefont {C.~M.}\ \bibnamefont {Wayman}},\ eds.,\ \href@noop
  {} {\emph {\bibinfo {title} {{Shape Memory Materials}}}}\ (\bibinfo
  {publisher} {Cambridge University Press},\ \bibinfo {address} {Cambridge},\
  \bibinfo {year} {1998})\BibitemShut {NoStop}%
\bibitem [{\citenamefont {Bhattacharya}(2003)}]{Bhattacharya2003}%
  \BibitemOpen
  \bibfield  {author} {\bibinfo {author} {\bibfnamefont {K.}~\bibnamefont
  {Bhattacharya}},\ }\href@noop {} {\emph {\bibinfo {title} {{Microestructure
  of Martensite: Why it Forms and How it Gives Rise to the Shape-Memory
  Effect}}}}\ (\bibinfo  {publisher} {Oxford University Press},\ \bibinfo
  {address} {Oxford},\ \bibinfo {year} {2003})\BibitemShut {NoStop}%
\bibitem [{\citenamefont {Balandraud}\ \emph {et~al.}(2015)\citenamefont
  {Balandraud}, \citenamefont {Barrera}, \citenamefont {Biscari}, \citenamefont
  {Gr{\'{e}}diac},\ and\ \citenamefont {Zanzotto}}]{Balandraud_PRB2015}%
  \BibitemOpen
  \bibfield  {author} {\bibinfo {author} {\bibfnamefont {X.}~\bibnamefont
  {Balandraud}}, \bibinfo {author} {\bibfnamefont {N.}~\bibnamefont {Barrera}},
  \bibinfo {author} {\bibfnamefont {P.}~\bibnamefont {Biscari}}, \bibinfo
  {author} {\bibfnamefont {M.}~\bibnamefont {Gr{\'{e}}diac}}, \ and\ \bibinfo
  {author} {\bibfnamefont {G.}~\bibnamefont {Zanzotto}},\ }\href {\doibase
  10.1103/PhysRevB.91.174111} {\bibfield  {journal} {\bibinfo  {journal} {Phys.
  Rev. B}\ }\textbf {\bibinfo {volume} {91}},\ \bibinfo {pages} {174111}
  (\bibinfo {year} {2015})}\BibitemShut {NoStop}%
\bibitem [{\citenamefont {Csikor}\ \emph {et~al.}(2007)\citenamefont {Csikor},
  \citenamefont {Motz}, \citenamefont {Weygand}, \citenamefont {Zaiser},\ and\
  \citenamefont {Zapperi}}]{Csikor_Science2007}%
  \BibitemOpen
  \bibfield  {author} {\bibinfo {author} {\bibfnamefont {F.~F.}\ \bibnamefont
  {Csikor}}, \bibinfo {author} {\bibfnamefont {C.}~\bibnamefont {Motz}},
  \bibinfo {author} {\bibfnamefont {D.}~\bibnamefont {Weygand}}, \bibinfo
  {author} {\bibfnamefont {M.}~\bibnamefont {Zaiser}}, \ and\ \bibinfo {author}
  {\bibfnamefont {S.}~\bibnamefont {Zapperi}},\ }\href {\doibase
  10.1126/science.1143719} {\bibfield  {journal} {\bibinfo  {journal}
  {Science}\ }\textbf {\bibinfo {volume} {318}},\ \bibinfo {pages} {251}
  (\bibinfo {year} {2007})}\BibitemShut {NoStop}%
\bibitem [{\citenamefont {Frick}\ \emph {et~al.}(2008)\citenamefont {Frick},
  \citenamefont {Clark}, \citenamefont {Orso}, \citenamefont {Schneider},\ and\
  \citenamefont {Arzt}}]{Frick_MatSciEngA2008}%
  \BibitemOpen
  \bibfield  {author} {\bibinfo {author} {\bibfnamefont {C.}~\bibnamefont
  {Frick}}, \bibinfo {author} {\bibfnamefont {B.}~\bibnamefont {Clark}},
  \bibinfo {author} {\bibfnamefont {S.}~\bibnamefont {Orso}}, \bibinfo {author}
  {\bibfnamefont {A.}~\bibnamefont {Schneider}}, \ and\ \bibinfo {author}
  {\bibfnamefont {E.}~\bibnamefont {Arzt}},\ }\href {\doibase
  10.1016/j.msea.2007.12.038} {\bibfield  {journal} {\bibinfo  {journal}
  {Mater. Sci. Eng. A}\ }\textbf {\bibinfo {volume} {489}},\ \bibinfo {pages}
  {319} (\bibinfo {year} {2008})}\BibitemShut {NoStop}%
\bibitem [{\citenamefont {Zaiser}(2013)}]{Zaiser_JMechBehavMater2013}%
  \BibitemOpen
  \bibfield  {author} {\bibinfo {author} {\bibfnamefont {M.}~\bibnamefont
  {Zaiser}},\ }\href
  {http://www.degruyter.com/view/j/jmbm.2013.22.issue-3-4/jmbm-2012-0006/jmbm-2012-0006.xml}
  {\bibfield  {journal} {\bibinfo  {journal} {J. Mech. Behav. Mater.}\ }\textbf
  {\bibinfo {volume} {22}},\ \bibinfo {pages} {89} (\bibinfo {year}
  {2013})}\BibitemShut {NoStop}%
\bibitem [{\citenamefont {Jensen}(1998)}]{Jensen1998}%
  \BibitemOpen
  \bibfield  {author} {\bibinfo {author} {\bibfnamefont {H.~J.}\ \bibnamefont
  {Jensen}},\ }\href@noop {} {\emph {\bibinfo {title} {{Self-organized
  criticality}}}}\ (\bibinfo  {publisher} {Cambridge University Press},\
  \bibinfo {address} {Cambridge},\ \bibinfo {year} {1998})\BibitemShut
  {NoStop}%
\bibitem [{\citenamefont {Sornette}(2000)}]{Sornette2000}%
  \BibitemOpen
  \bibfield  {author} {\bibinfo {author} {\bibfnamefont {D.}~\bibnamefont
  {Sornette}},\ }\href@noop {} {\emph {\bibinfo {title} {{Critical Phenomena in
  Natural Sciences}}}},\ Springer Series in Synergetics\ (\bibinfo  {publisher}
  {Springer Verlag},\ \bibinfo {address} {Berlin},\ \bibinfo {year}
  {2000})\BibitemShut {NoStop}%
\bibitem [{\citenamefont {Plenz}\ and\ \citenamefont
  {Niebur}(2014)}]{Plenz_BookCriticalityNeuralNetworks2014}%
  \BibitemOpen
  \bibinfo {editor} {\bibfnamefont {D.}~\bibnamefont {Plenz}}\ and\ \bibinfo
  {editor} {\bibfnamefont {E.}~\bibnamefont {Niebur}},\ eds.,\ \href {\doibase
  10.1002/9783527651009} {\emph {\bibinfo {title} {{Criticality in Neural
  Systems}}}}\ (\bibinfo  {publisher} {WILEY-VCH Verlag},\ \bibinfo {address}
  {Weinheim},\ \bibinfo {year} {2014})\BibitemShut {NoStop}%
\bibitem [{\citenamefont {Ginelli}\ \emph {et~al.}(2015)\citenamefont
  {Ginelli}, \citenamefont {Peruani}, \citenamefont {Pillot}, \citenamefont
  {Chat{\'{e}}}, \citenamefont {Theraulaz},\ and\ \citenamefont
  {Bon}}]{Ginelli2015}%
  \BibitemOpen
  \bibfield  {author} {\bibinfo {author} {\bibfnamefont {F.}~\bibnamefont
  {Ginelli}}, \bibinfo {author} {\bibfnamefont {F.}~\bibnamefont {Peruani}},
  \bibinfo {author} {\bibfnamefont {M.-H.}\ \bibnamefont {Pillot}}, \bibinfo
  {author} {\bibfnamefont {H.}~\bibnamefont {Chat{\'{e}}}}, \bibinfo {author}
  {\bibfnamefont {G.}~\bibnamefont {Theraulaz}}, \ and\ \bibinfo {author}
  {\bibfnamefont {R.}~\bibnamefont {Bon}},\ }\href {\doibase
  10.1073/pnas.1503749112} {\bibfield  {journal} {\bibinfo  {journal} {Proc.
  Natl. Acad. Sci.}\ }\textbf {\bibinfo {volume} {112}},\ \bibinfo {pages}
  {12729} (\bibinfo {year} {2015})}\BibitemShut {NoStop}%
\bibitem [{\citenamefont {Ben-Zion}(2008)}]{Ben-Zion_RevGeophy2008_Review}%
  \BibitemOpen
  \bibfield  {author} {\bibinfo {author} {\bibfnamefont {Y.}~\bibnamefont
  {Ben-Zion}},\ }\href {\doibase 10.1029/2008RG000260} {\bibfield  {journal}
  {\bibinfo  {journal} {Rev. Geophys.}\ }\textbf {\bibinfo {volume} {46}},\
  \bibinfo {pages} {RG4006} (\bibinfo {year} {2008})}\BibitemShut {NoStop}%
\bibitem [{\citenamefont
  {Aschwanden}(2011)}]{Aschwanden_BookCriticalityGalaxies}%
  \BibitemOpen
  \bibfield  {author} {\bibinfo {author} {\bibfnamefont {M.}~\bibnamefont
  {Aschwanden}},\ }\href {\doibase 978-3-642-15001-2} {\emph {\bibinfo {title}
  {{Self-Organized Criticality in Astrophysics}}}}\ (\bibinfo  {publisher}
  {Springer-Verlag Berlin Heidelberg},\ \bibinfo {address} {Berlin},\ \bibinfo
  {year} {2011})\BibitemShut {NoStop}%
\bibitem [{\citenamefont {Perkovi{\'{c}}}\ \emph {et~al.}(1995)\citenamefont
  {Perkovi{\'{c}}}, \citenamefont {Dahmen},\ and\ \citenamefont
  {Sethna}}]{Perkovic1995}%
  \BibitemOpen
  \bibfield  {author} {\bibinfo {author} {\bibfnamefont {O.}~\bibnamefont
  {Perkovi{\'{c}}}}, \bibinfo {author} {\bibfnamefont {K.~A.}\ \bibnamefont
  {Dahmen}}, \ and\ \bibinfo {author} {\bibfnamefont {J.~P.}\ \bibnamefont
  {Sethna}},\ }\href {\doibase 10.1103/PhysRevLett.75.4528} {\bibfield
  {journal} {\bibinfo  {journal} {Phys. Rev. Lett.}\ }\textbf {\bibinfo
  {volume} {75}},\ \bibinfo {pages} {4528} (\bibinfo {year} {1995})},\ \Eprint
  {http://arxiv.org/abs/9506111} {arXiv:9506111 [cond-mat]} \BibitemShut
  {NoStop}%
\bibitem [{\citenamefont {Fisher}(1998)}]{Fisher_PhysRep1998}%
  \BibitemOpen
  \bibfield  {author} {\bibinfo {author} {\bibfnamefont {D.~S.}\ \bibnamefont
  {Fisher}},\ }\href {\doibase 10.1016/S0370-1573(98)00008-8} {\bibfield
  {journal} {\bibinfo  {journal} {Phys. Rep.}\ }\textbf {\bibinfo {volume}
  {301}},\ \bibinfo {pages} {113} (\bibinfo {year} {1998})}\BibitemShut
  {NoStop}%
\bibitem [{\citenamefont {Dickman}\ \emph {et~al.}(2000)\citenamefont
  {Dickman}, \citenamefont {Mu{\~{n}}oz}, \citenamefont {Vespignani},\ and\
  \citenamefont {Zapperi}}]{Dickman2000}%
  \BibitemOpen
  \bibfield  {author} {\bibinfo {author} {\bibfnamefont {R.}~\bibnamefont
  {Dickman}}, \bibinfo {author} {\bibfnamefont {M.~A.}\ \bibnamefont
  {Mu{\~{n}}oz}}, \bibinfo {author} {\bibfnamefont {A.}~\bibnamefont
  {Vespignani}}, \ and\ \bibinfo {author} {\bibfnamefont {S.}~\bibnamefont
  {Zapperi}},\ }\href@noop {} {\bibfield  {journal} {\bibinfo  {journal} {Braz.
  J. Phys.}\ }\textbf {\bibinfo {volume} {30}},\ \bibinfo {pages} {27}
  (\bibinfo {year} {2000})}\BibitemShut {NoStop}%
\bibitem [{\citenamefont {Bertotti}\ \emph
  {et~al.}(2006{\natexlab{a}})\citenamefont {Bertotti}, \citenamefont
  {Mayergoyz}, \citenamefont {Durin},\ and\ \citenamefont
  {Zapperi}}]{Durin_review2004}%
  \BibitemOpen
  \bibfield  {author} {\bibinfo {author} {\bibfnamefont {G.}~\bibnamefont
  {Bertotti}}, \bibinfo {author} {\bibfnamefont {I.~D.}\ \bibnamefont
  {Mayergoyz}}, \bibinfo {author} {\bibfnamefont {G.}~\bibnamefont {Durin}}, \
  and\ \bibinfo {author} {\bibfnamefont {S.}~\bibnamefont {Zapperi}},\ }in\
  \href {\doibase 10.1016/B978-012480874-4/50014-2} {\emph {\bibinfo
  {booktitle} {Sci. Hysteresis}}}\ (\bibinfo  {publisher} {Elsevier},\ \bibinfo
  {year} {2006})\ Chap.~\bibinfo {chapter} {3}, pp.\ \bibinfo {pages}
  {181--267}\BibitemShut {NoStop}%
\bibitem [{\citenamefont {M{\"{u}}ller}\ and\ \citenamefont
  {Wyart}(2015)}]{Muller-Wyart_annurev-conmatphys2015}%
  \BibitemOpen
  \bibfield  {author} {\bibinfo {author} {\bibfnamefont {M.}~\bibnamefont
  {M{\"{u}}ller}}\ and\ \bibinfo {author} {\bibfnamefont {M.}~\bibnamefont
  {Wyart}},\ }\href {\doibase 10.1146/annurev-conmatphys-031214-014614}
  {\bibfield  {journal} {\bibinfo  {journal} {Annu. Rev. Condens. Matter
  Phys.}\ }\textbf {\bibinfo {volume} {6}},\ \bibinfo {pages} {177} (\bibinfo
  {year} {2015})}\BibitemShut {NoStop}%
\bibitem [{\citenamefont {Weiss}\ \emph {et~al.}(2015)\citenamefont {Weiss},
  \citenamefont {Rhouma}, \citenamefont {Richeton}, \citenamefont {Dechanel},
  \citenamefont {Louchet},\ and\ \citenamefont {Truskinovsky}}]{Weiss2015}%
  \BibitemOpen
  \bibfield  {author} {\bibinfo {author} {\bibfnamefont {J.}~\bibnamefont
  {Weiss}}, \bibinfo {author} {\bibfnamefont {W.~B.}\ \bibnamefont {Rhouma}},
  \bibinfo {author} {\bibfnamefont {T.}~\bibnamefont {Richeton}}, \bibinfo
  {author} {\bibfnamefont {S.}~\bibnamefont {Dechanel}}, \bibinfo {author}
  {\bibfnamefont {F.}~\bibnamefont {Louchet}}, \ and\ \bibinfo {author}
  {\bibfnamefont {L.}~\bibnamefont {Truskinovsky}},\ }\href {\doibase
  10.1103/PhysRevLett.114.105504} {\bibfield  {journal} {\bibinfo  {journal}
  {Phys. Rev. Lett.}\ }\textbf {\bibinfo {volume} {114}},\ \bibinfo {pages}
  {105504} (\bibinfo {year} {2015})}\BibitemShut {NoStop}%
\bibitem [{\citenamefont {Stanley}(1983)}]{Stanley1971}%
  \BibitemOpen
  \bibfield  {author} {\bibinfo {author} {\bibfnamefont {H.~E.}\ \bibnamefont
  {Stanley}},\ }\href@noop {} {\emph {\bibinfo {title} {{Introduction to Phase
  Transitions and Critical Phenomena}}}}\ (\bibinfo  {publisher} {Oxford
  University Press},\ \bibinfo {address} {New York},\ \bibinfo {year}
  {1983})\BibitemShut {NoStop}%
\bibitem [{\citenamefont {Sethna}\ \emph {et~al.}(2001)\citenamefont {Sethna},
  \citenamefont {Dahmen},\ and\ \citenamefont {Myers}}]{Sethna2001}%
  \BibitemOpen
  \bibfield  {author} {\bibinfo {author} {\bibfnamefont {J.~P.}\ \bibnamefont
  {Sethna}}, \bibinfo {author} {\bibfnamefont {K.~A.}\ \bibnamefont {Dahmen}},
  \ and\ \bibinfo {author} {\bibfnamefont {C.~R.}\ \bibnamefont {Myers}},\
  }\href {\doibase 10.1038/35065675} {\bibfield  {journal} {\bibinfo  {journal}
  {Nature}\ }\textbf {\bibinfo {volume} {410}},\ \bibinfo {pages} {242}
  (\bibinfo {year} {2001})}\BibitemShut {NoStop}%
\bibitem [{\citenamefont {Bertotti}\ \emph
  {et~al.}(2006{\natexlab{b}})\citenamefont {Bertotti}, \citenamefont
  {Mayergoyz}, \citenamefont {Sethna}, \citenamefont {Dahmen},\ and\
  \citenamefont {Perkovic}}]{Sethna_review2004}%
  \BibitemOpen
  \bibfield  {author} {\bibinfo {author} {\bibfnamefont {G.}~\bibnamefont
  {Bertotti}}, \bibinfo {author} {\bibfnamefont {I.~D.}\ \bibnamefont
  {Mayergoyz}}, \bibinfo {author} {\bibfnamefont {J.~P.}\ \bibnamefont
  {Sethna}}, \bibinfo {author} {\bibfnamefont {K.~A.}\ \bibnamefont {Dahmen}},
  \ and\ \bibinfo {author} {\bibfnamefont {O.}~\bibnamefont {Perkovic}},\ }in\
  \href {\doibase 10.1016/B978-012480874-4/50013-0} {\emph {\bibinfo
  {booktitle} {Sci. Hysteresis}}}\ (\bibinfo  {publisher} {Elsevier},\ \bibinfo
  {year} {2006})\ Chap.~\bibinfo {chapter} {2}, pp.\ \bibinfo {pages}
  {107--179}\BibitemShut {NoStop}%
\bibitem [{\citenamefont {P{\'{e}}rez-Reche}\ and\ \citenamefont
  {Vives}(2004)}]{PerezReche2004RFIMField}%
  \BibitemOpen
  \bibfield  {author} {\bibinfo {author} {\bibfnamefont {F.~J.}\ \bibnamefont
  {P{\'{e}}rez-Reche}}\ and\ \bibinfo {author} {\bibfnamefont {E.}~\bibnamefont
  {Vives}},\ }\href {\doibase 10.1103/PhysRevB.70.214422} {\bibfield  {journal}
  {\bibinfo  {journal} {Phys. Rev. B}\ }\textbf {\bibinfo {volume} {70}},\
  \bibinfo {pages} {214422} (\bibinfo {year} {2004})}\BibitemShut {NoStop}%
\bibitem [{\citenamefont {Handford}\ \emph {et~al.}(2013)\citenamefont
  {Handford}, \citenamefont {P{\'{e}}rez-Reche},\ and\ \citenamefont
  {Taraskin}}]{Handford_PRE2013a}%
  \BibitemOpen
  \bibfield  {author} {\bibinfo {author} {\bibfnamefont {T.~P.}\ \bibnamefont
  {Handford}}, \bibinfo {author} {\bibfnamefont {F.~J.}\ \bibnamefont
  {P{\'{e}}rez-Reche}}, \ and\ \bibinfo {author} {\bibfnamefont {S.~N.}\
  \bibnamefont {Taraskin}},\ }\href {\doibase 10.1103/PhysRevE.87.062122}
  {\bibfield  {journal} {\bibinfo  {journal} {Phys. Rev. E}\ }\textbf {\bibinfo
  {volume} {87}},\ \bibinfo {pages} {062122} (\bibinfo {year} {2013})},\
  \Eprint {http://arxiv.org/abs/1302.7027} {arXiv:1302.7027} \BibitemShut
  {NoStop}%
\bibitem [{\citenamefont {Sethna}\ \emph {et~al.}(1993)\citenamefont {Sethna},
  \citenamefont {Dahmen}, \citenamefont {Kartha}, \citenamefont {Krumhansl},
  \citenamefont {Roberts},\ and\ \citenamefont {Shore}}]{Sethna1993}%
  \BibitemOpen
  \bibfield  {author} {\bibinfo {author} {\bibfnamefont {J.~P.}\ \bibnamefont
  {Sethna}}, \bibinfo {author} {\bibfnamefont {K.}~\bibnamefont {Dahmen}},
  \bibinfo {author} {\bibfnamefont {S.}~\bibnamefont {Kartha}}, \bibinfo
  {author} {\bibfnamefont {J.~A.}\ \bibnamefont {Krumhansl}}, \bibinfo {author}
  {\bibfnamefont {B.~W.}\ \bibnamefont {Roberts}}, \ and\ \bibinfo {author}
  {\bibfnamefont {J.~D.}\ \bibnamefont {Shore}},\ }\href {\doibase
  10.1103/PhysRevLett.70.3347} {\bibfield  {journal} {\bibinfo  {journal}
  {Phys. Rev. Lett.}\ }\textbf {\bibinfo {volume} {70}},\ \bibinfo {pages}
  {3347} (\bibinfo {year} {1993})},\ \Eprint {http://arxiv.org/abs/9210018}
  {arXiv:9210018 [cond-mat]} \BibitemShut {NoStop}%
\bibitem [{\citenamefont {Carrillo}\ \emph {et~al.}(1998)\citenamefont
  {Carrillo}, \citenamefont {Ma{\~{n}}osa}, \citenamefont {Ort{\'{i}}n},
  \citenamefont {Planes},\ and\ \citenamefont {Vives}}]{Carrillo1998}%
  \BibitemOpen
  \bibfield  {author} {\bibinfo {author} {\bibfnamefont {L.}~\bibnamefont
  {Carrillo}}, \bibinfo {author} {\bibfnamefont {L.}~\bibnamefont
  {Ma{\~{n}}osa}}, \bibinfo {author} {\bibfnamefont {J.}~\bibnamefont
  {Ort{\'{i}}n}}, \bibinfo {author} {\bibfnamefont {A.}~\bibnamefont {Planes}},
  \ and\ \bibinfo {author} {\bibfnamefont {E.}~\bibnamefont {Vives}},\ }\href
  {\doibase 10.1103/PhysRevLett.81.1889} {\bibfield  {journal} {\bibinfo
  {journal} {Phys. Rev. Lett.}\ }\textbf {\bibinfo {volume} {81}},\ \bibinfo
  {pages} {1889} (\bibinfo {year} {1998})}\BibitemShut {NoStop}%
\bibitem [{\citenamefont {P{\'{e}}rez-Reche}\ \emph
  {et~al.}(2004{\natexlab{b}})\citenamefont {P{\'{e}}rez-Reche}, \citenamefont
  {Stipcich}, \citenamefont {Vives}, \citenamefont {Ma{\~{n}}osa},
  \citenamefont {Planes},\ and\ \citenamefont {Morin}}]{PerezReche2004Cyc}%
  \BibitemOpen
  \bibfield  {author} {\bibinfo {author} {\bibfnamefont {F.-J.}\ \bibnamefont
  {P{\'{e}}rez-Reche}}, \bibinfo {author} {\bibfnamefont {M.}~\bibnamefont
  {Stipcich}}, \bibinfo {author} {\bibfnamefont {E.}~\bibnamefont {Vives}},
  \bibinfo {author} {\bibfnamefont {L.}~\bibnamefont {Ma{\~{n}}osa}}, \bibinfo
  {author} {\bibfnamefont {A.}~\bibnamefont {Planes}}, \ and\ \bibinfo {author}
  {\bibfnamefont {M.}~\bibnamefont {Morin}},\ }\href {\doibase
  10.1103/PhysRevB.69.064101} {\bibfield  {journal} {\bibinfo  {journal} {Phys.
  Rev. B}\ }\textbf {\bibinfo {volume} {69}},\ \bibinfo {pages} {064101}
  (\bibinfo {year} {2004}{\natexlab{b}})}\BibitemShut {NoStop}%
\bibitem [{\citenamefont {Krauss}(1963)}]{Krauss_ActaMetall1963}%
  \BibitemOpen
  \bibfield  {author} {\bibinfo {author} {\bibfnamefont {G.}~\bibnamefont
  {Krauss}},\ }\href {\doibase 10.1016/0001-6160(63)90085-3} {\bibfield
  {journal} {\bibinfo  {journal} {Acta Metall.}\ }\textbf {\bibinfo {volume}
  {11}},\ \bibinfo {pages} {499} (\bibinfo {year} {1963})}\BibitemShut
  {NoStop}%
\bibitem [{\citenamefont {Pons}\ \emph {et~al.}(1990)\citenamefont {Pons},
  \citenamefont {Lovey},\ and\ \citenamefont
  {Cesari}}]{Pons_ActaMetallMater1990}%
  \BibitemOpen
  \bibfield  {author} {\bibinfo {author} {\bibfnamefont {J.}~\bibnamefont
  {Pons}}, \bibinfo {author} {\bibfnamefont {F.}~\bibnamefont {Lovey}}, \ and\
  \bibinfo {author} {\bibfnamefont {E.}~\bibnamefont {Cesari}},\ }\href
  {\doibase 10.1016/0956-7151(90)90287-Q} {\bibfield  {journal} {\bibinfo
  {journal} {Acta Metall. Mater.}\ }\textbf {\bibinfo {volume} {38}},\ \bibinfo
  {pages} {2733} (\bibinfo {year} {1990})}\BibitemShut {NoStop}%
\bibitem [{\citenamefont {Simon}\ \emph {et~al.}(2010)\citenamefont {Simon},
  \citenamefont {Kr{\"{o}}ger}, \citenamefont {Somsen}, \citenamefont
  {Dlouhy},\ and\ \citenamefont {Eggeler}}]{Simon_ActaMater2010}%
  \BibitemOpen
  \bibfield  {author} {\bibinfo {author} {\bibfnamefont {T.}~\bibnamefont
  {Simon}}, \bibinfo {author} {\bibfnamefont {A.}~\bibnamefont {Kr{\"{o}}ger}},
  \bibinfo {author} {\bibfnamefont {C.}~\bibnamefont {Somsen}}, \bibinfo
  {author} {\bibfnamefont {A.}~\bibnamefont {Dlouhy}}, \ and\ \bibinfo {author}
  {\bibfnamefont {G.}~\bibnamefont {Eggeler}},\ }\href {\doibase
  10.1016/j.actamat.2009.11.028} {\bibfield  {journal} {\bibinfo  {journal}
  {Acta Mater.}\ }\textbf {\bibinfo {volume} {58}},\ \bibinfo {pages} {1850}
  (\bibinfo {year} {2010})}\BibitemShut {NoStop}%
\bibitem [{\citenamefont {Gil}\ and\ \citenamefont
  {Sornette}(1996)}]{Gil-Sornette_PRL1996}%
  \BibitemOpen
  \bibfield  {author} {\bibinfo {author} {\bibfnamefont {L.}~\bibnamefont
  {Gil}}\ and\ \bibinfo {author} {\bibfnamefont {D.}~\bibnamefont {Sornette}},\
  }\href {\doibase 10.1103/PhysRevLett.76.3991} {\bibfield  {journal} {\bibinfo
   {journal} {Phys. Rev. Lett.}\ }\textbf {\bibinfo {volume} {76}},\ \bibinfo
  {pages} {3991} (\bibinfo {year} {1996})}\BibitemShut {NoStop}%
\bibitem [{\citenamefont {Lin}\ \emph {et~al.}(2015)\citenamefont {Lin},
  \citenamefont {Gueudr{\'{e}}}, \citenamefont {Rosso},\ and\ \citenamefont
  {Wyart}}]{Lin-Wyart_PRL2015}%
  \BibitemOpen
  \bibfield  {author} {\bibinfo {author} {\bibfnamefont {J.}~\bibnamefont
  {Lin}}, \bibinfo {author} {\bibfnamefont {T.}~\bibnamefont {Gueudr{\'{e}}}},
  \bibinfo {author} {\bibfnamefont {A.}~\bibnamefont {Rosso}}, \ and\ \bibinfo
  {author} {\bibfnamefont {M.}~\bibnamefont {Wyart}},\ }\href {\doibase
  10.1103/PhysRevLett.115.168001} {\bibfield  {journal} {\bibinfo  {journal}
  {Phys. Rev. Lett.}\ }\textbf {\bibinfo {volume} {115}},\ \bibinfo {pages}
  {168001} (\bibinfo {year} {2015})}\BibitemShut {NoStop}%
\bibitem [{\citenamefont {Nakayama}\ \emph {et~al.}(2015)\citenamefont
  {Nakayama}, \citenamefont {Yoshino},\ and\ \citenamefont
  {Zamponi}}]{Nakayama_2015}%
  \BibitemOpen
  \bibfield  {author} {\bibinfo {author} {\bibfnamefont {D.}~\bibnamefont
  {Nakayama}}, \bibinfo {author} {\bibfnamefont {H.}~\bibnamefont {Yoshino}}, \
  and\ \bibinfo {author} {\bibfnamefont {F.}~\bibnamefont {Zamponi}},\ }\href
  {http://arxiv.org/abs/1512.06544} {\enquote {\bibinfo {title}
  {{Protocol-dependent shear modulus of amorphous solids}},}\ } (\bibinfo
  {year} {2015}),\ \Eprint {http://arxiv.org/abs/1512.06544} {arXiv:1512.06544}
  \BibitemShut {NoStop}%
\bibitem [{\citenamefont {Shenoy}\ \emph {et~al.}(2005)\citenamefont {Shenoy},
  \citenamefont {Lookman},\ and\ \citenamefont
  {Saxena}}]{Shenoy_BookBenasque2005}%
  \BibitemOpen
  \bibfield  {author} {\bibinfo {author} {\bibfnamefont {S.~R.}\ \bibnamefont
  {Shenoy}}, \bibinfo {author} {\bibfnamefont {T.}~\bibnamefont {Lookman}}, \
  and\ \bibinfo {author} {\bibfnamefont {A.}~\bibnamefont {Saxena}},\ }in\
  \href {\doibase 10.1007/3-540-31631-0_2} {\emph {\bibinfo {booktitle} {Magn.
  Struct. Funct. Mater.}}},\ \bibinfo {editor} {edited by\ \bibinfo {editor}
  {\bibfnamefont {A.}~\bibnamefont {Planes}}, \bibinfo {editor} {\bibfnamefont
  {L.}~\bibnamefont {Ma{\~{n}}osa}}, \ and\ \bibinfo {editor} {\bibfnamefont
  {A.}~\bibnamefont {Saxena}}}\ (\bibinfo  {publisher} {Springer Berlin
  Heidelberg},\ \bibinfo {year} {2005}),\ pp.\ \bibinfo {pages} {3--25}\BibitemShut {NoStop}%
\bibitem [{\citenamefont {Conti}\ and\ \citenamefont
  {Zanzotto}(2004)}]{Conti2004}%
  \BibitemOpen
  \bibfield  {author} {\bibinfo {author} {\bibfnamefont {S.}~\bibnamefont
  {Conti}}\ and\ \bibinfo {author} {\bibfnamefont {G.}~\bibnamefont
  {Zanzotto}},\ }\href {\doibase 10.1007/s00205-004-0311-z} {\bibfield
  {journal} {\bibinfo  {journal} {Arch. Ration. Mech. Anal.}\ }\textbf
  {\bibinfo {volume} {173}},\ \bibinfo {pages} {69} (\bibinfo {year}
  {2004})}\BibitemShut {NoStop}%
\bibitem [{\citenamefont {Jensen}(1995)}]{Jensen_JPhysA1995}%
  \BibitemOpen
  \bibfield  {author} {\bibinfo {author} {\bibfnamefont {H.~J.}\ \bibnamefont
  {Jensen}},\ }\href {http://stacks.iop.org/0305-4470/28/i=7/a=010} {\bibfield
  {journal} {\bibinfo  {journal} {J. Phys. A. Math. Gen.}\ }\textbf {\bibinfo
  {volume} {28}},\ \bibinfo {pages} {1861} (\bibinfo {year}
  {1995})}\BibitemShut {NoStop}%
\bibitem [{\citenamefont {Miguel}\ \emph {et~al.}(2002)\citenamefont {Miguel},
  \citenamefont {Vespignani}, \citenamefont {Zaiser},\ and\ \citenamefont
  {Zapperi}}]{Miguel2002}%
  \BibitemOpen
  \bibfield  {author} {\bibinfo {author} {\bibfnamefont {M.-C.}\ \bibnamefont
  {Miguel}}, \bibinfo {author} {\bibfnamefont {A.}~\bibnamefont {Vespignani}},
  \bibinfo {author} {\bibfnamefont {M.}~\bibnamefont {Zaiser}}, \ and\ \bibinfo
  {author} {\bibfnamefont {S.}~\bibnamefont {Zapperi}},\ }\href {\doibase
  10.1103/PhysRevLett.89.165501} {\bibfield  {journal} {\bibinfo  {journal}
  {Phys. Rev. Lett.}\ }\textbf {\bibinfo {volume} {89}},\ \bibinfo {pages}
  {165501} (\bibinfo {year} {2002})}\BibitemShut {NoStop}%
\bibitem [{\citenamefont {{M.-C. Miguel, A. Vespignani, S. Zapperi, J.
  Weiss}}(2001)}]{Miguel2001}%
  \BibitemOpen
  \bibfield  {author} {\bibinfo {author} {\bibfnamefont {M.-C.}\
  \bibnamefont {{Miguel, A. Vespignani, S. Zapperi, J. Weiss}}},\ }\href
  {http://www.nature.com/nature/journal/v410/n6829/full/410667a0.html}
  {\bibfield  {journal} {\bibinfo  {journal} {Nature}\ }\textbf {\bibinfo
  {volume} {410}},\ \bibinfo {pages} {667} (\bibinfo {year}
  {2001})}\BibitemShut {NoStop}%
\bibitem [{\citenamefont {Salman}\ and\ \citenamefont
  {Truskinovsky}(2011)}]{Salman-Truskinovsky_PRL2011}%
  \BibitemOpen
  \bibfield  {author} {\bibinfo {author} {\bibfnamefont {O.~U.}\ \bibnamefont
  {Salman}}\ and\ \bibinfo {author} {\bibfnamefont {L.}~\bibnamefont
  {Truskinovsky}},\ }\href {\doibase 10.1103/PhysRevLett.106.175503} {\bibfield
   {journal} {\bibinfo  {journal} {Phys. Rev. Lett.}\ }\textbf {\bibinfo
  {volume} {106}},\ \bibinfo {pages} {175503} (\bibinfo {year}
  {2011})}\BibitemShut {NoStop}%
\bibitem [{\citenamefont {Salman}\ and\ \citenamefont
  {Truskinovsky}(2012)}]{Salman-Truskinovsky_IntJEngSci2012}%
  \BibitemOpen
  \bibfield  {author} {\bibinfo {author} {\bibfnamefont {O.~U.}\ \bibnamefont
  {Salman}}\ and\ \bibinfo {author} {\bibfnamefont {L.}~\bibnamefont
  {Truskinovsky}},\ }\href {\doibase 10.1016/j.ijengsci.2012.03.012} {\bibfield
   {journal} {\bibinfo  {journal} {Int. J. Eng. Sci.}\ }\textbf {\bibinfo
  {volume} {59}},\ \bibinfo {pages} {219} (\bibinfo {year} {2012})},\ \Eprint
  {http://arxiv.org/abs/arXiv:1202.4753v1} {arXiv:arXiv:1202.4753v1}
  \BibitemShut {NoStop}%
\bibitem [{Note1()}]{Note1}%
  \BibitemOpen
  \bibinfo {note} {In general, a complete description of the deformation of a
  2D solid requires three strain variables~\cite
  {PitteriZanzotto2003,Kartha1995}.}\BibitemShut {Stop}%
\bibitem [{\citenamefont {Kartha}\ \emph {et~al.}(1995)\citenamefont {Kartha},
  \citenamefont {Krumhansl}, \citenamefont {Sethna},\ and\ \citenamefont
  {Wickham}}]{Kartha1995}%
  \BibitemOpen
  \bibfield  {author} {\bibinfo {author} {\bibfnamefont {S.}~\bibnamefont
  {Kartha}}, \bibinfo {author} {\bibfnamefont {J.~A.}\ \bibnamefont
  {Krumhansl}}, \bibinfo {author} {\bibfnamefont {J.~P.}\ \bibnamefont
  {Sethna}}, \ and\ \bibinfo {author} {\bibfnamefont {L.~K.}\ \bibnamefont
  {Wickham}},\ }\href {\doibase 10.1103/PhysRevB.52.803} {\bibfield  {journal}
  {\bibinfo  {journal} {Phys. Rev. B}\ }\textbf {\bibinfo {volume} {52}},\
  \bibinfo {pages} {803} (\bibinfo {year} {1995})}\BibitemShut {NoStop}%
\bibitem [{\citenamefont {Shenoy}\ and\ \citenamefont
  {Lookman}(2008)}]{Shenoy_PRB2008}%
  \BibitemOpen
  \bibfield  {author} {\bibinfo {author} {\bibfnamefont {S.~R.}\ \bibnamefont
  {Shenoy}}\ and\ \bibinfo {author} {\bibfnamefont {T.}~\bibnamefont
  {Lookman}},\ }\href {\doibase 10.1103/PhysRevB.78.144103} {\bibfield
  {journal} {\bibinfo  {journal} {Phys. Rev. B}\ }\textbf {\bibinfo {volume}
  {78}},\ \bibinfo {pages} {144103} (\bibinfo {year} {2008})}\BibitemShut
  {NoStop}%
\bibitem [{\citenamefont {Bhattacharya}\ \emph {et~al.}(2004)\citenamefont
  {Bhattacharya}, \citenamefont {Conti}, \citenamefont {Zanzotto},\ and\
  \citenamefont {Zimmer}}]{Bhattacharya2004}%
  \BibitemOpen
  \bibfield  {author} {\bibinfo {author} {\bibfnamefont {K.}~\bibnamefont
  {Bhattacharya}}, \bibinfo {author} {\bibfnamefont {S.}~\bibnamefont {Conti}},
  \bibinfo {author} {\bibfnamefont {G.}~\bibnamefont {Zanzotto}}, \ and\
  \bibinfo {author} {\bibfnamefont {J.}~\bibnamefont {Zimmer}},\ }\href
  {\doibase doi:10.1038/nature02378} {\bibfield  {journal} {\bibinfo  {journal}
  {Nature}\ }\textbf {\bibinfo {volume} {428}},\ \bibinfo {pages} {55}
  (\bibinfo {year} {2004})}\BibitemShut {NoStop}%
\bibitem [{\citenamefont {Gonz{\`{a}}lez-Comas}\ \emph
  {et~al.}(1997)\citenamefont {Gonz{\`{a}}lez-Comas}, \citenamefont
  {Ma{\~{n}}osa}, \citenamefont {Planes}, \citenamefont {Lovey}, \citenamefont
  {Pelegrina},\ and\ \citenamefont {Gu{\'{e}}nin}}]{Gonzalez-Comas1997}%
  \BibitemOpen
  \bibfield  {author} {\bibinfo {author} {\bibfnamefont {A.}~\bibnamefont
  {Gonz{\`{a}}lez-Comas}}, \bibinfo {author} {\bibfnamefont {L.}~\bibnamefont
  {Ma{\~{n}}osa}}, \bibinfo {author} {\bibfnamefont {A.}~\bibnamefont
  {Planes}}, \bibinfo {author} {\bibfnamefont {F.~C.}\ \bibnamefont {Lovey}},
  \bibinfo {author} {\bibfnamefont {J.~L.}\ \bibnamefont {Pelegrina}}, \ and\
  \bibinfo {author} {\bibfnamefont {G.}~\bibnamefont {Gu{\'{e}}nin}},\ }\href
  {\doibase 10.1103/PhysRevB.56.5200} {\bibfield  {journal} {\bibinfo
  {journal} {Phys. Rev. B}\ }\textbf {\bibinfo {volume} {56}},\ \bibinfo
  {pages} {5200} (\bibinfo {year} {1997})}\BibitemShut {NoStop}%
\bibitem [{\citenamefont {Lloveras}\ \emph {et~al.}(2008)\citenamefont
  {Lloveras}, \citenamefont {Cast{\'{a}}n}, \citenamefont {Porta},
  \citenamefont {Planes},\ and\ \citenamefont {Saxena}}]{Lloveras_PRL2008}%
  \BibitemOpen
  \bibfield  {author} {\bibinfo {author} {\bibfnamefont {P.}~\bibnamefont
  {Lloveras}}, \bibinfo {author} {\bibfnamefont {T.}~\bibnamefont
  {Cast{\'{a}}n}}, \bibinfo {author} {\bibfnamefont {M.}~\bibnamefont {Porta}},
  \bibinfo {author} {\bibfnamefont {A.}~\bibnamefont {Planes}}, \ and\ \bibinfo
  {author} {\bibfnamefont {A.}~\bibnamefont {Saxena}},\ }\href {\doibase
  10.1103/PhysRevLett.100.165707} {\bibfield  {journal} {\bibinfo  {journal}
  {Phys. Rev. Lett.}\ }\textbf {\bibinfo {volume} {100}},\ \bibinfo {pages}
  {165707} (\bibinfo {year} {2008})}\BibitemShut {NoStop}%
\bibitem [{\citenamefont {Ericksen}(1980)}]{Ericksen_ArchRationalMechAnal1980}%
  \BibitemOpen
  \bibfield  {author} {\bibinfo {author} {\bibfnamefont {J.~L.}\ \bibnamefont
  {Ericksen}},\ }\href {\doibase 10.1007/BF00258233} {\bibfield  {journal}
  {\bibinfo  {journal} {Arch. Ration. Mech. Anal.}\ }\textbf {\bibinfo {volume}
  {73}},\ \bibinfo {pages} {99} (\bibinfo {year} {1980})}\BibitemShut {NoStop}%
\bibitem [{\citenamefont {Pitteri}\ and\ \citenamefont
  {Zanzotto}(2003)}]{PitteriZanzotto2003}%
  \BibitemOpen
  \bibfield  {author} {\bibinfo {author} {\bibfnamefont {M.}~\bibnamefont
  {Pitteri}}\ and\ \bibinfo {author} {\bibfnamefont {G.}~\bibnamefont
  {Zanzotto}},\ }\href@noop {} {\emph {\bibinfo {title} {{Continuum Models for
  Phase Transitions and Twinning in Crystals}}}}\ (\bibinfo  {publisher}
  {Chapman {\&} Hall/CRC},\ \bibinfo {address} {Boca Raton},\ \bibinfo {year}
  {2003})\BibitemShut {NoStop}%
\bibitem [{\citenamefont {P{\'{e}}rez-Reche}\ \emph {et~al.}(2001)\citenamefont
  {P{\'{e}}rez-Reche}, \citenamefont {Vives}, \citenamefont {Ma{\~{n}}osa},\
  and\ \citenamefont {Planes}}]{PerezReche2001}%
  \BibitemOpen
  \bibfield  {author} {\bibinfo {author} {\bibfnamefont {F.~J.}\ \bibnamefont
  {P{\'{e}}rez-Reche}}, \bibinfo {author} {\bibfnamefont {E.}~\bibnamefont
  {Vives}}, \bibinfo {author} {\bibfnamefont {L.}~\bibnamefont {Ma{\~{n}}osa}},
  \ and\ \bibinfo {author} {\bibfnamefont {A.}~\bibnamefont {Planes}},\ }\href
  {\doibase 10.1103/PhysRevLett.87.195701} {\bibfield  {journal} {\bibinfo
  {journal} {Phys. Rev. Lett.}\ }\textbf {\bibinfo {volume} {87}},\ \bibinfo
  {pages} {195701} (\bibinfo {year} {2001})},\ \Eprint
  {http://arxiv.org/abs/0109019} {arXiv:0109019 [cond-mat]} \BibitemShut
  {NoStop}%
\bibitem [{\citenamefont {Puglisi}\ and\ \citenamefont
  {Truskinovsky}(2005)}]{Puglisi2005a}%
  \BibitemOpen
  \bibfield  {author} {\bibinfo {author} {\bibfnamefont {G.}~\bibnamefont
  {Puglisi}}\ and\ \bibinfo {author} {\bibfnamefont {L.}~\bibnamefont
  {Truskinovsky}},\ }\href {\doibase 10.1016/j.jmps.2004.08.004} {\bibfield
  {journal} {\bibinfo  {journal} {J. Mech. Phys. Solids}\ }\textbf {\bibinfo
  {volume} {53}},\ \bibinfo {pages} {655} (\bibinfo {year} {2005})}\BibitemShut
  {NoStop}%
\bibitem [{\citenamefont {Kuntz}\ \emph {et~al.}(1999)\citenamefont {Kuntz},
  \citenamefont {Perkovi{\'{c}}}, \citenamefont {Dahmen}, \citenamefont
  {Roberts},\ and\ \citenamefont {Sethna}}]{Kuntz1999}%
  \BibitemOpen
  \bibfield  {author} {\bibinfo {author} {\bibfnamefont {M.~C.}\ \bibnamefont
  {Kuntz}}, \bibinfo {author} {\bibfnamefont {O.}~\bibnamefont
  {Perkovi{\'{c}}}}, \bibinfo {author} {\bibfnamefont {K.~A.}\ \bibnamefont
  {Dahmen}}, \bibinfo {author} {\bibfnamefont {B.~W.}\ \bibnamefont {Roberts}},
  \ and\ \bibinfo {author} {\bibfnamefont {J.~P.}\ \bibnamefont {Sethna}},\
  }\href@noop {} {\bibfield  {journal} {\bibinfo  {journal} {Comput. Sci.
  Eng.}\ }\textbf {\bibinfo {volume} {1}},\ \bibinfo {pages} {73} (\bibinfo
  {year} {1999})}\BibitemShut {NoStop}%
\bibitem [{\citenamefont {Hinrichsen}(2000)}]{Hinrichsen_AdvPhys2000}%
  \BibitemOpen
  \bibfield  {author} {\bibinfo {author} {\bibfnamefont {H.}~\bibnamefont
  {Hinrichsen}},\ }\href {\doibase 10.1080/00018730050198152} {\bibfield
  {journal} {\bibinfo  {journal} {Adv. Phys.}\ }\textbf {\bibinfo {volume}
  {49}},\ \bibinfo {pages} {815} (\bibinfo {year} {2000})}\BibitemShut
  {NoStop}%
\bibitem [{\citenamefont {Picard}\ \emph {et~al.}(2004)\citenamefont {Picard},
  \citenamefont {Ajdari}, \citenamefont {Lequeux},\ and\ \citenamefont
  {Bocquet}}]{Picard_EPJE2004}%
  \BibitemOpen
  \bibfield  {author} {\bibinfo {author} {\bibfnamefont {G.}~\bibnamefont
  {Picard}}, \bibinfo {author} {\bibfnamefont {A.}~\bibnamefont {Ajdari}},
  \bibinfo {author} {\bibfnamefont {F.}~\bibnamefont {Lequeux}}, \ and\
  \bibinfo {author} {\bibfnamefont {L.}~\bibnamefont {Bocquet}},\ }\href
  {\doibase 10.1140/epje/i2004-10054-8} {\bibfield  {journal} {\bibinfo
  {journal} {Eur. Phys. J. E. Soft Matter}\ }\textbf {\bibinfo {volume} {15}},\
  \bibinfo {pages} {371} (\bibinfo {year} {2004})}\BibitemShut {NoStop}%
\bibitem [{\citenamefont {Talamali}\ \emph {et~al.}(2012)\citenamefont
  {Talamali}, \citenamefont {Pet{\"{a}}j{\"{a}}}, \citenamefont
  {Vandembroucq},\ and\ \citenamefont {Roux}}]{Talamali-Roux_CRMecanique2012}%
  \BibitemOpen
  \bibfield  {author} {\bibinfo {author} {\bibfnamefont {M.}~\bibnamefont
  {Talamali}}, \bibinfo {author} {\bibfnamefont {V.}~\bibnamefont
  {Pet{\"{a}}j{\"{a}}}}, \bibinfo {author} {\bibfnamefont {D.}~\bibnamefont
  {Vandembroucq}}, \ and\ \bibinfo {author} {\bibfnamefont {S.}~\bibnamefont
  {Roux}},\ }\href {\doibase http://dx.doi.org/10.1016/j.crme.2012.02.010}
  {\bibfield  {journal} {\bibinfo  {journal} {Comptes Rendus Mec.}\ }\textbf
  {\bibinfo {volume} {340}},\ \bibinfo {pages} {275} (\bibinfo {year}
  {2012})}\BibitemShut {NoStop}%
\bibitem [{\citenamefont {Kapetanou}\ \emph {et~al.}(2015)\citenamefont
  {Kapetanou}, \citenamefont {Weygand},\ and\ \citenamefont
  {Zaiser}}]{Kapetanou_JSTAT2015}%
  \BibitemOpen
  \bibfield  {author} {\bibinfo {author} {\bibfnamefont {O.}~\bibnamefont
  {Kapetanou}}, \bibinfo {author} {\bibfnamefont {D.}~\bibnamefont {Weygand}},
  \ and\ \bibinfo {author} {\bibfnamefont {M.}~\bibnamefont {Zaiser}},\ }\href
  {\doibase 10.1088/1742-5468/2015/08/P08009} {\bibfield  {journal} {\bibinfo
  {journal} {J. Stat. Mech. Theory Exp.}\ }\textbf {\bibinfo {volume} {2015}},\
  \bibinfo {pages} {P08009} (\bibinfo {year} {2015})}\BibitemShut {NoStop}%
\bibitem [{\citenamefont {Ren}\ and\ \citenamefont
  {Truskinovsky}(2000)}]{ren2000}%
  \BibitemOpen
  \bibfield  {author} {\bibinfo {author} {\bibfnamefont {X.}~\bibnamefont
  {Ren}}\ and\ \bibinfo {author} {\bibfnamefont {L.}~\bibnamefont
  {Truskinovsky}},\ }\href@noop {} {\bibfield  {journal} {\bibinfo  {journal}
  {J. Elast.}\ }\textbf {\bibinfo {volume} {59}},\ \bibinfo {pages} {319}
  (\bibinfo {year} {2000})}\BibitemShut {NoStop}%
\bibitem [{\citenamefont {Salje}(1993)}]{Salje1993}%
  \BibitemOpen
  \bibfield  {author} {\bibinfo {author} {\bibfnamefont {E.~K.~H.}\
  \bibnamefont {Salje}},\ }\href@noop {} {\emph {\bibinfo {title} {{Phase
  transitions in ferroelastic and co-elastic crystals}}}}\ (\bibinfo
  {publisher} {Cambridge University Press},\ \bibinfo {address} {Cambridge},\
  \bibinfo {year} {1993})\BibitemShut {NoStop}%
\bibitem [{\citenamefont {Selke}(1992)}]{Selke1992}%
  \BibitemOpen
  \bibfield  {author} {\bibinfo {author} {\bibfnamefont {W.}~\bibnamefont
  {Selke}},\ }\href@noop {} {\emph {\bibinfo {title} {{Spatially modulated
  structures in systems with competing interactions}}}},\ edited by\ \bibinfo
  {editor} {\bibfnamefont {C.}~\bibnamefont {Domb}}\ and\ \bibinfo {editor}
  {\bibfnamefont {J.~L.}\ \bibnamefont {Lebowitz}},\ \bibinfo {series} {Phase
  Transitions and Critical Phenomena}, Vol.~\bibinfo {volume} {15}\ (\bibinfo
  {publisher} {Academic Press},\ \bibinfo {address} {London},\ \bibinfo {year}
  {1992})\ pp.\ \bibinfo {pages} {1--72}\BibitemShut {NoStop}%
\bibitem [{\citenamefont {Balandraud}\ and\ \citenamefont
  {Zanzotto}(2007)}]{Balandraud_Zanzotto2007}%
  \BibitemOpen
  \bibfield  {author} {\bibinfo {author} {\bibfnamefont {X.}~\bibnamefont
  {Balandraud}}\ and\ \bibinfo {author} {\bibfnamefont {G.}~\bibnamefont
  {Zanzotto}},\ }\href {\doibase 10.1016/j.jmps.2006.03.009} {\bibfield
  {journal} {\bibinfo  {journal} {J. Mech. Phys. Solids}\ }\textbf {\bibinfo
  {volume} {55}},\ \bibinfo {pages} {194} (\bibinfo {year} {2007})}\BibitemShut
  {NoStop}%
\bibitem [{\citenamefont {T{\'{e}}l}\ and\ \citenamefont
  {Lai}(2008)}]{Tel-YingChengLai_PhysRep2008}%
  \BibitemOpen
  \bibfield  {author} {\bibinfo {author} {\bibfnamefont {T.}~\bibnamefont
  {T{\'{e}}l}}\ and\ \bibinfo {author} {\bibfnamefont {Y.-C.}\ \bibnamefont
  {Lai}},\ }\href {\doibase http://dx.doi.org/10.1016/j.physrep.2008.01.001}
  {\bibfield  {journal} {\bibinfo  {journal} {Phys. Rep.}\ }\textbf {\bibinfo
  {volume} {460}},\ \bibinfo {pages} {245} (\bibinfo {year}
  {2008})}\BibitemShut {NoStop}%
\bibitem [{\citenamefont {Politi}\ and\ \citenamefont
  {Torcini}(2010)}]{Politi-Torcini_StableChaos2010}%
  \BibitemOpen
  \bibfield  {author} {\bibinfo {author} {\bibfnamefont {A.}~\bibnamefont
  {Politi}}\ and\ \bibinfo {author} {\bibfnamefont {A.}~\bibnamefont
  {Torcini}},\ }in\ \href {\doibase 10.1007/978-3-642-04629-2} {\emph {\bibinfo
  {booktitle} {Nonlinear Dyn. Chaos Adv. Perspect.}}},\ \bibinfo {series and
  number} {Understanding Complex Systems},\ \bibinfo {editor} {edited by\
  \bibinfo {editor} {\bibfnamefont {M.}~\bibnamefont {Thiel}}, \bibinfo
  {editor} {\bibfnamefont {J.}~\bibnamefont {Kurths}}, \bibinfo {editor}
  {\bibfnamefont {M.~C.}\ \bibnamefont {Romano}}, \bibinfo {editor}
  {\bibfnamefont {G.}~\bibnamefont {K{\'{a}}rolyi}}, \ and\ \bibinfo {editor}
  {\bibfnamefont {A.}~\bibnamefont {Moura}}}\ (\bibinfo  {publisher} {Springer
  Berlin Heidelberg},\ \bibinfo {address} {Berlin, Heidelberg},\ \bibinfo
  {year} {2010})\ pp.\ \bibinfo {pages} {103--129}\BibitemShut {NoStop}%
\bibitem [{\citenamefont {Regev}\ \emph {et~al.}(2013)\citenamefont {Regev},
  \citenamefont {Lookman},\ and\ \citenamefont
  {Reichhardt}}]{Regev-Lookman-Charles_PRE2013}%
  \BibitemOpen
  \bibfield  {author} {\bibinfo {author} {\bibfnamefont {I.}~\bibnamefont
  {Regev}}, \bibinfo {author} {\bibfnamefont {T.}~\bibnamefont {Lookman}}, \
  and\ \bibinfo {author} {\bibfnamefont {C.}~\bibnamefont {Reichhardt}},\
  }\href {\doibase 10.1103/PhysRevE.88.062401} {\bibfield  {journal} {\bibinfo
  {journal} {Phys. Rev. E}\ }\textbf {\bibinfo {volume} {88}},\ \bibinfo
  {pages} {062401} (\bibinfo {year} {2013})},\ \Eprint
  {http://arxiv.org/abs/1301.7479} {arXiv:1301.7479} \BibitemShut {NoStop}%
\bibitem [{\citenamefont {Fiocco}\ \emph {et~al.}(2013)\citenamefont {Fiocco},
  \citenamefont {Foffi},\ and\ \citenamefont
  {Sastry}}]{Fiocco-Foffi-Sastry_PRE2013_CyclicGlass}%
  \BibitemOpen
  \bibfield  {author} {\bibinfo {author} {\bibfnamefont {D.}~\bibnamefont
  {Fiocco}}, \bibinfo {author} {\bibfnamefont {G.}~\bibnamefont {Foffi}}, \
  and\ \bibinfo {author} {\bibfnamefont {S.}~\bibnamefont {Sastry}},\ }\href
  {\doibase 10.1103/PhysRevE.88.020301} {\bibfield  {journal} {\bibinfo
  {journal} {Phys. Rev. E}\ }\textbf {\bibinfo {volume} {88}},\ \bibinfo
  {pages} {020301(R)} (\bibinfo {year} {2013})},\ \Eprint
  {http://arxiv.org/abs/1302.6518} {arXiv:1302.6518} \BibitemShut {NoStop}%
\bibitem [{\citenamefont {Hof}\ \emph {et~al.}(2006)\citenamefont {Hof},
  \citenamefont {Westerweel}, \citenamefont {Schneider},\ and\ \citenamefont
  {Eckhardt}}]{Hof2006}%
  \BibitemOpen
  \bibfield  {author} {\bibinfo {author} {\bibfnamefont {B.}~\bibnamefont
  {Hof}}, \bibinfo {author} {\bibfnamefont {J.}~\bibnamefont {Westerweel}},
  \bibinfo {author} {\bibfnamefont {T.~M.}\ \bibnamefont {Schneider}}, \ and\
  \bibinfo {author} {\bibfnamefont {B.}~\bibnamefont {Eckhardt}},\ }\href
  {\doibase 10.1038/nature05089} {\bibfield  {journal} {\bibinfo  {journal}
  {Nature}\ }\textbf {\bibinfo {volume} {443}},\ \bibinfo {pages} {59}
  (\bibinfo {year} {2006})}\BibitemShut {NoStop}%
\bibitem [{\citenamefont {Riley}\ \emph {et~al.}(2006)\citenamefont {Riley},
  \citenamefont {Hobson},\ and\ \citenamefont
  {Bence}}]{Riley_Hobson_Bence_MathsBook2006}%
  \BibitemOpen
  \bibfield  {author} {\bibinfo {author} {\bibfnamefont {K.~F.}\ \bibnamefont
  {Riley}}, \bibinfo {author} {\bibfnamefont {M.~P.}\ \bibnamefont {Hobson}}, \
  and\ \bibinfo {author} {\bibfnamefont {S.~J.}\ \bibnamefont {Bence}},\
  }\href@noop {} {\emph {\bibinfo {title} {{Mathematical Methods for Physics
  and Engineering}}}},\ \bibinfo {edition} {3rd}\ ed.\ (\bibinfo  {publisher}
  {Cambridge University Press},\ \bibinfo {year} {2006})\BibitemShut {NoStop}%
\bibitem [{\citenamefont {Gr{\"{o}}ger}\ \emph {et~al.}(2008)\citenamefont
  {Gr{\"{o}}ger}, \citenamefont {Lookman},\ and\ \citenamefont
  {Saxena}}]{Groger_PRB2008}%
  \BibitemOpen
  \bibfield  {author} {\bibinfo {author} {\bibfnamefont {R.}~\bibnamefont
  {Gr{\"{o}}ger}}, \bibinfo {author} {\bibfnamefont {T.}~\bibnamefont
  {Lookman}}, \ and\ \bibinfo {author} {\bibfnamefont {A.}~\bibnamefont
  {Saxena}},\ }\href {\doibase 10.1103/PhysRevB.78.184101} {\bibfield
  {journal} {\bibinfo  {journal} {Phys. Rev. B}\ }\textbf {\bibinfo {volume}
  {78}},\ \bibinfo {pages} {184101} (\bibinfo {year} {2008})}\BibitemShut
  {NoStop}%
\bibitem [{\citenamefont {Picard}\ \emph {et~al.}(2005)\citenamefont {Picard},
  \citenamefont {Ajdari}, \citenamefont {Lequeux},\ and\ \citenamefont
  {Bocquet}}]{Picard_PRE2005}%
  \BibitemOpen
  \bibfield  {author} {\bibinfo {author} {\bibfnamefont {G.}~\bibnamefont
  {Picard}}, \bibinfo {author} {\bibfnamefont {A.}~\bibnamefont {Ajdari}},
  \bibinfo {author} {\bibfnamefont {F.}~\bibnamefont {Lequeux}}, \ and\
  \bibinfo {author} {\bibfnamefont {L.}~\bibnamefont {Bocquet}},\ }\href
  {\doibase 10.1103/PhysRevE.71.010501} {\bibfield  {journal} {\bibinfo
  {journal} {Phys. Rev. E}\ }\textbf {\bibinfo {volume} {71}},\ \bibinfo
  {pages} {010501} (\bibinfo {year} {2005})}\BibitemShut {NoStop}%
\bibitem [{\citenamefont {Bar{\'{o}}}\ and\ \citenamefont
  {Vives}(2012)}]{Baro_PRE2012_PLML}%
  \BibitemOpen
  \bibfield  {author} {\bibinfo {author} {\bibfnamefont {J.}~\bibnamefont
  {Bar{\'{o}}}}\ and\ \bibinfo {author} {\bibfnamefont {E.}~\bibnamefont
  {Vives}},\ }\href {\doibase 10.1103/PhysRevE.85.066121} {\bibfield  {journal}
  {\bibinfo  {journal} {Phys. Rev. E}\ }\textbf {\bibinfo {volume} {85}},\
  \bibinfo {pages} {066121} (\bibinfo {year} {2012})}\BibitemShut {NoStop}%
\bibitem [{\citenamefont {Bray}(1986)}]{Bray_JPhysC1986_LongRangeRF}%
  \BibitemOpen
  \bibfield  {author} {\bibinfo {author} {\bibfnamefont {A.~J.}\ \bibnamefont
  {Bray}},\ }\href {http://stacks.iop.org/0022-3719/19/i=31/a=017} {\bibfield
  {journal} {\bibinfo  {journal} {J. Phys. C Solid State Phys.}\ }\textbf
  {\bibinfo {volume} {19}},\ \bibinfo {pages} {6225} (\bibinfo {year}
  {1986})}\BibitemShut {NoStop}%
\bibitem [{\citenamefont {Dahmen}\ \emph {et~al.}(2009)\citenamefont {Dahmen},
  \citenamefont {Ben-Zion},\ and\ \citenamefont {Uhl}}]{Dahmen2009}%
  \BibitemOpen
  \bibfield  {author} {\bibinfo {author} {\bibfnamefont {K.~A.}\ \bibnamefont
  {Dahmen}}, \bibinfo {author} {\bibfnamefont {Y.}~\bibnamefont {Ben-Zion}}, \
  and\ \bibinfo {author} {\bibfnamefont {J.~T.}\ \bibnamefont {Uhl}},\ }\href
  {\doibase 10.1103/PhysRevLett.102.175501} {\bibfield  {journal} {\bibinfo
  {journal} {Phys. Rev. Lett.}\ }\textbf {\bibinfo {volume} {102}},\ \bibinfo
  {pages} {175501} (\bibinfo {year} {2009})}\BibitemShut {NoStop}%
\bibitem [{\citenamefont {Sabhapandit}\ \emph {et~al.}(2000)\citenamefont
  {Sabhapandit}, \citenamefont {Shukla},\ and\ \citenamefont
  {Dhar}}]{Sabhapandit2000}%
  \BibitemOpen
  \bibfield  {author} {\bibinfo {author} {\bibfnamefont {S.}~\bibnamefont
  {Sabhapandit}}, \bibinfo {author} {\bibfnamefont {P.}~\bibnamefont {Shukla}},
  \ and\ \bibinfo {author} {\bibfnamefont {D.}~\bibnamefont {Dhar}},\ }\href
  {\doibase 10.1023/A:1018622805347} {\bibfield  {journal} {\bibinfo  {journal}
  {J. Stat. Phys.}\ }\textbf {\bibinfo {volume} {98}},\ \bibinfo {pages} {103}
  (\bibinfo {year} {2000})},\ \Eprint {http://arxiv.org/abs/9905236}
  {arXiv:9905236 [cond-mat]} \BibitemShut {NoStop}%
\bibitem [{\citenamefont {Br{\"{o}}ker}\ and\ \citenamefont
  {Grassberger}(1997)}]{Broker-Grassberger_PRE1997}%
  \BibitemOpen
  \bibfield  {author} {\bibinfo {author} {\bibfnamefont {H.-M.}\ \bibnamefont
  {Br{\"{o}}ker}}\ and\ \bibinfo {author} {\bibfnamefont {P.}~\bibnamefont
  {Grassberger}},\ }\href {\doibase 10.1103/PhysRevE.56.3944} {\bibfield
  {journal} {\bibinfo  {journal} {Phys. Rev. E}\ }\textbf {\bibinfo {volume}
  {56}},\ \bibinfo {pages} {3944} (\bibinfo {year} {1997})}\BibitemShut
  {NoStop}%
\bibitem [{\citenamefont {Kinouchi}\ and\ \citenamefont
  {Prado}(1999)}]{Kinouchi_PRE1999}%
  \BibitemOpen
  \bibfield  {author} {\bibinfo {author} {\bibfnamefont {O.}~\bibnamefont
  {Kinouchi}}\ and\ \bibinfo {author} {\bibfnamefont {C.~P.~C.}\ \bibnamefont
  {Prado}},\ }\href {\doibase 10.1103/PhysRevE.59.4964} {\bibfield  {journal}
  {\bibinfo  {journal} {Phys. Rev. E}\ }\textbf {\bibinfo {volume} {59}},\
  \bibinfo {pages} {4964} (\bibinfo {year} {1999})}\BibitemShut {NoStop}%
\bibitem [{\citenamefont {P{\'{a}}zm{\'{a}}ndi}\ \emph
  {et~al.}(1999)\citenamefont {P{\'{a}}zm{\'{a}}ndi}, \citenamefont
  {Zar{\'{a}}nd},\ and\ \citenamefont {Zim{\'{a}}nyi}}]{Pazmandy_PRL1999}%
  \BibitemOpen
  \bibfield  {author} {\bibinfo {author} {\bibfnamefont {F.}~\bibnamefont
  {P{\'{a}}zm{\'{a}}ndi}}, \bibinfo {author} {\bibfnamefont {G.}~\bibnamefont
  {Zar{\'{a}}nd}}, \ and\ \bibinfo {author} {\bibfnamefont {G.~T.}\
  \bibnamefont {Zim{\'{a}}nyi}},\ }\href {\doibase 10.1103/PhysRevLett.83.1034}
  {\bibfield  {journal} {\bibinfo  {journal} {Phys. Rev. Lett.}\ }\textbf
  {\bibinfo {volume} {83}},\ \bibinfo {pages} {1034} (\bibinfo {year}
  {1999})}\BibitemShut {NoStop}%
\bibitem [{\citenamefont {{Le Doussal}}\ \emph {et~al.}(2012)\citenamefont {{Le
  Doussal}}, \citenamefont {M{\"{u}}ller},\ and\ \citenamefont
  {Wiese}}]{LeDoussal_PRB2012_SK-Analytics}%
  \BibitemOpen
  \bibfield  {author} {\bibinfo {author} {\bibfnamefont {P.}~\bibnamefont {{Le
  Doussal}}}, \bibinfo {author} {\bibfnamefont {M.}~\bibnamefont
  {M{\"{u}}ller}}, \ and\ \bibinfo {author} {\bibfnamefont {K.~J.}\
  \bibnamefont {Wiese}},\ }\href {\doibase 10.1103/PhysRevB.85.214402}
  {\bibfield  {journal} {\bibinfo  {journal} {Phys. Rev. B}\ }\textbf {\bibinfo
  {volume} {85}},\ \bibinfo {pages} {214402} (\bibinfo {year}
  {2012})}\BibitemShut {NoStop}%
\bibitem [{\citenamefont {Talamali}\ \emph {et~al.}(2011)\citenamefont
  {Talamali}, \citenamefont {Pet{\"{a}}j{\"{a}}}, \citenamefont
  {Vandembroucq},\ and\ \citenamefont {Roux}}]{Talamali-Roux_PRE2011}%
  \BibitemOpen
  \bibfield  {author} {\bibinfo {author} {\bibfnamefont {M.}~\bibnamefont
  {Talamali}}, \bibinfo {author} {\bibfnamefont {V.}~\bibnamefont
  {Pet{\"{a}}j{\"{a}}}}, \bibinfo {author} {\bibfnamefont {D.}~\bibnamefont
  {Vandembroucq}}, \ and\ \bibinfo {author} {\bibfnamefont {S.}~\bibnamefont
  {Roux}},\ }\href {\doibase 10.1103/PhysRevE.84.016115} {\bibfield  {journal}
  {\bibinfo  {journal} {Phys. Rev. E}\ }\textbf {\bibinfo {volume} {84}},\
  \bibinfo {pages} {016115} (\bibinfo {year} {2011})}\BibitemShut {NoStop}%
\bibitem [{\citenamefont {Planes}\ \emph {et~al.}(2013)\citenamefont {Planes},
  \citenamefont {Ma{\~{n}}osa},\ and\ \citenamefont
  {Vives}}]{Planes_JAlloysCompounds2013}%
  \BibitemOpen
  \bibfield  {author} {\bibinfo {author} {\bibfnamefont {A.}~\bibnamefont
  {Planes}}, \bibinfo {author} {\bibfnamefont {L.}~\bibnamefont
  {Ma{\~{n}}osa}}, \ and\ \bibinfo {author} {\bibfnamefont {E.}~\bibnamefont
  {Vives}},\ }\href {\doibase 10.1016/j.jallcom.2011.10.082} {\bibfield
  {journal} {\bibinfo  {journal} {J. Alloys Compd.}\ }\textbf {\bibinfo
  {volume} {577}},\ \bibinfo {pages} {S699} (\bibinfo {year}
  {2013})}\BibitemShut {NoStop}%
\bibitem [{\citenamefont {Bonnot}\ \emph {et~al.}(2008)\citenamefont {Bonnot},
  \citenamefont {Ma{\~{n}}osa}, \citenamefont {Planes}, \citenamefont
  {Soto-Parra}, \citenamefont {Vives}, \citenamefont {Ludwig}, \citenamefont
  {Strothkaemper}, \citenamefont {Fukuda},\ and\ \citenamefont
  {Kakeshita}}]{Bonnot_PRB2008_AE-FePd}%
  \BibitemOpen
  \bibfield  {author} {\bibinfo {author} {\bibfnamefont {E.}~\bibnamefont
  {Bonnot}}, \bibinfo {author} {\bibfnamefont {L.}~\bibnamefont
  {Ma{\~{n}}osa}}, \bibinfo {author} {\bibfnamefont {A.}~\bibnamefont
  {Planes}}, \bibinfo {author} {\bibfnamefont {D.}~\bibnamefont {Soto-Parra}},
  \bibinfo {author} {\bibfnamefont {E.}~\bibnamefont {Vives}}, \bibinfo
  {author} {\bibfnamefont {B.}~\bibnamefont {Ludwig}}, \bibinfo {author}
  {\bibfnamefont {C.}~\bibnamefont {Strothkaemper}}, \bibinfo {author}
  {\bibfnamefont {T.}~\bibnamefont {Fukuda}}, \ and\ \bibinfo {author}
  {\bibfnamefont {T.}~\bibnamefont {Kakeshita}},\ }\href {\doibase
  10.1103/PhysRevB.78.184103} {\bibfield  {journal} {\bibinfo  {journal} {Phys.
  Rev. B}\ }\textbf {\bibinfo {volume} {78}},\ \bibinfo {pages} {184103}
  (\bibinfo {year} {2008})}\BibitemShut {NoStop}%
\bibitem [{\citenamefont {Lin}\ \emph {et~al.}(2014)\citenamefont {Lin},
  \citenamefont {Saade}, \citenamefont {Lerner}, \citenamefont {Rosso},\ and\
  \citenamefont {Wyart}}]{Lin-Wyart_EPL2014}%
  \BibitemOpen
  \bibfield  {author} {\bibinfo {author} {\bibfnamefont {J.}~\bibnamefont
  {Lin}}, \bibinfo {author} {\bibfnamefont {A.}~\bibnamefont {Saade}}, \bibinfo
  {author} {\bibfnamefont {E.}~\bibnamefont {Lerner}}, \bibinfo {author}
  {\bibfnamefont {A.}~\bibnamefont {Rosso}}, \ and\ \bibinfo {author}
  {\bibfnamefont {M.}~\bibnamefont {Wyart}},\ }\href {\doibase
  10.1209/0295-5075/105/26003} {\bibfield  {journal} {\bibinfo  {journal} {Europhys. Lett.}\ }\textbf {\bibinfo {volume} {105}},\ \bibinfo {pages}
  {26003} (\bibinfo {year} {2014})},\ \Eprint {http://arxiv.org/abs/1307.1646}
  {arXiv:1307.1646} \BibitemShut {NoStop}%
\bibitem [{\citenamefont {{Le Doussal}}\ \emph {et~al.}(2010)\citenamefont {{Le
  Doussal}}, \citenamefont {M{\"{u}}ller},\ and\ \citenamefont
  {Wiese}}]{LeDoussal-Muller-Wiese_EPL2010}%
  \BibitemOpen
  \bibfield  {author} {\bibinfo {author} {\bibfnamefont {P.}~\bibnamefont {{Le
  Doussal}}}, \bibinfo {author} {\bibfnamefont {M.}~\bibnamefont
  {M{\"{u}}ller}}, \ and\ \bibinfo {author} {\bibfnamefont {K.~J.}\
  \bibnamefont {Wiese}},\ }\href {\doibase 10.1209/0295-5075/91/57004}
  {\bibfield  {journal} {\bibinfo  {journal} {Europhys. Lett.}\ }\textbf
  {\bibinfo {volume} {91}},\ \bibinfo {pages} {57004} (\bibinfo {year}
  {2010})}\BibitemShut {NoStop}%
\bibitem [{\citenamefont {Papanikolaou}\ \emph {et~al.}(2012)\citenamefont
  {Papanikolaou}, \citenamefont {Dimiduk}, \citenamefont {Choi}, \citenamefont
  {Sethna}, \citenamefont {Uchic}, \citenamefont {Woodward},\ and\
  \citenamefont {Zapperi}}]{Papanikolaou_Nature2012}%
  \BibitemOpen
  \bibfield  {author} {\bibinfo {author} {\bibfnamefont {S.}~\bibnamefont
  {Papanikolaou}}, \bibinfo {author} {\bibfnamefont {D.~M.}\ \bibnamefont
  {Dimiduk}}, \bibinfo {author} {\bibfnamefont {W.}~\bibnamefont {Choi}},
  \bibinfo {author} {\bibfnamefont {J.~P.}\ \bibnamefont {Sethna}}, \bibinfo
  {author} {\bibfnamefont {M.~D.}\ \bibnamefont {Uchic}}, \bibinfo {author}
  {\bibfnamefont {C.~F.}\ \bibnamefont {Woodward}}, \ and\ \bibinfo {author}
  {\bibfnamefont {S.}~\bibnamefont {Zapperi}},\ }\href
  {http://dx.doi.org/10.1038/nature11568} {\bibfield  {journal} {\bibinfo
  {journal} {Nature}\ }\textbf {\bibinfo {volume} {490}},\ \bibinfo {pages}
  {517} (\bibinfo {year} {2012})}\BibitemShut {NoStop}%
\bibitem [{\citenamefont {Andersen}\ \emph {et~al.}(1997)\citenamefont
  {Andersen}, \citenamefont {Sornette},\ and\ \citenamefont
  {Leung}}]{Andersen1997}%
  \BibitemOpen
  \bibfield  {author} {\bibinfo {author} {\bibfnamefont {J.~V.}\ \bibnamefont
  {Andersen}}, \bibinfo {author} {\bibfnamefont {D.}~\bibnamefont {Sornette}},
  \ and\ \bibinfo {author} {\bibfnamefont {K.-t.}\ \bibnamefont {Leung}},\
  }\href {\doibase 10.1103/PhysRevLett.78.2140} {\bibfield  {journal} {\bibinfo
   {journal} {Phys. Rev. Lett.}\ }\textbf {\bibinfo {volume} {78}},\ \bibinfo
  {pages} {2140} (\bibinfo {year} {1997})}\BibitemShut {NoStop}%
\end{thebibliography}

%

\end{document}